\documentclass{vgtc}                          

\graphicspath{{figures/}{pictures/}{images/}{./}} 

\usepackage{wrapfig}

\usepackage{times}                     

\usepackage{booktabs}                  
\usepackage{lipsum}                    
\usepackage{mwe}                       
\usepackage{stfloats}
\usepackage{listings}
\usepackage{graphicx}
\usepackage{comment}
\usepackage[caption=false,font=footnotesize]{subfig}
\usepackage{pdflscape}
\usepackage[toc,page]{appendix}
\usepackage{placeins}
\usepackage{amsmath}

\usepackage{tabularx}
\usepackage{longtable}
\usepackage{array} 

\usepackage{ragged2e}
\newcolumntype{Y}{>{\RaggedRight\arraybackslash}X}
\newcommand{\eg}{e.g.,\ }
\newcommand{\ie}{i.e.,\xspace}

\newcommand{\QArow}[4]{%
  #1 & #2 & #3 & \includegraphics[height=15mm,keepaspectratio]{#4}%
  \\%
}

\usepackage{mathptmx}                  

\onlineid{0}

\vgtccategory{Research}

\vgtcinsertpkg

\title{
GPT-5 Model Corrected GPT-4V's Chart Reading Errors, Not Prompting
}

\author{Kaichun Yang\thanks{e-mail: ky27@illinois.edu}\\ %
        \scriptsize University of Illinois Urbana-Champaign %
\and Jian Chen\thanks{e-mail: chen.8028@osu.edu}\\ %
     \scriptsize The Ohio State University %
}

\abstract{
We present a quantitative evaluation to understand the effect of zero-shot large-language model (LLMs) and prompting uses on chart reading tasks.
We 
asked LLMs to answer 107 visualization questions to compare  inference accuracies between the agentic GPT-5 and multimodal GPT-4V,
for 
\textit{difficult} image instances, where GPT-4V failed to produce correct answers. 
Our results show that model architecture dominates the inference accuracy: GPT-5 largely improved accuracy, while prompt variants yielded only small
effects. Pre-registration of this work is available \href{https://osf.io/u78td/?view_only=6b075584311f48e991c39335c840ded3}{here}; 
the Google Drive materials are \href{https://drive.google.com/file/d/1ll8WWZDf7cCNcfNWrLViWt8GwDNSvVrp/view?usp=drive_link}{here}.
} 


\begin{document}


\firstsection{Introduction}

\maketitle

Benchmarking visual literacy, \ie ``the ability and skill to read and interpret visually represented data and to extract information from data visualizations''~\cite{Lee:2017:Vlat} shapes progress in measuring AI's ability in handling visualization images. Often, 
the same tasks as designed to assess visual literacy questions traditionally performed by human observers are now being assigned to algorithms.

    
    Following this trend, our goal in this paper is to quantify the new GPT-5's ability to read charts. 
    Specifically, we 
    used 
    questions where GPT-4V
    failed and other LLMs achieved only low accuracy, as reported in Verma et al.'s CHART-6 benchmark~\cite{Verma:2025:CHART6}. 
    

    We also controlled prompt useage and compared LLM performance across (i) CHART-6 instruction, (ii) question-only (instruction removed), and (iii) instruction replaced by a GPT-5 chart description.

    Our results showed that model choice dominates: GPT-5 consistently reduces error relative to GPT-4o, whereas prompt variants (question-only and chart-description) produce small and inconsistent changes with no reliable main effects; in some cases (\eg GPT-5 with chart description) performance even declines. 
    Thus, for chart understanding on the selected CHART-6 subset, 
    GPT-5's agentic reasoning  does seem to outperform the multimodel GPT-4 family for popular chart readings.

\section{Related Work}
This section briefly describe works that have inspired our experiment. 

\subsection{Quantifying Observers by Visualization Literacy}

Research on human-level chart comprehension has mainly focused on three primary approaches for evaluating LLMs.
First, visual literacy (\eg Graph Literacy(GGR)~\cite{Galesic:2011:Graph} and Visualization Literacy Assessment Test (VLAT)~\cite{Lee:2017:Vlat}), which are commonly used to quantify human chart-reading abilities have also been 
employed to quantify unsupervised zero-shot LLMs.
The second line of work
emphasized the detection of so-called ``VisLies'', or misleading visual design 
(\eg Critical Thinking Assessment for Literacy in Visualizations (CALVI) and HOLF~\cite{Ge:2023:Calvi,Huey:2023:Communicative}) to measure whether models can be easily fooled.
Finally, a third approach is to 
conduct parallel experiments on both humans and 
LLMs to compare their differences to study the gap~\cite{Verma:2025:CHART6} or consistency~\cite{jiang2025rigorous} between observers. 
For example, CHART-6 organizes multiple types of chart-based questions and procedures for model evaluation~\cite{Verma:2025:CHART6}, showing that LLMs still exhibited a substantial accuracy gap compared to human participants.

\subsection{Quantifying LLM's Prompt Design}

Existing work shows that the description of a task, aka prompt engineering can significantly affect model behavior prior to GPT-5.  
To GPT-4V, concrete prompting strategies exhibit notable yet task dependent effects: adding a short reasoning cue such as ``Let’s think step by step'' can trigger zero shot chain of thought and markedly improve multi step reasoning accuracy on arithmetic and logic benchmarks~\cite{Kojima:2022:ZeroShotCoT}. Sampling multiple reasoning paths and selecting a self-consistent answer further improves performance on the same families of tasks \cite{Wang:2022:SelfConsistency}. Overall, prior work indicates that long and sometimes unnatural language in prompt design 
specifying task goals and output formats, can elicit intermediate reasoning, and reduce ambiguity, therefore improve LLMs prior to GPT-5. These factors help extraction style and structured reasoning tasks, although the gains vary with task complexity and the level of distraction.

In contrast,
GPT-5, the new agentic reasoning model, only need  simple, direct instructions to perform task-oriented workflow~\cite{openai:o1-preview,openai:reasoning}. 
In particular, the GPT-5 system card describes a unified system with ``a smart and fast model for everyday tasks, a deeper reasoning model for harder problems, and a real-time router that quickly decides which model to use … (including when you say `think hard about this' in the prompt)''~\cite{openai:gpt5-systemcard}. Taken together with the Agents and prompting guidance~\cite{openai:agents,openai:textgen}, this indicates that concise, task-focused instructions are preferable for reasoning models, such as GPT-5 whereas verbose prompts can be unnecessary.

\section{Methodology}

This section reports the main contribution of this paper: answering how LLMs respond when we instructed them differently and whether or not the most recent LLM implementations (GPT-5 and GPT-4o) are better than GPT-4V, for the most difficult problems reported 
in Verma et al.'s CHART-6~\cite{Verma:2025:CHART6}. 

\subsection{Dataset and Sampling}

We use 107 questions selected from the publicly available 
CHART-6 benchmark~\cite{Verma:2025:CHART6}.
We identified questions within the CHART-6 benchmark, on which GPT-4V produced incorrect responses, and other tested models (Blip2-FlanT5-4B, Blip2-FlanT5-11B, LLaVA1.5-Vicuna-7B, LLaVA1.5-Vicuna-13B, LLaVA1.6-Yi-34B, Pix2Struct-0.3B, MatCha-0.3B) showed particularly low accuracy~\cite{Verma:2025:CHART6}.
We used GPT-4V because this model achieved the highest mean accuracy in CHART-6 benchmark. 
We first filtered all question–image pairs where GPT-4V’s correctness was 0 across \textit{all} trials. 
We then ranked these items by 
mean accuracy across 
all questions for each model 
reported in the dataset and selected a total of 107 questions across the five datasets.
This curated subset serves as the evaluation set in the present study.

\autoref{tab:subset_selection} has the selection criteria and example questions.
\autoref{app:107_questions} contains the full question list.

\begin{table*}[t]
    \centering
    \caption{Our task selection criteria, for a set of 107  difficult questions, chosen from Verma et al.~\cite{Verma:2025:CHART6}.}
    \small
    \setlength{\tabcolsep}{6pt}
    \begin{tabularx}{\textwidth}{l>{\raggedright\arraybackslash}X c >{\raggedright\arraybackslash}X}
    \toprule
    \textbf{Dataset} & \textbf{Selection rule} & \textbf{Selected/Total} & \textbf{Example Question \& Image} \\
    \midrule
    GGR        & Excluded 2 partial-correct questions of GPT-4V, kept remaining
               & 11/13
               & What is the percentage of cancer patients who die after chemotherapy? Image: \autoref{fig:ggr_item13} \\
    VLAT       & Ordered by decreasing accuracy of other models
               & 11/53
               & What was the unemployment rate for Indiana (IN) in 2015? Image: \autoref{fig:Vlat_item55} \\
    CALVI      & Top 20 from trick, top 5 from standard
               & 25/60
               & Does cell phone brand A have more than half of the total market share? Image: \autoref{fig:CALVI_T28} \\
    HOLF       & 5 lowest-accuracy images each with 6 questions
               & 30/384
               & On average, how many miles per hour faster was Hurricane Alberto compared to Hurricane Claudette? Image: \autoref{fig:holf_bar20} \\
    HOLF-Multi & 4 bars, 3 lines, 3 scatters (lowest accuracy in each types)
               & 30/216
               & What is the average number of fatalities in northeast + midwest states with populations of 1M-1.5M people? Image: \autoref{fig:holf2_bar1} \\
    \bottomrule
    \end{tabularx}
    \label{tab:subset_selection}
\end{table*}

\begin{figure*}[t]
    \centering
    \resizebox{1\textwidth}{!}{
        \setlength{\tabcolsep}{6pt}
        \begin{tabular}{cccc}
            \subfloat[Bar chart in GGR\label{fig:ggr_item13}]{
              \includegraphics[width=0.24\textwidth]{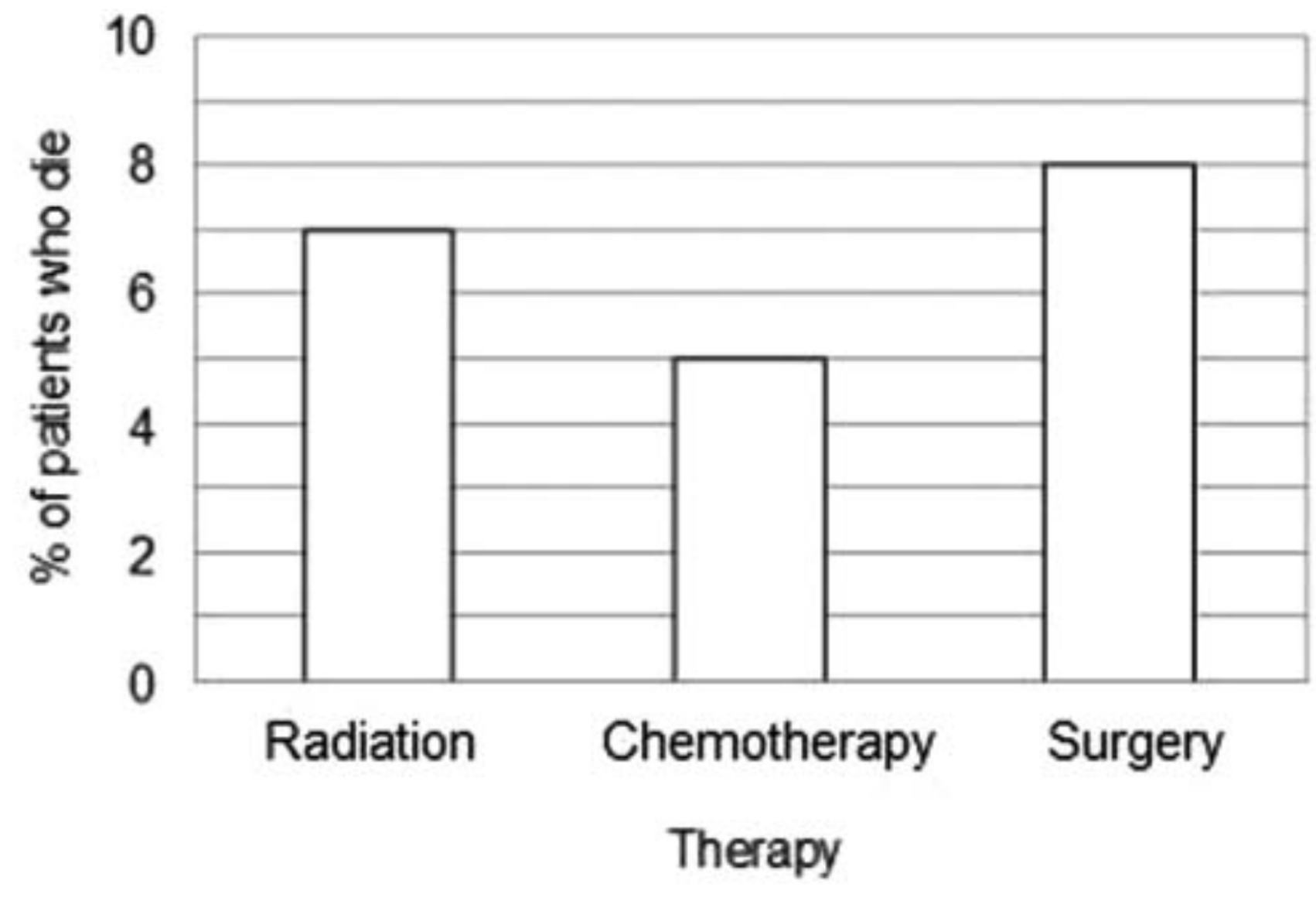}
            } &
            \subfloat[Choropleth map in VLAT\label{fig:Vlat_item55}]{
              \includegraphics[width=0.24\textwidth]{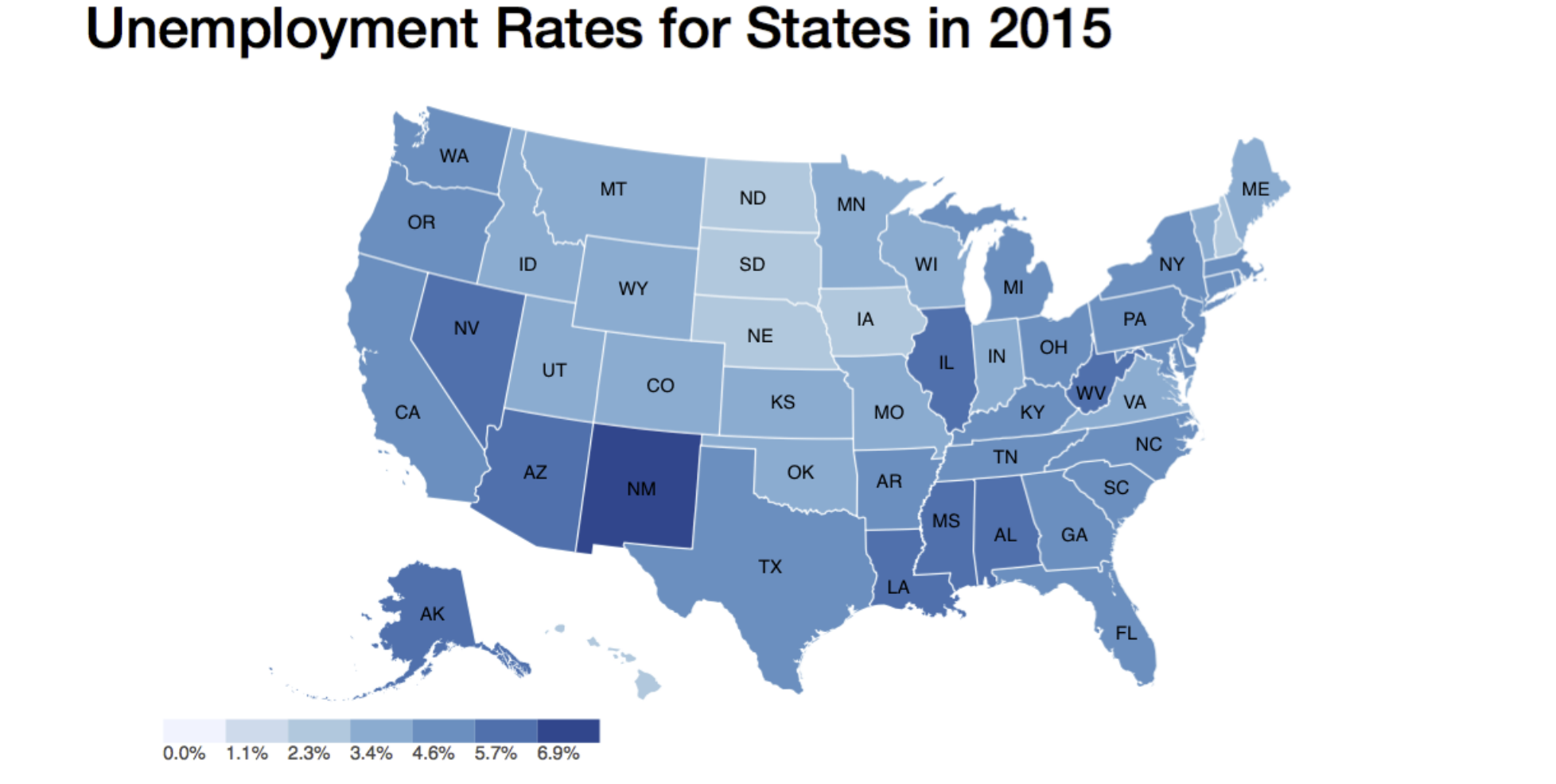}
            } &
            \subfloat[Pie chart in CALVI\label{fig:CALVI_T28}]{
              \includegraphics[width=0.24\textwidth]{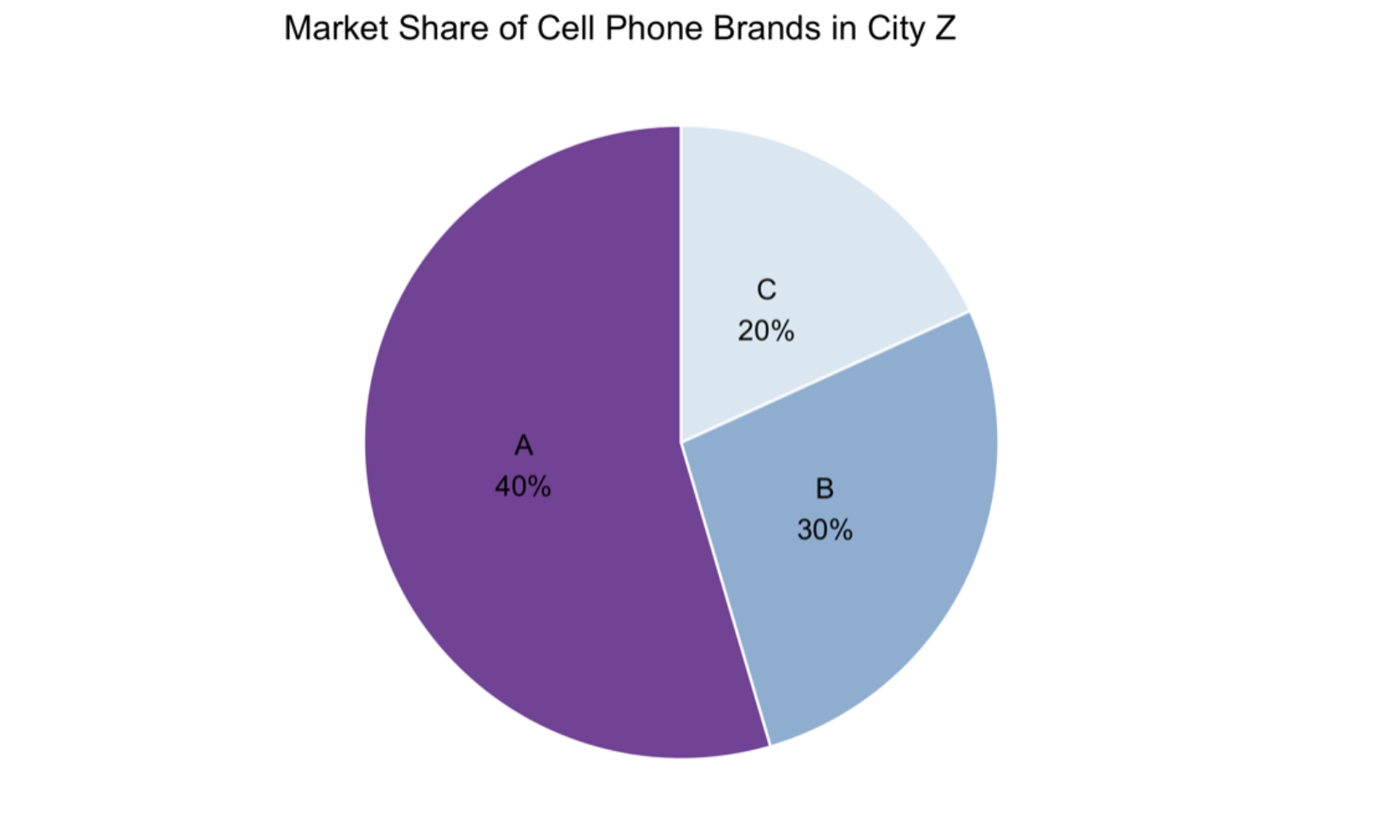}
            } &
            \subfloat[Bar chart in HOLF\label{fig:holf_bar20}]{
              \includegraphics[width=0.24\textwidth]{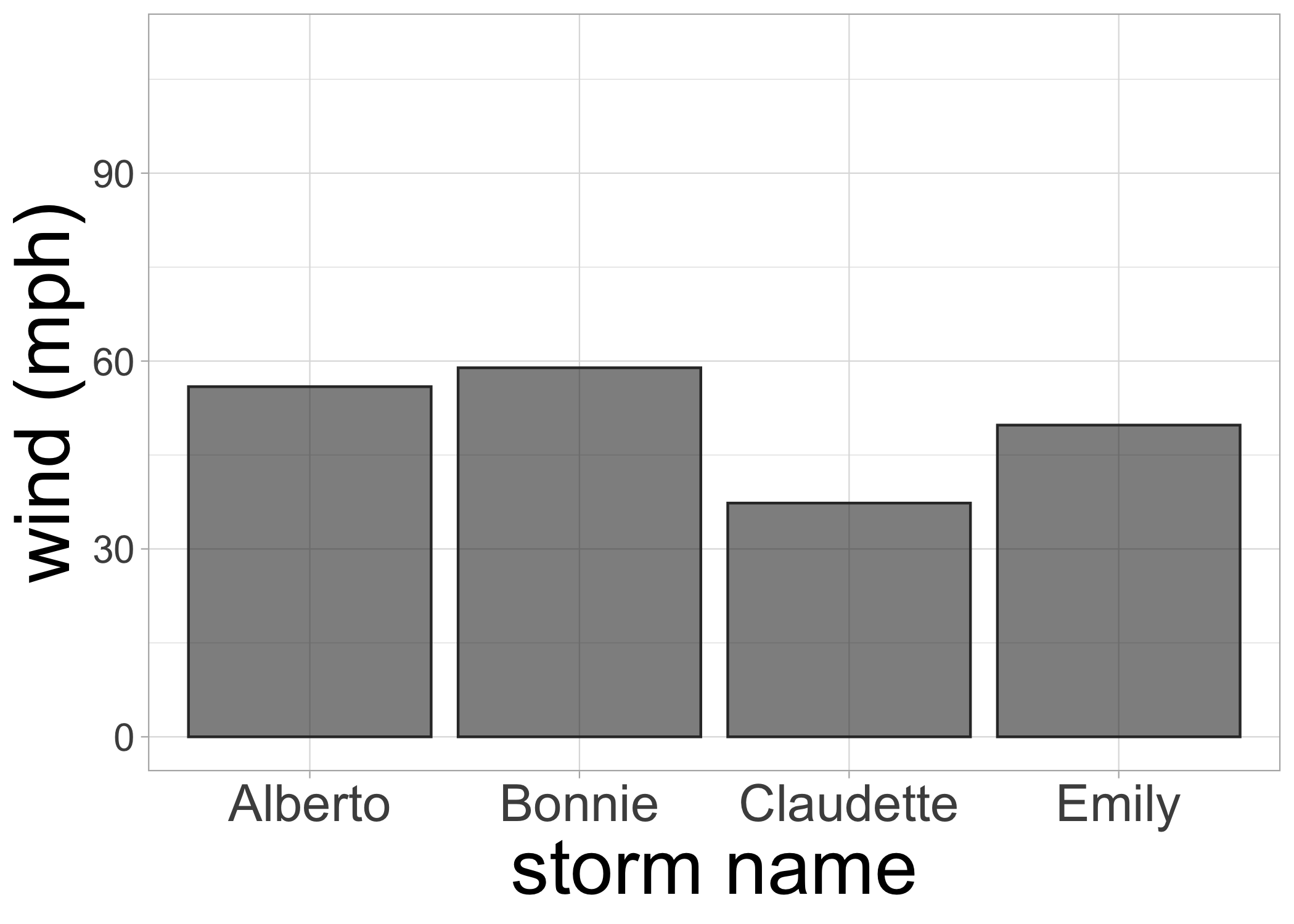}
            }
        \end{tabular}
    }
    
    \vspace{5pt}
    
    \resizebox{0.85\textwidth}{!}{
        \setlength{\tabcolsep}{8pt}
        \begin{tabular}{ccc}
            \subfloat[Bar chart in HOLF-Multi\label{fig:holf2_bar1}]{
              \includegraphics[width=0.30\textwidth]{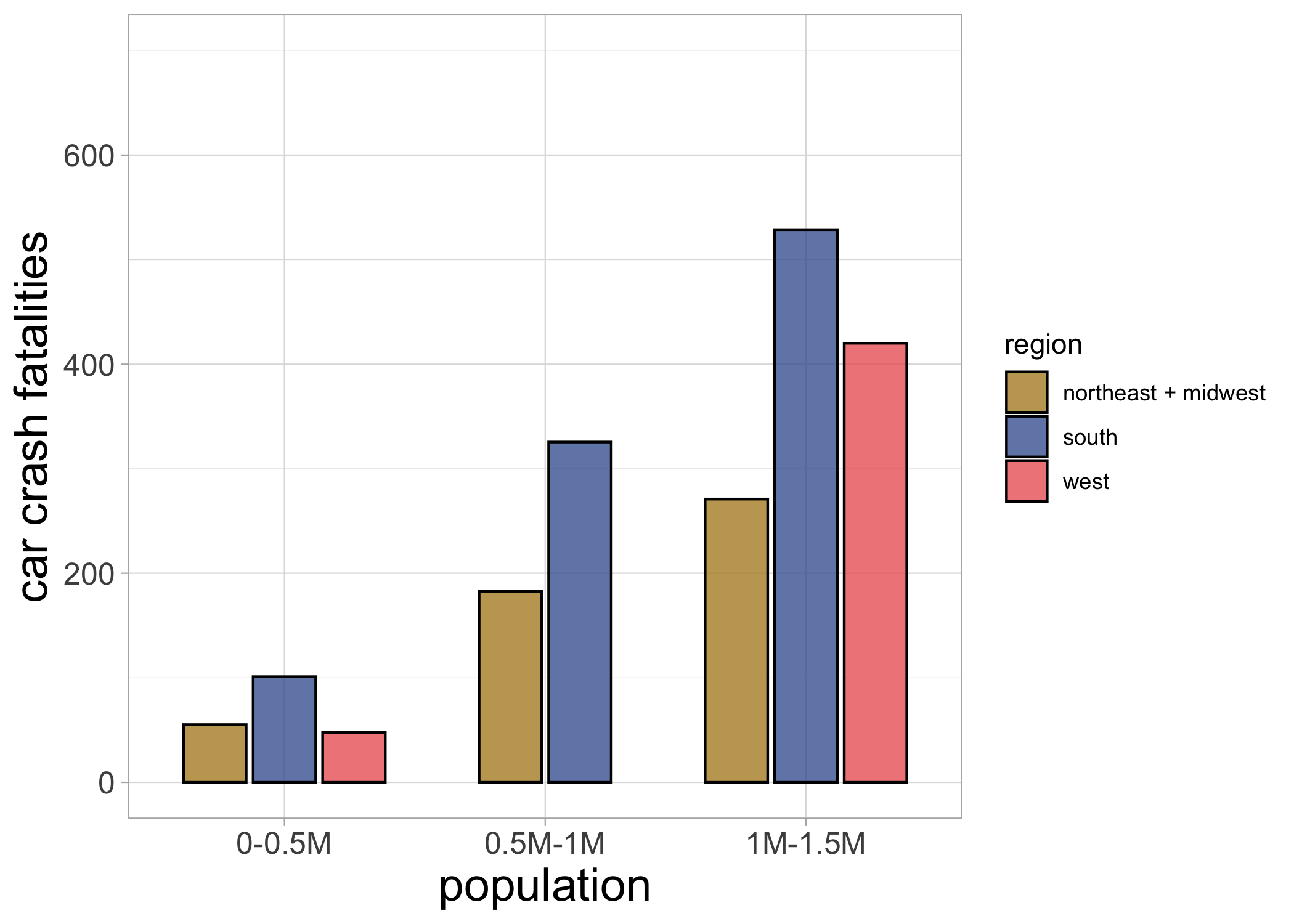}
            } &
            \subfloat[Line chart in HOLF-Multi\label{fig:holf2_line1}]{
              \includegraphics[width=0.30\textwidth]{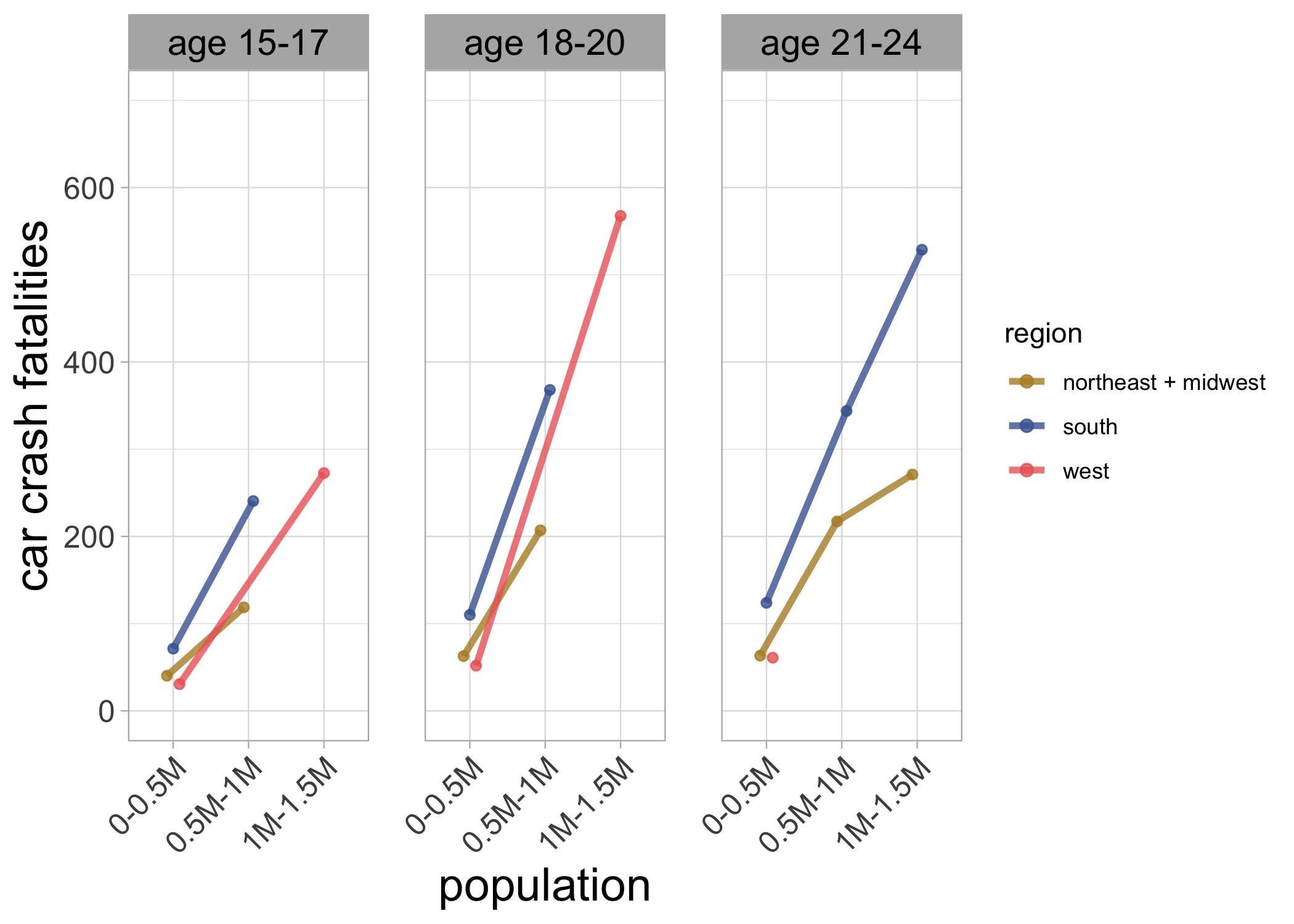}
            } &
            \subfloat[Scatter plot in HOLF-Multi\label{fig:holf2_scatter1}]{
              \includegraphics[width=0.30\textwidth]{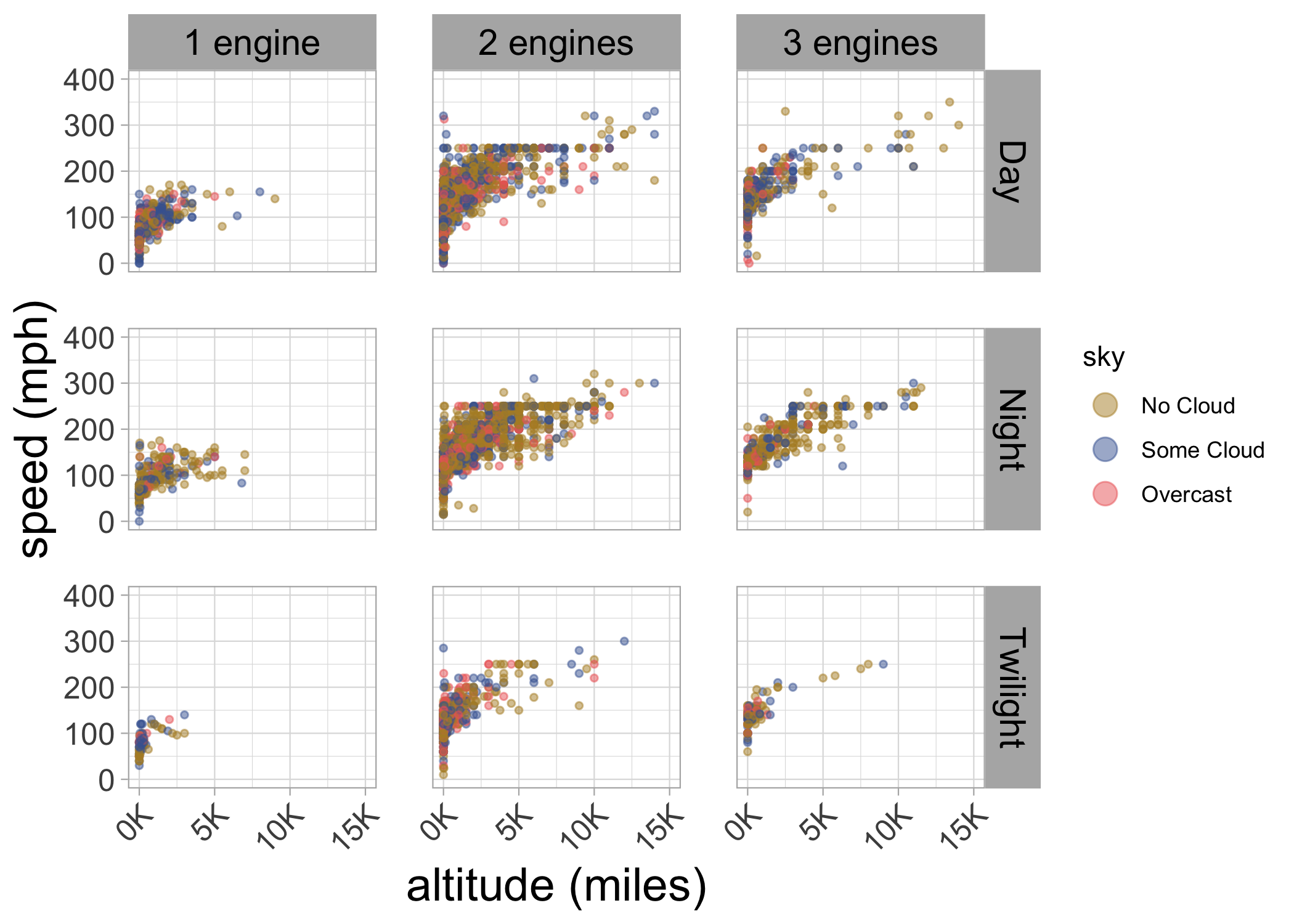}
            }
        \end{tabular}
    }
    \caption{Selected image samples 
    in our experiment. 
    }
    \label{fig:holf2_eg}
\end{figure*}

\subsection{Independent variables}
We had two independent variables: LLM models and prompt. 
\subsubsection{LLMs}
We used three levels: GPT-5, GPT-4o and GPT-4V. 
We first examined the results of CHART-6~\cite{Verma:2025:CHART6} and observed that the GPT-4V model achieved the highest accuracy across most datasets. 
However, GPT-the GPT-4 Turbo snapshot gpt-4-1106-vision-preview used in CHART-6 has been deprecated, we replaced it 
with the closest available snapshot, gpt-4o-2024-05-13.
In addition, we also used the latest GPT-5 agentic reasoning model, version gpt-5-2025-08-07, as the control group for GPT-4o to observe the impact of improvements in LLM performance on accuracy.


The configuration of 
GPT-4o and GPT-5 are the following:
GPT-4o used temperature 0.2 and top-p 1.0 matching GPT-4V; GPT-5 used its default reasoning settings. To avoid truncation on long numeric items (HOLF/HOLF-Multi), we set the maximum output length up to 3000 tokens. 

\begin{table}[t]
\centering
\caption{Independent 
and
dependent variables.
}
\small
\begin{tabular}{p{0.42\linewidth} p{0.52\linewidth}}
\toprule
\textbf{Independent variables} & \textbf{Dependent variables} \\
\midrule
Model: \{GPT-4V, GPT-4o, GPT-5\} & Correctness (0/1)\,–\,for GGR, VLAT, CALVI \\
Prompt condition: \{CHART-6 instruction, Question-only, GPT-5 description\} & 
LRAE\,for HOLF, HOLF-Multi\\
\bottomrule
\end{tabular}
\label{tab:iv-dv}
\end{table}

\subsubsection{Prompting Conditions}
We had three levels of prompt useage.   
Our baseline prompt is Verma et al.'s CHART-6~\cite{Verma:2025:CHART6}, which
consisted the combination of ''\textit{general task instructions}'' and the task question. An example of general task instructions from CHART-6: "The first part of this study consists of 13 multiple choice questions associated with visualizations, and you will be asked to choose the best answer for each question. You are required to provide an answer to the current question. Your answer must be one of the choices provided." In order to investigate whether the general task instructions contribute substantially to model performance, 
we designed two additional prompting conditions: 
(1) 
a Question-only setting, where the general task instructions were removed, and 
(2) a description-augmented setting, where the general task instructions were replaced with a paragraph describing the chart, generated by GPT-5 using a designed prompt. An example of chart description of \autoref{fig:ggr_item13}: "Bar chart with no title. Axes: x-axis labeled “Therapy” with categories “Radiation,” “Chemotherapy,” and “Surgery”; y-axis labeled “\% of patients who de.” Marks: three vertical bars with white fill and black outlines. Legend: none. Gridlines \& ticks: horizontal gridlines span the plot area; tick marks are present. Overall styling: monochrome, simple layout with bold category labels."

\paragraph{Generation of Graph Description}
We used GPT-5 to generate exclusive descriptions for each chart, presenting its visual structure and elements such as chart type, title, axis, legend, and style. It 
is worth noting
that this prompt explicitly prohibits GPT-5 from generating any information related to values or establishing semantic connections between different chart elements. This ensures that the generated description only depicts the appearance features of the chart, without providing additional hints for the question itself.

\subsection{Experimental Setup and Data Analysis}

We evaluated six combinations in total, consisting of two LLMs (GPT-4o and GPT-5) under three prompting conditions: (i) generic task instruction from Verma et al.~\cite{Verma:2025:CHART6}, (ii) 
no task instruction, use question only, (iii) replace task instruction with GPT5-generated chart description. 
For each question in the dataset, we submitted five repeated queries and recorded the model responses.

After we get the response, we 
followed Verma et al.~\cite{Verma:2025:CHART6}
to first clean the outputs by removing auxiliary text and irrelevant symbols. For numerical questions, only the extracted floating-point values were retained; for questions with string answer, the output was normalized to a string that strictly matches the standard answer format. For multiple choice questions, if the cleaned output was not in the options, the response will be changed to \texttt{NaN}.



\section{Results}

\begin{figure*}[!t]
    \centering
    \includegraphics[width=0.9\textwidth]{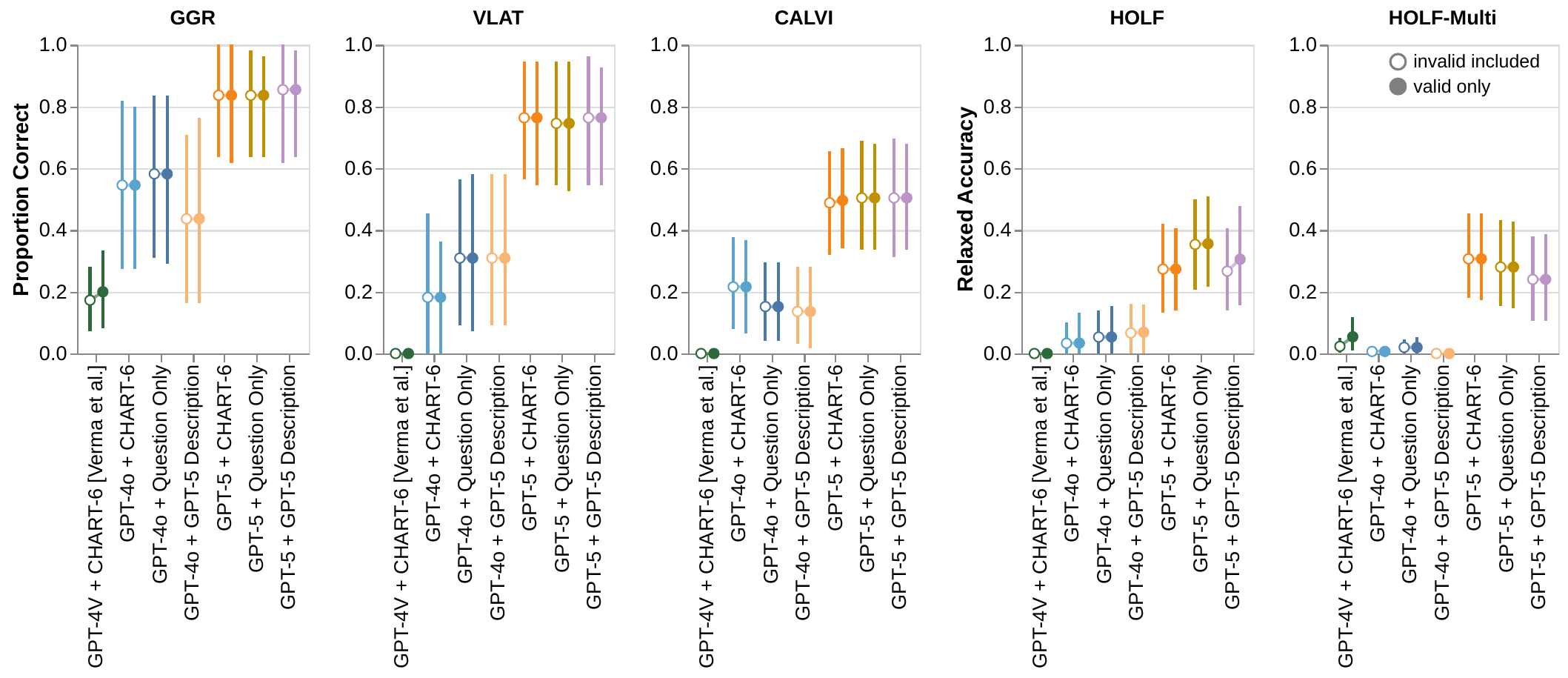}
    \caption{Performance comparison results. 
    }
    \label{fig:CI_graph}
\end{figure*}

Our core finding is GPT-5 is much better than GPT-4o across all datasets, while differences between prompting conditions are smaller. ~\autoref{fig:CI_graph} summarizes mean accuracy with 95\% CIs for all six experimental conditions and the baseline Verma et al.'s CHART-6 (GPT-4V+CHART-6).


\paragraph{Statistical analysis}
For \textbf{GGR}, \textbf{VLAT}, and \textbf{CALVI}, we treat each model output as \emph{correct/incorrect} and report accuracy. For \textbf{HOLF} and \textbf{HOLF-Multi}, we follow CHART-6 and report \emph{relaxed accuracy}, counting a response as correct if it is within $\pm 5\%$ of the ground truth. For model-based inference on numeric responses, we also compute the \emph{log-ratio absolute error (LRAE)}:
\[
\mathrm{LRAE} \;=\; \bigl| \log(\hat{y}) - \log(y) \bigr|\,,
\]
and assign invalid outputs the maximum observed error within the dataset to keep scales comparable.

\paragraph{Confidence intervals.}
We form \emph{item-level bootstrap} confidence intervals for accuracy. In each bootstrap replicate, we resample items with replacement (the same number of unique items as in the original set) and, for every resampled item, keep all five trials per condition. We then recompute accuracy. Repeating this $B=1000$ times yields the empirical distribution of the mean; the 2.5th and 97.5th percentiles give the 95\% CI.

\paragraph{Hypothesis tests / regression.}
To assess the effects of \emph{model} (GPT-4o vs.\ GPT-5), \emph{prompt condition} (three levels), and their interaction while accounting for repeated measures per item:
\begin{itemize}
    \setlength{\itemsep}{0pt}
    \item For the \textbf{binary outcomes} (GGR, VLAT, CALVI relaxed-accuracy indicators), we fit \emph{GEE} with a \emph{binomial family} and \emph{logit link}, clustering by \emph{item} and using an \emph{exchangeable} using cluster-robust standard errors. We report \emph{log-odds coefficients} with robust standard errors.
    \item For the \textbf{numeric error} datasets (HOLF, HOLF-Multi), we fit a \emph{LMM} on LRAE, with fixed effects for model, prompt, and their interaction, and a random intercept for item to capture item-level heterogeneity.
\end{itemize}

\subsection{Results}
\paragraph{GPT-5 outperformed GPT-4o and GPT-4V}
We observed that GPT-5 consistently outperformed GPT-4o across all the datasets, confirming that model capability was an important factor in visualization question answering. The average performance gap between GPT-5 and GPT-4o was approximately from 20 to 40 percentage points. In contrast, differences between prompting conditions were relatively small. 

\paragraph{Prompts with chart descriptions did not provide benefits over baseline or question-only prompts, and in some cases the question-only condition performed slightly better than baseline.} These results indicate that prompt modifications can only marginally affect performance and are negligible compared to the improvements brought by different LLMs. 

\paragraph{Performance by Dataset}
On the simpler datasets, GGR and VLAT, both GPT-5 and GPT-4o achieved performance far beyond that of the earlier GPT-4V. GPT-5 reached over 80\% accuracy on GGR and nearly 80\% on VLAT. Notably, GPT-4o showed a decrease in accuracy on GGR under the description of chart prompting condition.
    

\begin{wrapfigure}[11]{r}{0.3\columnwidth}
    \centering
    \vspace{-14pt}
    \hspace{-15pt}
    \begin{minipage}{0.3\columnwidth}
        \centering
        \includegraphics[width=\linewidth]{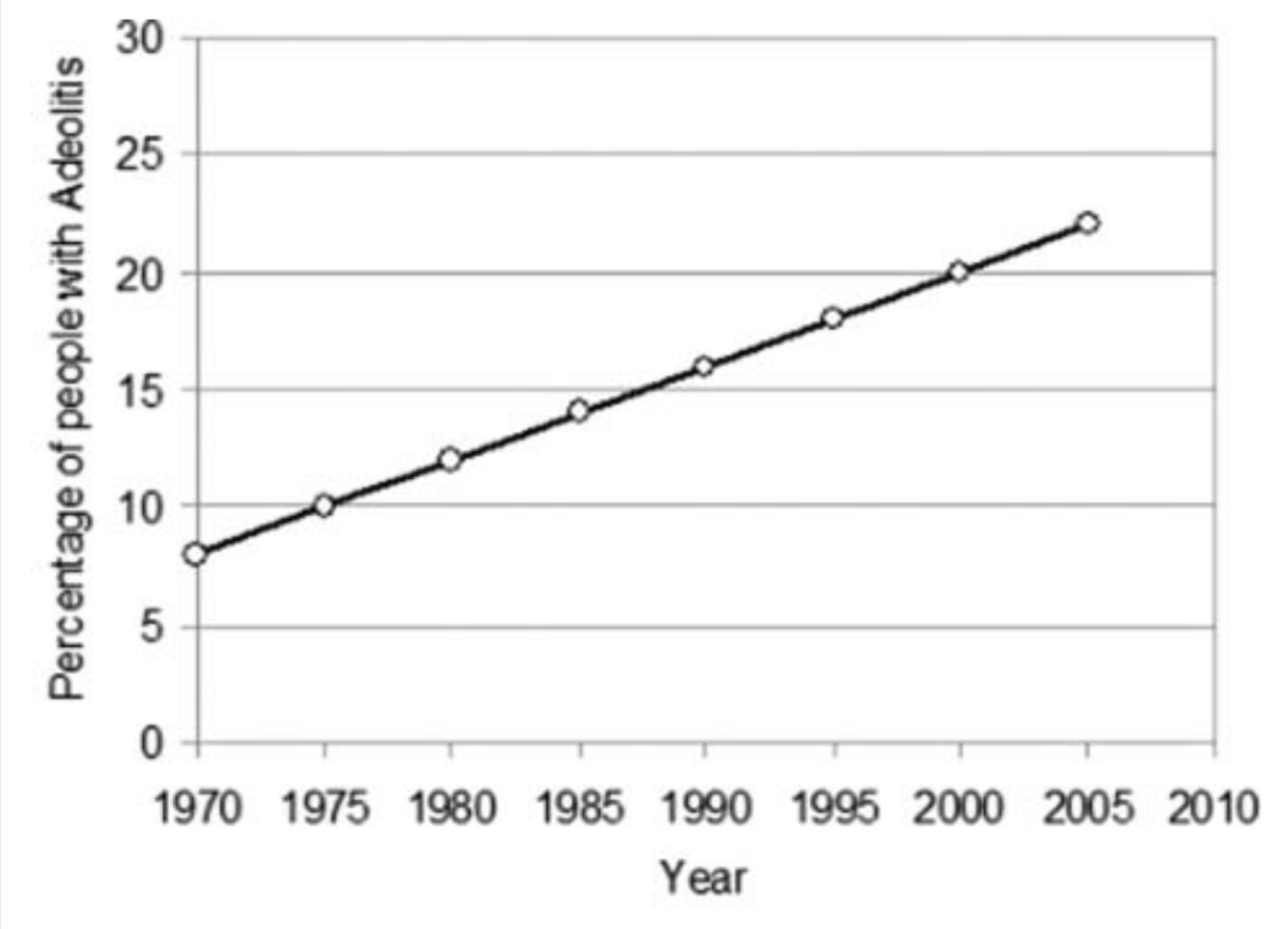}\\ [4pt]
        \includegraphics[width=\linewidth]{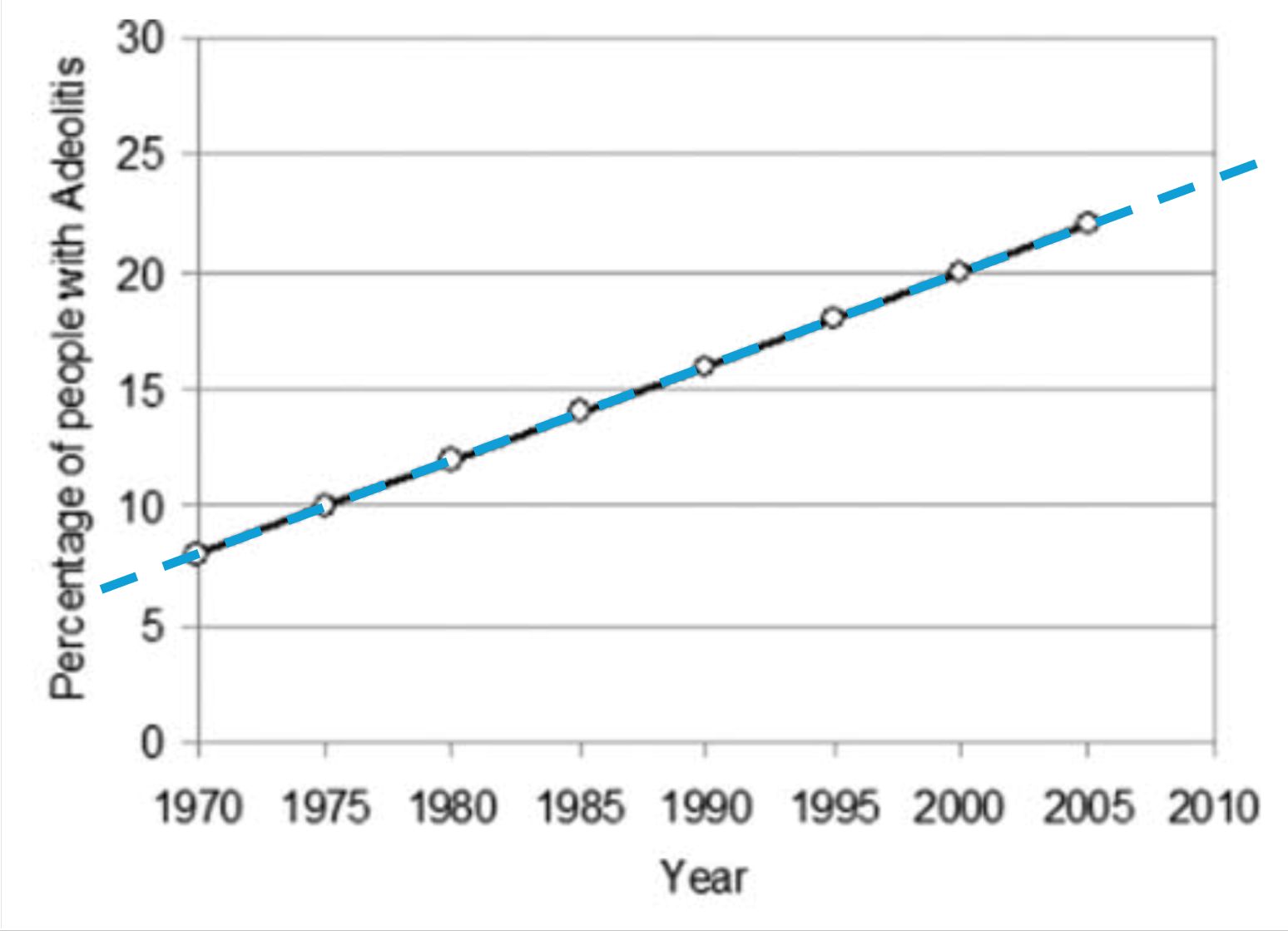}
    \end{minipage}
\end{wrapfigure}

After case-by-case investigation, 
we found that 
the only question that GPT-5 consistently failed in GGR was a consequence of its precision. The question from GGR dataset: ''\textit{Approximately what percentage of people who die from cancer die from colon cancer, breast cancer, and prostate cancer taken together?}'' (\textit{top figure}).
The correct answer is 25, but GPT-5 gave the answer 24, which more closely matches the value one would obtain by extending the bar with a reference line (\textit{bottom figure}).

On CALVI and HOLF, the models generally performed poorly. GPT-5 still achieved much higher accuracy than GPT-4o and GPT-4V, while GPT-4o, except for performance on HOLF-Multi which was similar to GPT-4V, still achieved higher accuracy than GPT-4V. It is worth noting that on the more complex HOLF dataset, long chart descriptions almost always had a negative effect, with accuracy much lower than when only the question itself was used.

\paragraph{Effects of Model and Prompting Condition}

\begin{figure}[!t]
    \centering
    \includegraphics[width=\columnwidth]{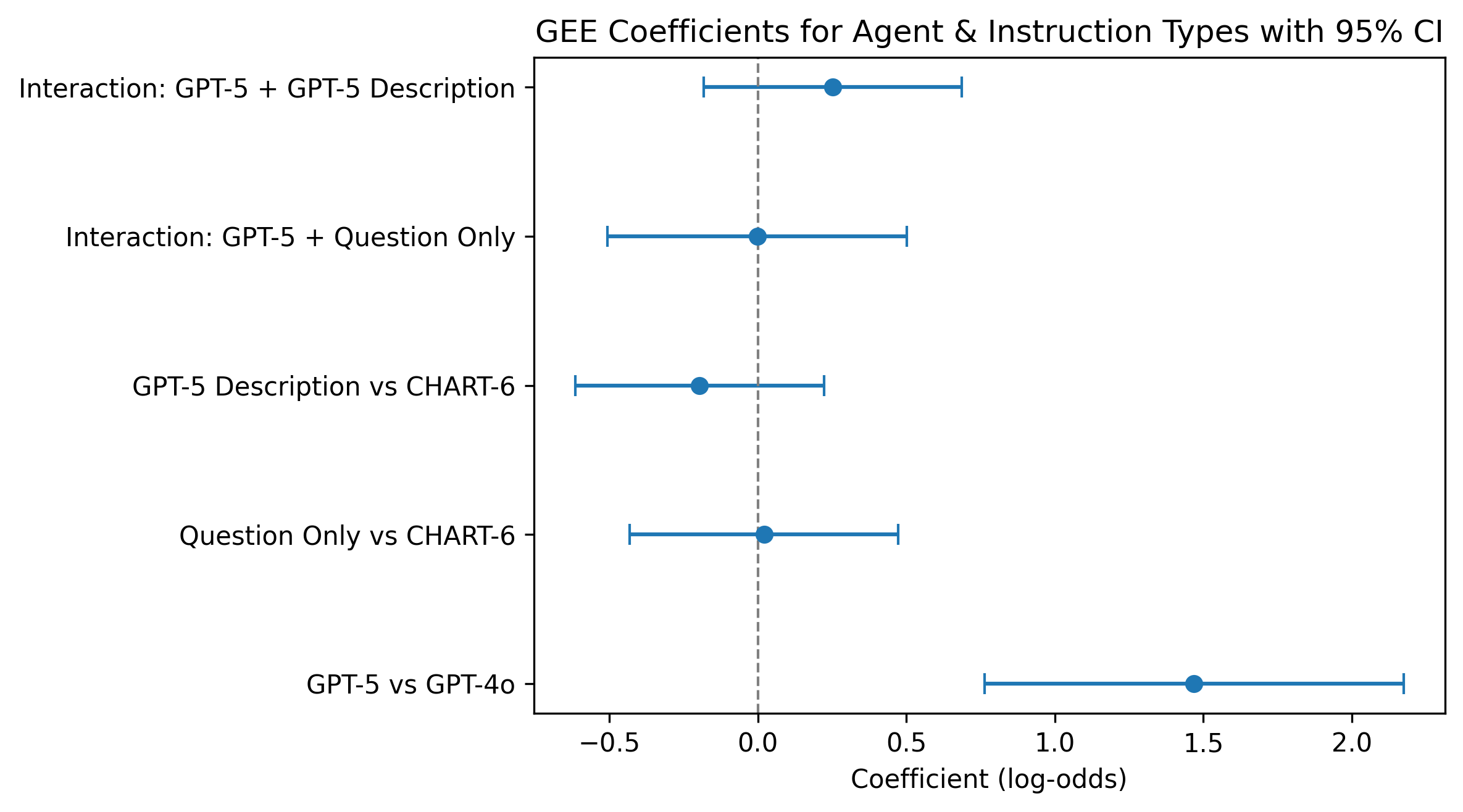}
    \caption{GEE estimated coefficients (log-odds) with 95\% confidence intervals for model and prompt effects across multiple-choice datasets(GGR, VLAT, CALVI).}
    \label{fig:gee_coef}
\end{figure}

We used odd-ratio test statistics of correctness
and found that model type is  
a significant main effect~(\autoref{fig:gee_coef}). GPT-5 had a large positive 
coefficient compared to GPT-4o, indicating higher odds of producing correct responses ($\beta = 1.4689,\ p < 0.001$). In contrast, the coefficients for prompting conditions Question Only vs. CHART-6 ($\beta = 0.0208, \ p = 0.928$) and GPT-5 Description vs. CHART-6 instruction ($\beta = -0.1964, \ p = 0.358$) were small.

\begin{figure}[!t]
    \centering
    \includegraphics[width=\columnwidth]{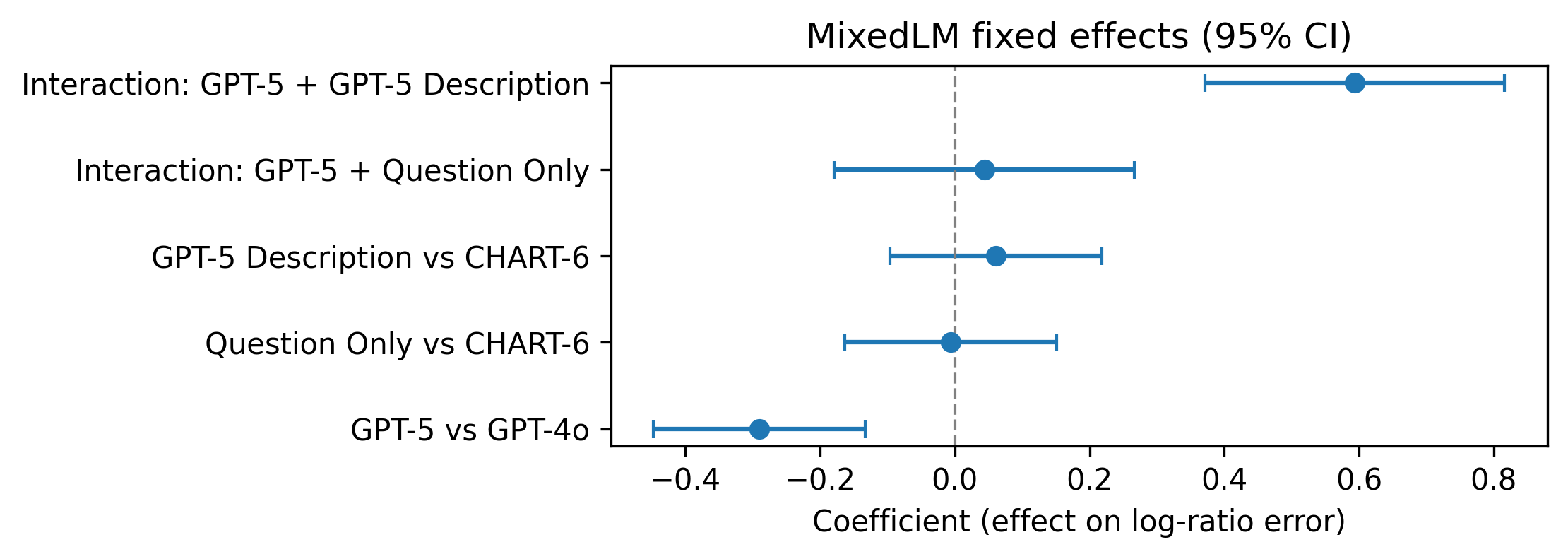}
    \caption{LMM fixed-effect coefficients with 95\% confidence intervals for model and prompt effects across continuous-response datasets (HOLF and HOLF-Multi).}
    \label{fig:coef_LMM}
\end{figure}

To analyze performance on the continuous-response datasets: HOLF and HOLF-Multi, we fitted a linear mixed-effects model (LMM) with random intercepts for each item, which accounts for the repeated responses associated with the same question and shared variance within items. The dependent variable was the log-ratio absolute error 
which quantifies the deviation of a model’s numeric response from the correct answer on a logarithmic scale. Smaller 
log-errors 
indicate higher accuracy.
Invalid responses were replaced with the maximum observed error within the corresponding dataset to ensure comparability across agents and prompts~\cite{Bates:2015:Fitting}.

We analyzed the fixed effects.
~\autoref{fig:coef_LMM} shows that model has a strong effect on LRAE. GPT-5 achieved significantly lower LRAE compared to GPT-4o ($\beta=-0.244$, $p<0.001$). Prompting conditions Question Only vs. CHART-6 ($\beta = 1.4689, \ p=0.671$) and GPT-5 Description vs. CHART-6 ($\beta = -0.009, \ p=0.824$) had no significant main effect. However, we observed a significant positive interaction between GPT-5 with GPT-5 Description condition ($\beta=0.219$, $p<0.001$), indicating that GPT-5’s error increased when combined with GPT-5 descriptions.

\section{Discussion}

The most striking finding is that GPT-5 greatly outperformed GPT-4V for some difficult tasks in Verma et al.~\cite{Verma:2025:CHART6}, 
a  paper being in its preprint stage, 
which highlights the rapid advances of the AI field. This outcome may further highlight that evaluation focusing on behavior quantification~\cite{jiang2025rigorous, geirhos2018imagenettrained} may be more important than relying solely on benchmark performance, if our goal is to explain the model inference accuracy. 

We expected that adding chart descriptions would improve accuracy, since the descriptions will give more information of the charts to the LLM. However, our exploratory experiment here
did not support this hypothesis. 
Across all datasets, the accuracy differences between prompting conditions were relatively small, and prompts with GPT-5 self-descriptions did not 
improve compared to baseline or question-only prompts. 

LLM
was a significant main effect. 
GPT-5 consistently outperformed GPT-4o, with nearly 80\% accuracy on simple questions in datasets such as GGR and VLAT. In contrast, both models still faced difficulties on more complex datasets such as CALVI and HOLF.
It seems that recent agent-based model GPT-5 could have greatly improved inference accuracy at least for the few tasks we quantified here. 
Further testing on other datasets and reasoning tasks is necessary to generalize its performance, for example, for the simple value retrievals in Bendeck and Stasko (they used GPT-4)~\cite{bendeck2024empirical}, for the data interpretation in Hong et al.~\cite{hong2025llms}, and for the bar chart takeaways in Wang et al.~\cite{wang2024aligned}.



\textbf{Mechanistic evaluation beyond benchmarking.}
(i) \emph{Ablation of description components}: title-only / axes-only / legend-only / style-only, to map which elements (if any) shift answers. 
(ii) \emph{Answer consistency and perturbation curves}: for each chart, estimate each model's consistency of the same question and tolerance to incremental description length.

\textbf{Use  difficult charts and more reasoning and difficult tasks to evaluate LLM abilities.}
Quantifying LLM is crucial for real-world users but such evaluation is considerably difficult~\cite{liang2022helm, chang2023survey}. We plan to add more reasoning tasks and 
more \textit{difficult}~\cite{truong2025reliable}: questions that may test frontier models' ability to reason and understand charts.
This is in sharp contrast to visual literacy tests, where the most popular charts are often used.


\section{Conclusion}
We evaluated the impact of model type and prompt conditions on visual question answering using a subset of the CHART-6 benchmark. In all datasets, GPT-5 consistently outperforms GPT-4o, confirming that LLM capability plays a greater role in answering questions correctly. However, the differences between the prompt conditions were relatively small, and the prompts with chart description did not 
improve compared to baseline or question only prompts. 
These results may suggest that the use of generic chart descriptions 
needs further investigation 
to guide model reasoning.
These findings indicate that while prompt modifications could influence superficial aspects of model outputs, the primary determinant of visualization understanding remains the underlying model architecture.

\bibliographystyle{abbrv-doi}

\bibliography{template}

\clearpage

\begin{appendices}

\section{Prompt for chart descriptions}
\begin{quote}
    {\small
    You are an assistant that describes the visual structure of chart images.
    Rules:
    - Output must be one paragraph, less than 100 words, in English.
    - Follow the template structure: chart type \& title; axes; marks; legend; gridlines \& ticks; overall styling.
    - Allowed:
      • Chart type (bar chart, line chart, scatter plot, pie chart, etc.).
      • Title text (quote exactly as written).
      • Axis titles and units, if present (no tick values).
      • Category labels (list exactly as shown).
      • Marks: shape (bars, slices, dots, lines), outline/fill style, colors.
      • Legend: state if present and list labels, otherwise state absence.
      • Gridlines: mention orientation (horizontal/vertical/both).
      • Tick marks: acknowledge existence, not values.
      • Overall styling (monochrome, bold outline, shading, etc.).
    - Forbidden:
      • Reporting or paraphrasing any tick values, data values, or scale values.
      • Counts of marks, comparisons, relative size, proportions, trends, correlations, distributions.
      • Words like evenly, uniformly, represent, correspond, show, increase, decrease, cluster, spread, dense, scattered, outlier, variability.
      • Any interpretation of meaning.
    - Return only the paragraph, no lists or bullets.
    }
\end{quote}

\FloatBarrier

\onecolumn
\section{Total 107 questions}
\label{app:107_questions}

\setlength\LTleft{0pt}
\setlength\LTright{0pt}
\begin{longtable}{%
  >{\raggedright\arraybackslash}p{18mm}
  >{\raggedright\arraybackslash}p{0.54\textwidth}
  >{\raggedright\arraybackslash}p{18mm}
  >{\centering\arraybackslash}p{26mm}
}
\caption{107 questions with answers and images.}\label{tab:qa_fullwidth_simple}\\
\toprule
\textbf{Dataset} & \textbf{Question} & \textbf{Answer} & \textbf{Image} \\
\midrule
\endfirsthead

\toprule
\textbf{Dataset} & \textbf{Question} & \textbf{Answer} & \textbf{Image} \\
\midrule
\endhead

\midrule
\multicolumn{4}{r}{\footnotesize Continued on next page} \\
\endfoot

\bottomrule
\endlastfoot

\QArow{GGR}{What percentage of patients recovered after chemotherapy?}{35}{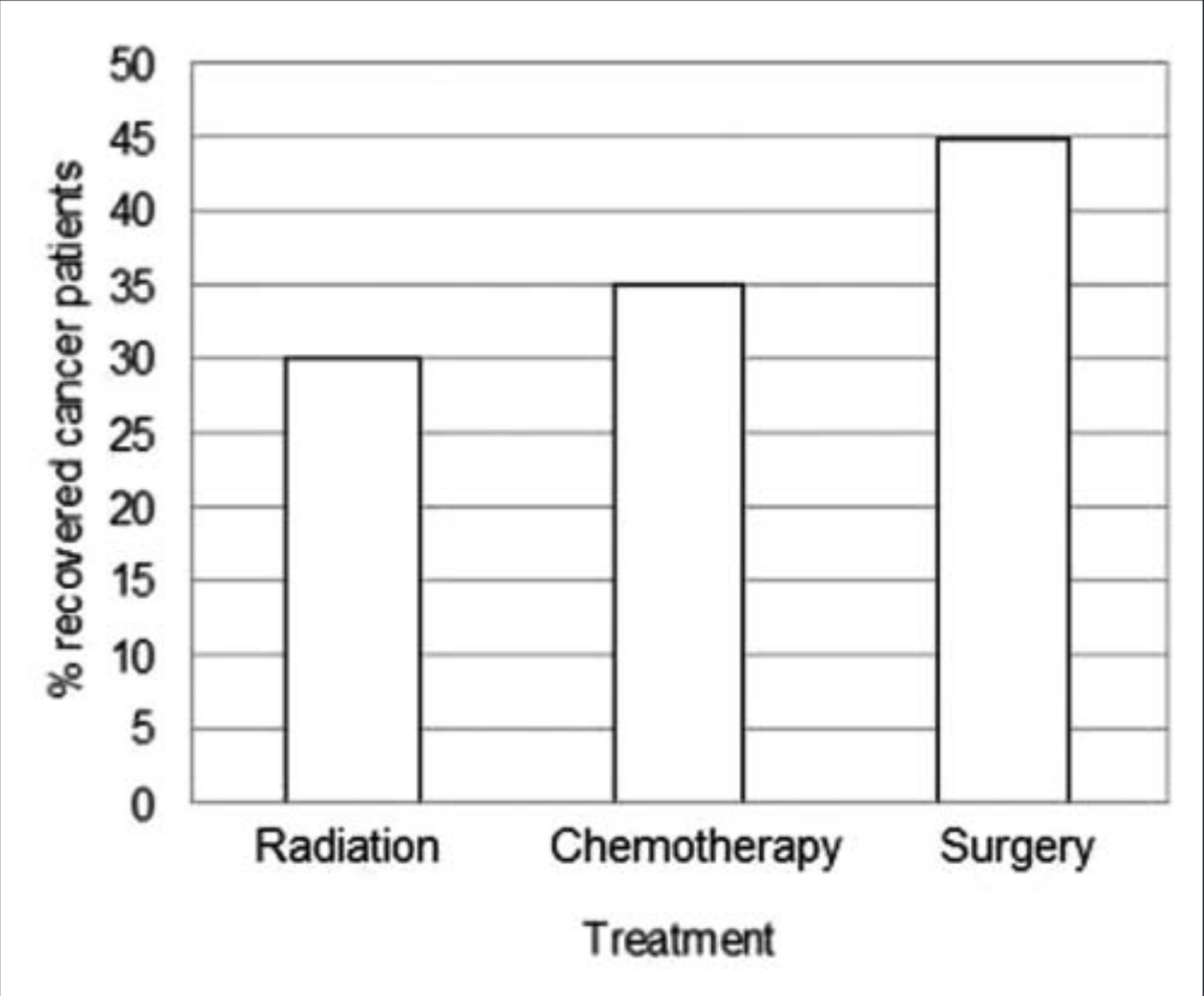}
\QArow{GGR}{What is the difference between the percentage of patients who recovered after a surgery and the percentage of patients who recovered after radiation therapy?}{15}{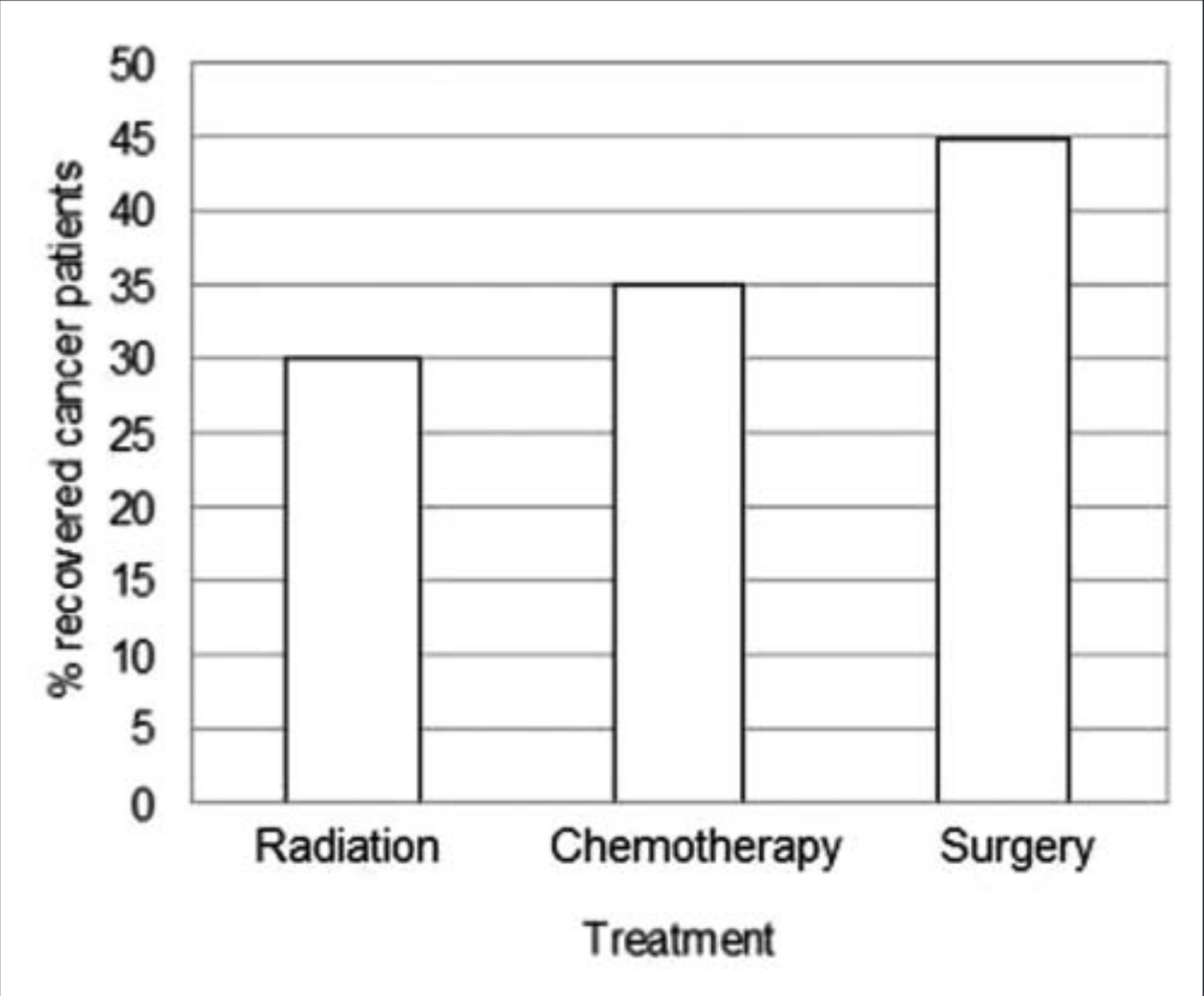}
\QArow{GGR}{"Of all the people who die from cancer, approximately what percentage dies from lung cancer?"}{25}{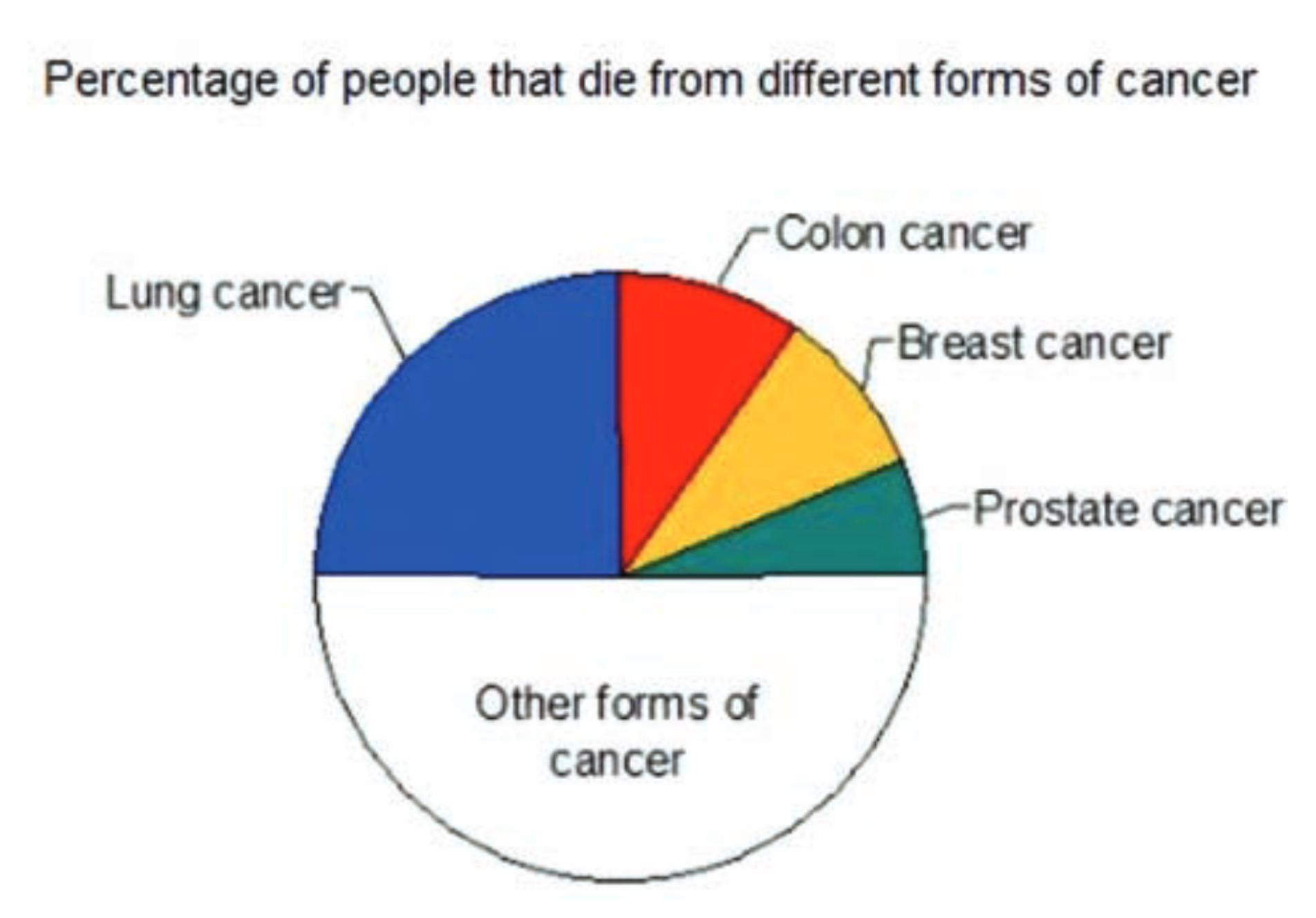}
\QArow{GGR}{"Approximately what percentage of people who die from cancer die from colon cancer, breast cancer, and prostate cancer taken together?"}{25}{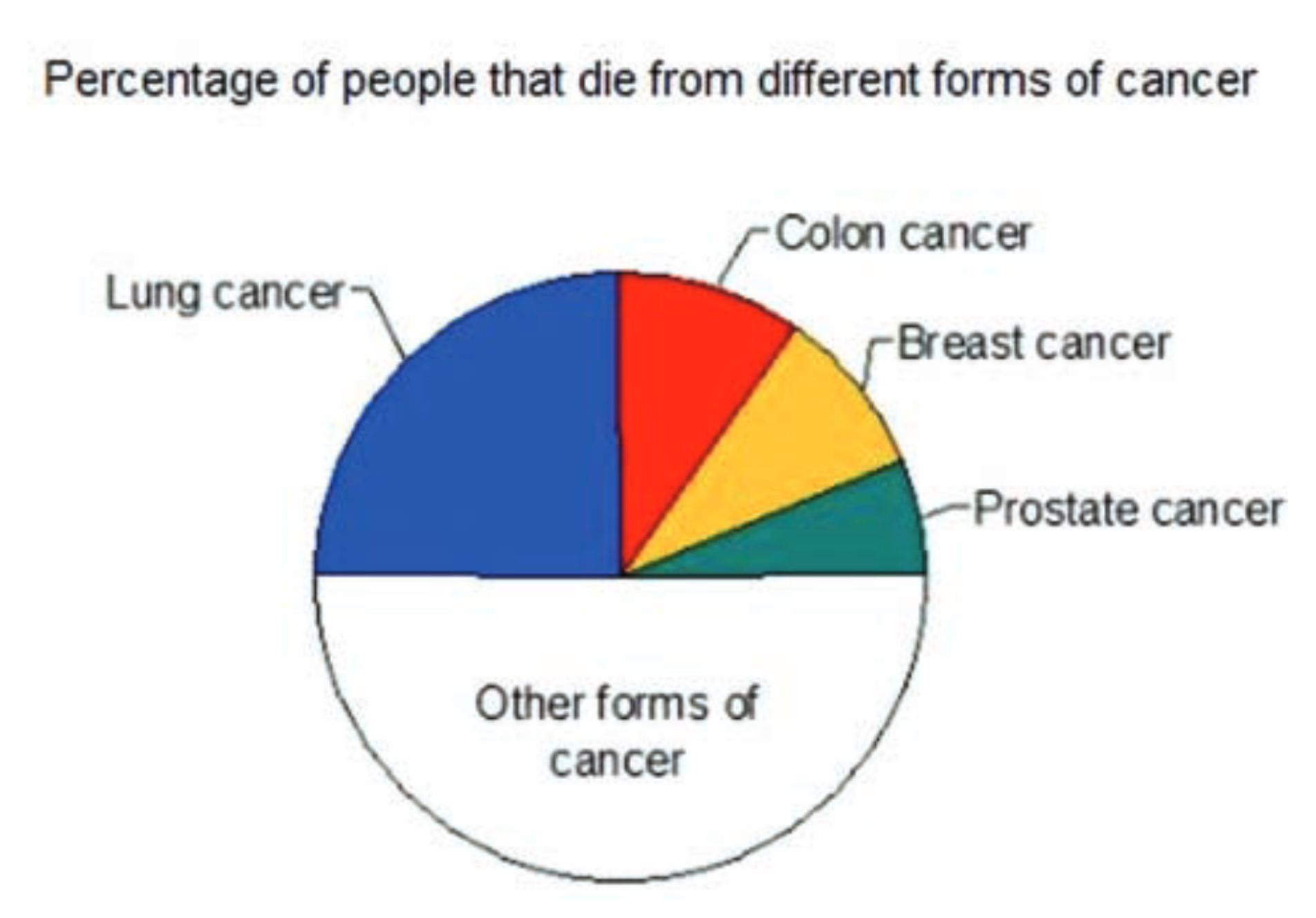}
\QArow{GGR}{Approximately what percentage of people had Adeolitis in the year 2000?}{20}{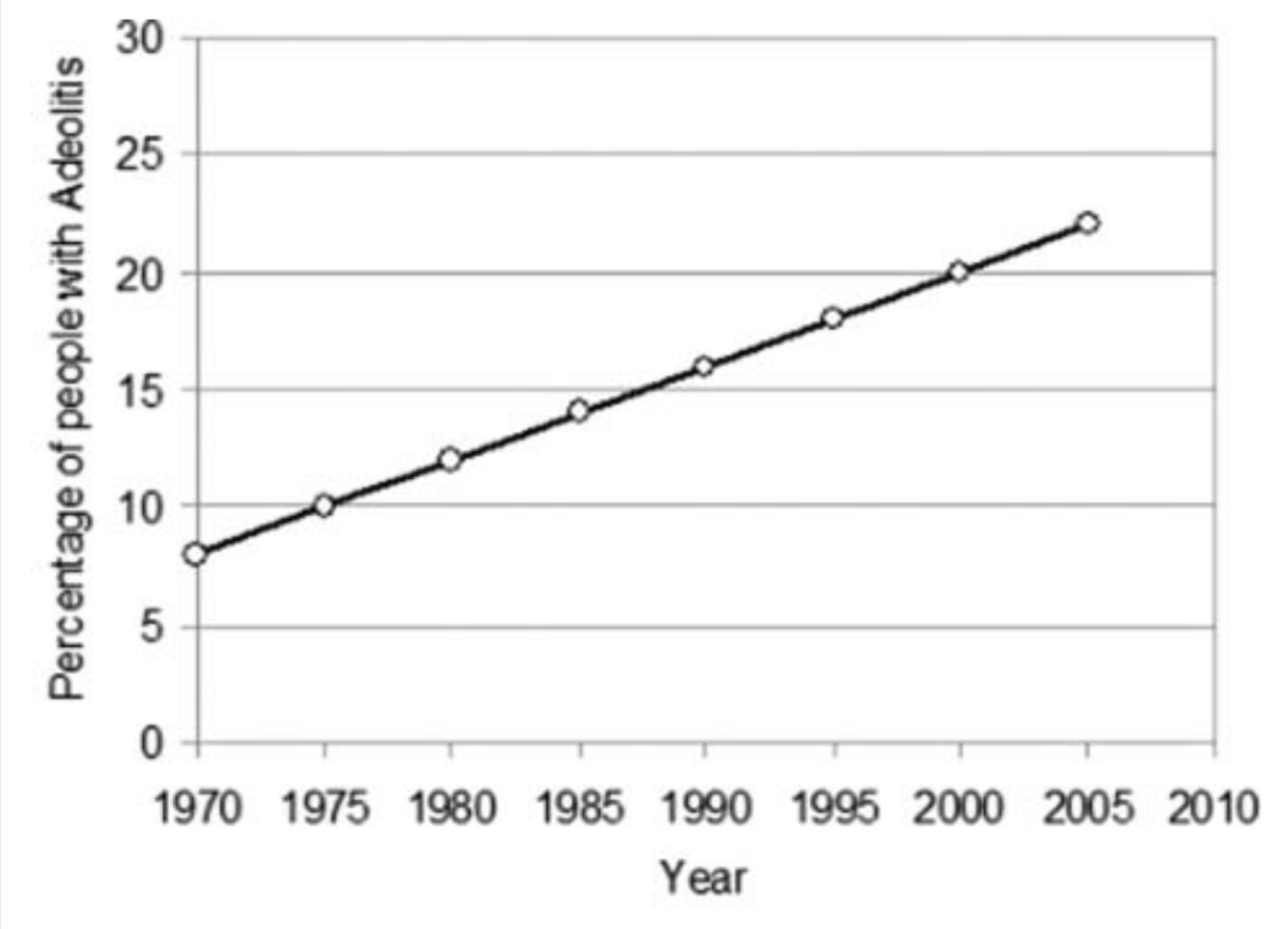}
\QArow{GGR}{"According to your best guess, what will the percentage of people with Adeolitis be in the year 2010?"}{25}{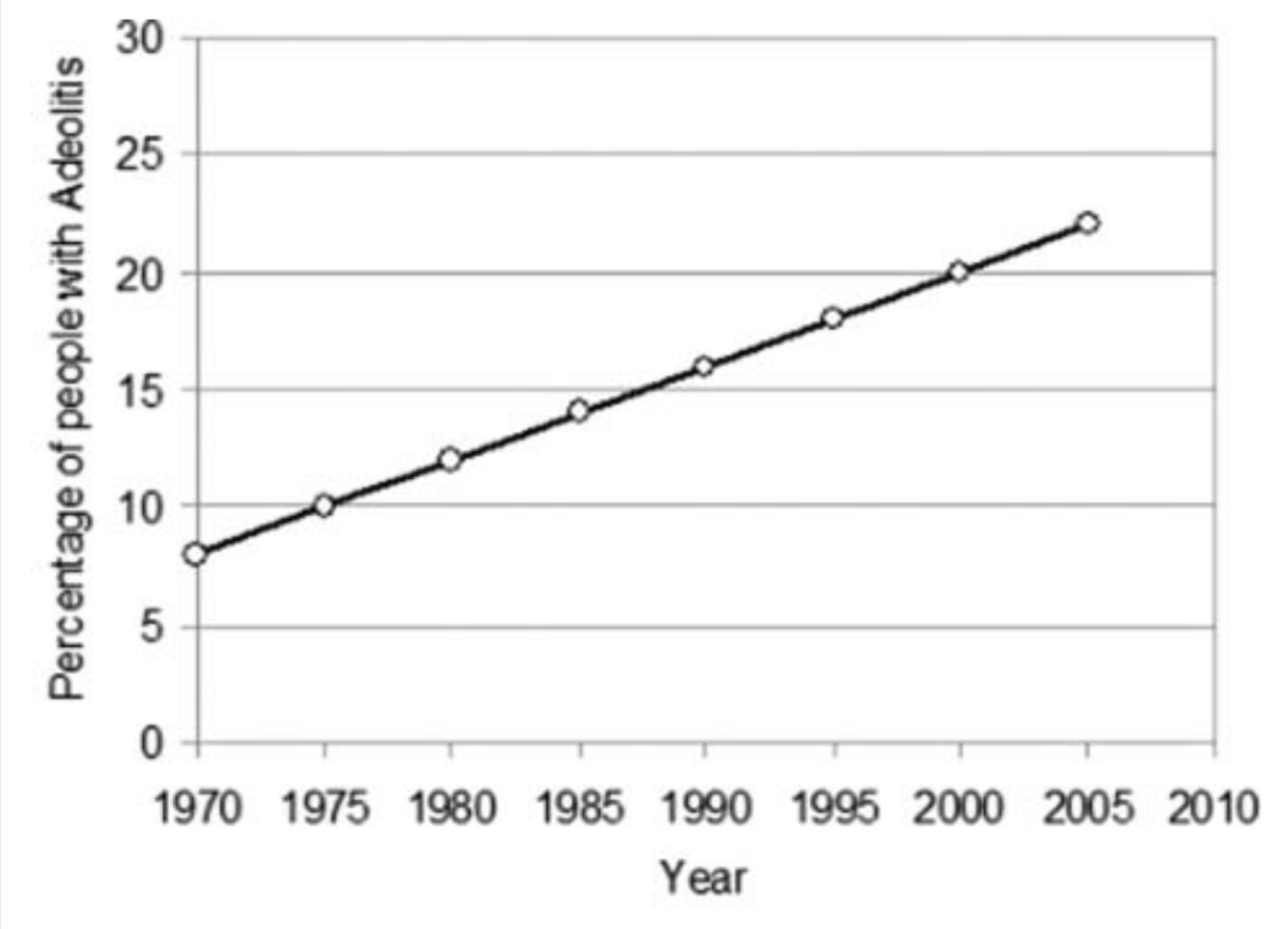}
\QArow{GGR}{"Of 100 patients with disease X, how many are women?"}{40}{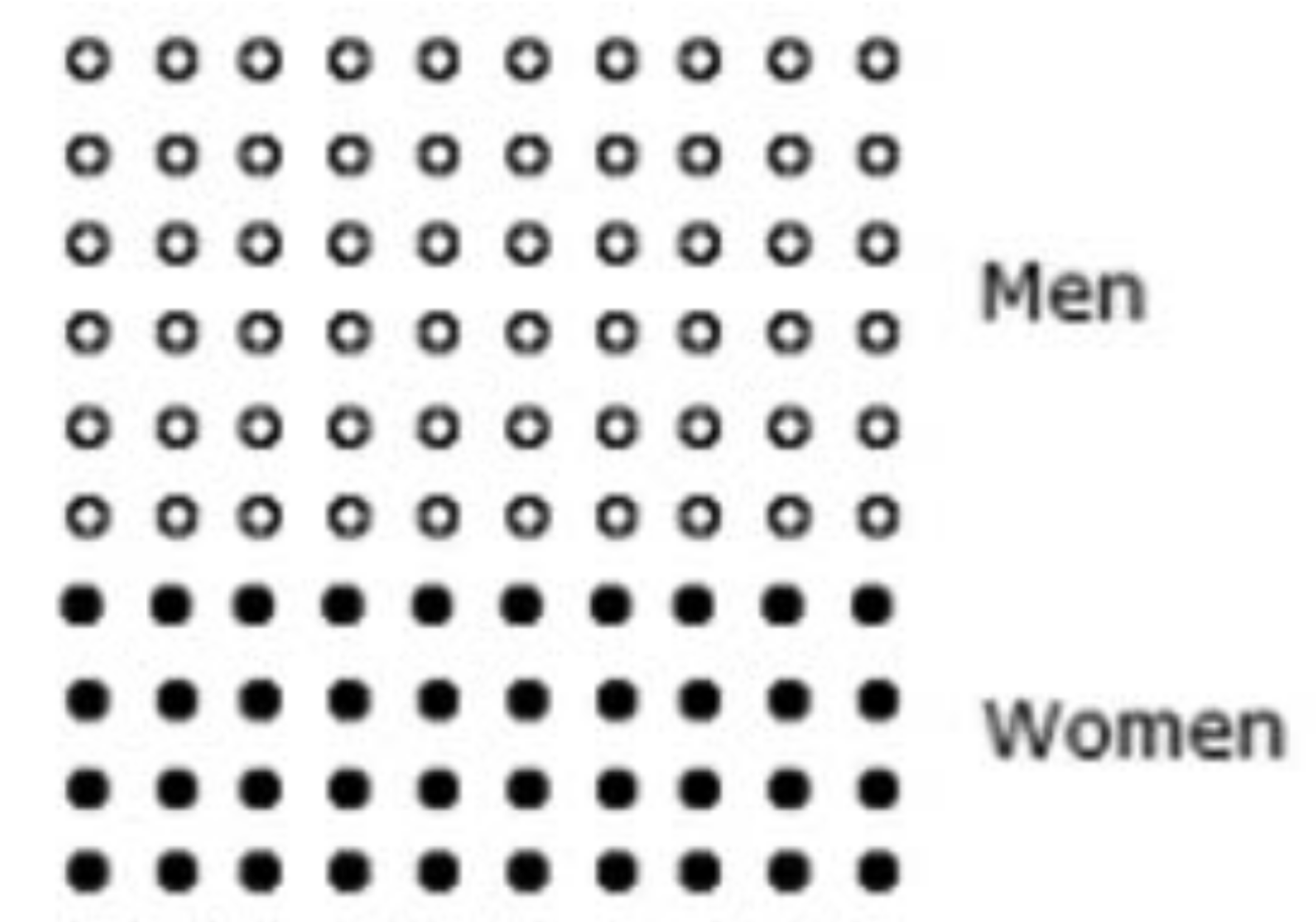}
\QArow{GGR}{How many more men than women are there among 100 patients with disease X?}{20}{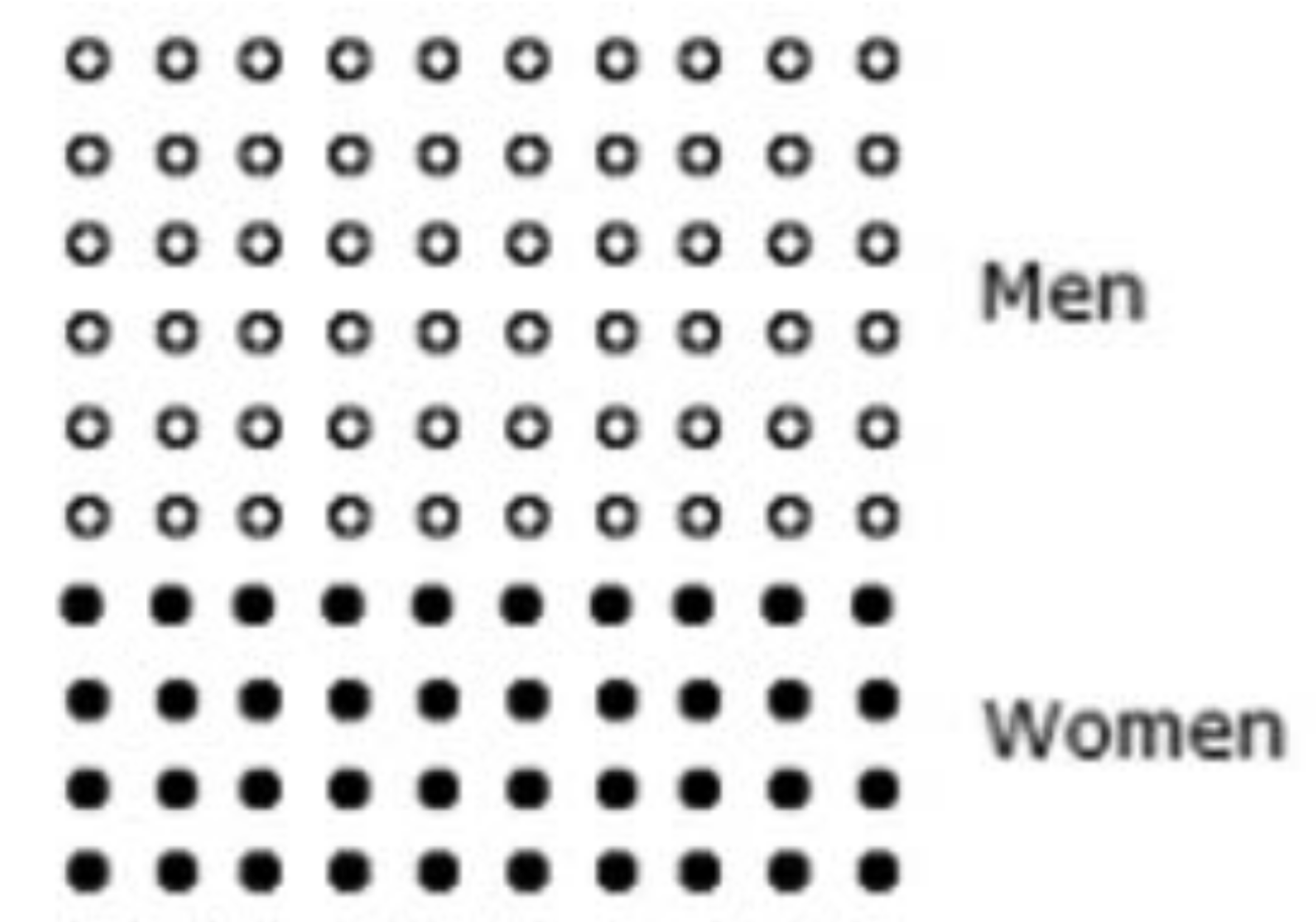}
\QArow{GGR}{"Compared to the placebo, which treatment leads to a larger decrease in the percentage of patients who die?"}{They are equal}{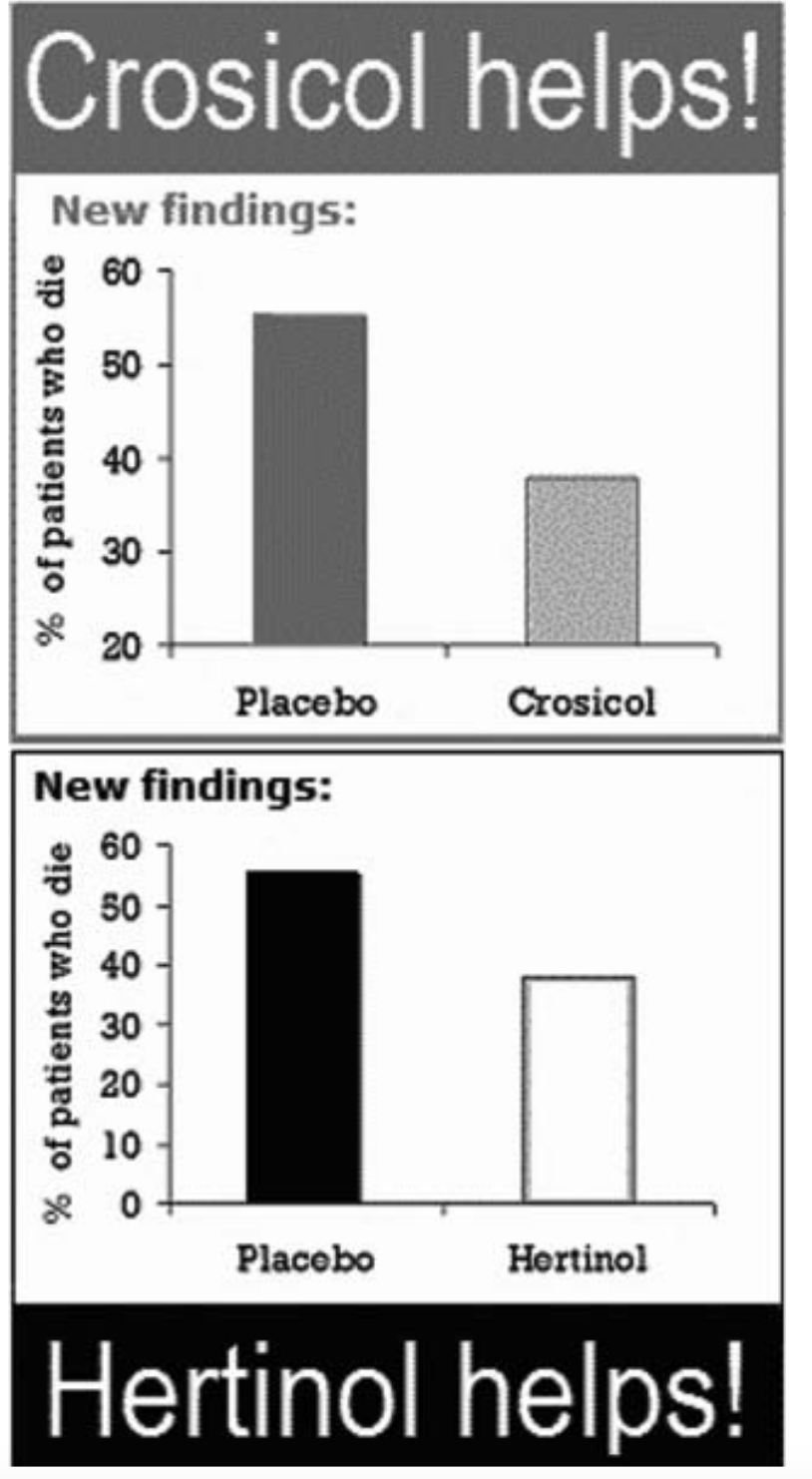}
\QArow{GGR}{Which of the treatments contributes to a larger decrease in the percentage of sick patients?}{Can't say}{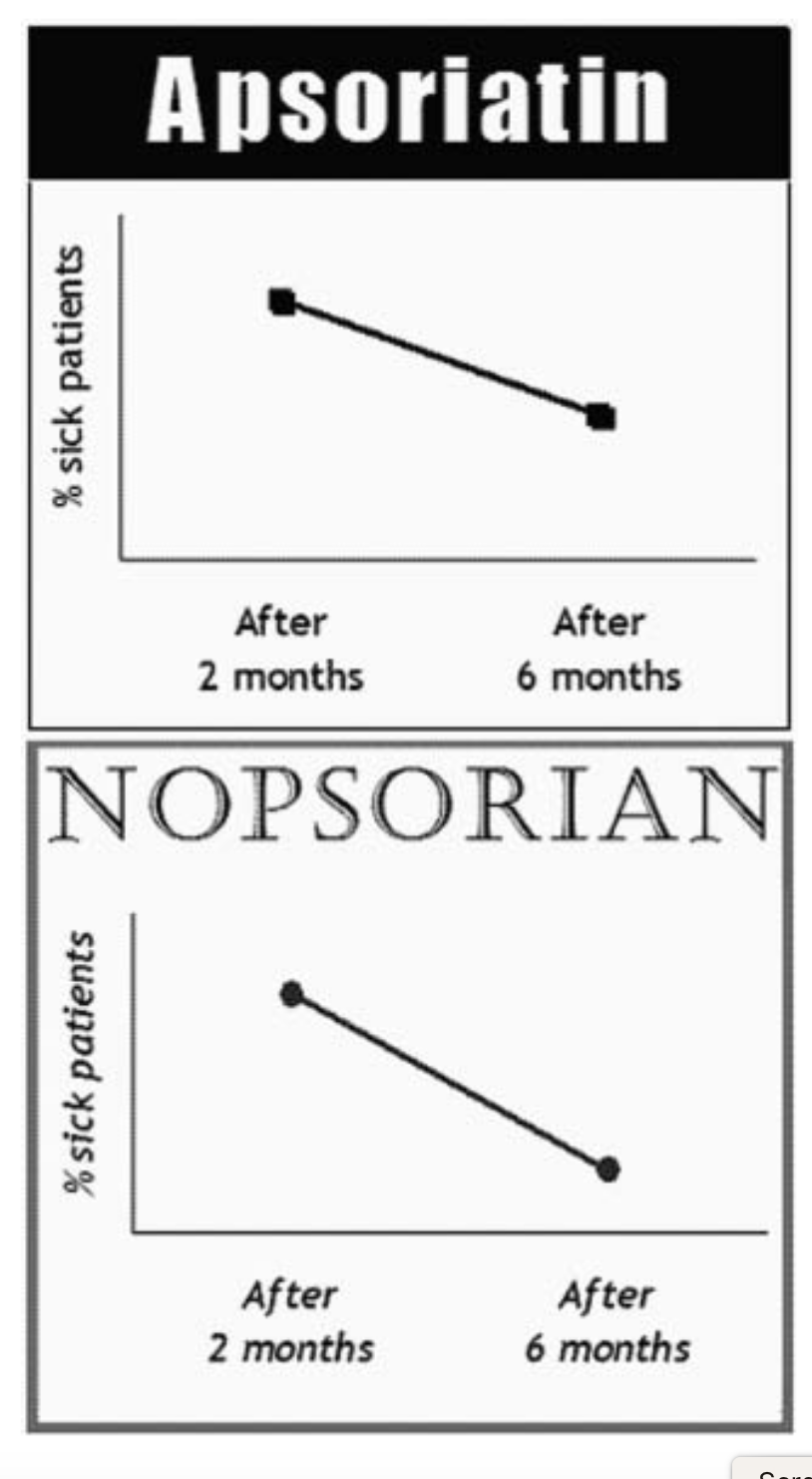}
\QArow{GGR}{What is the percentage of cancer patients who die after chemotherapy?}{5}{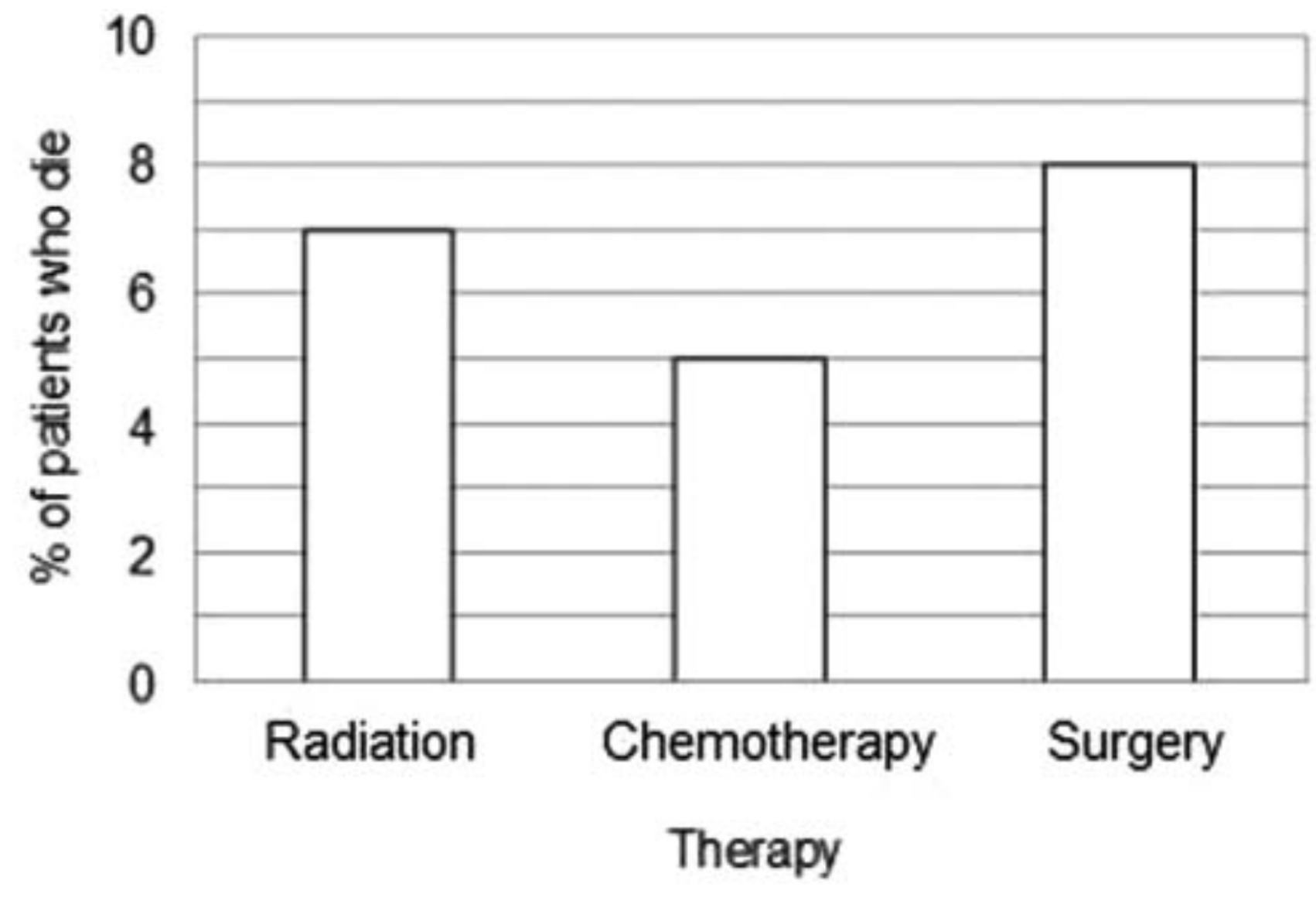}

\QArow{VLAT}{About what is the ratio of the cost of a sandwich to the total cost of room service in Seattle?}{4 to 10}{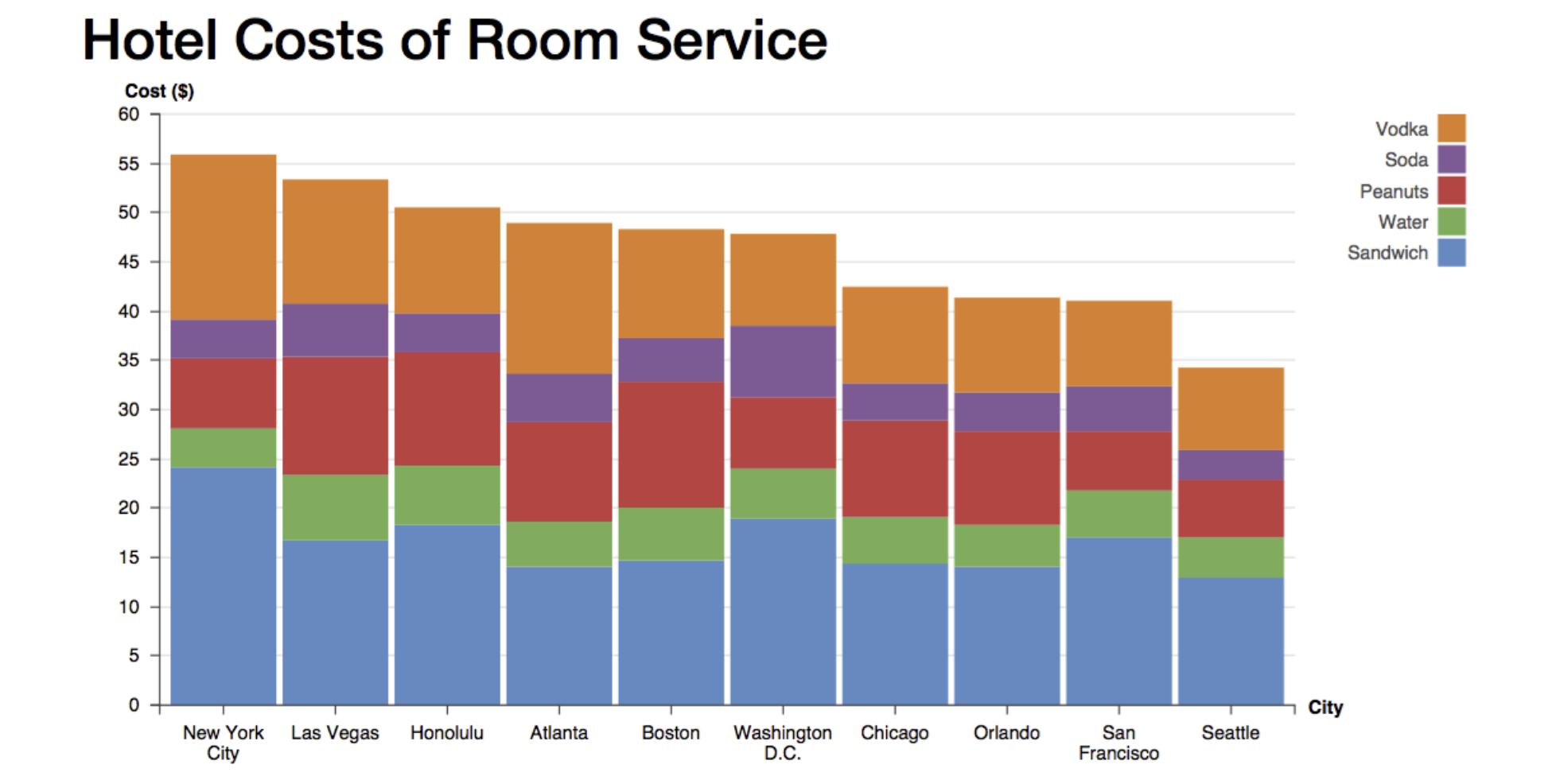}
\QArow{VLAT}{In which city is the cost of soda the highest?}{Washington D.C.}{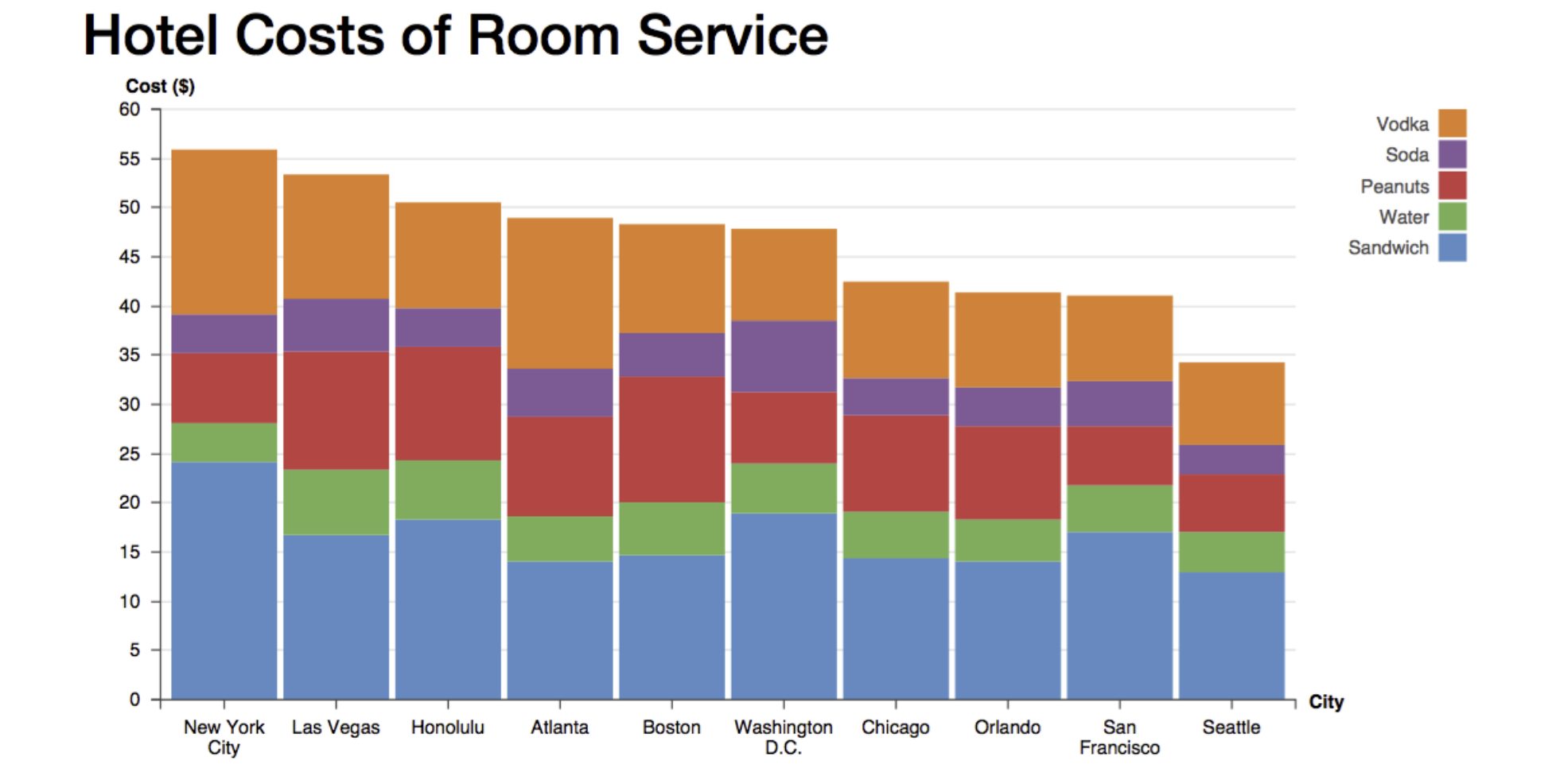}
\QArow{VLAT}{The cost of vodka in Atlanta is higher than that of Honolulu.}{True}{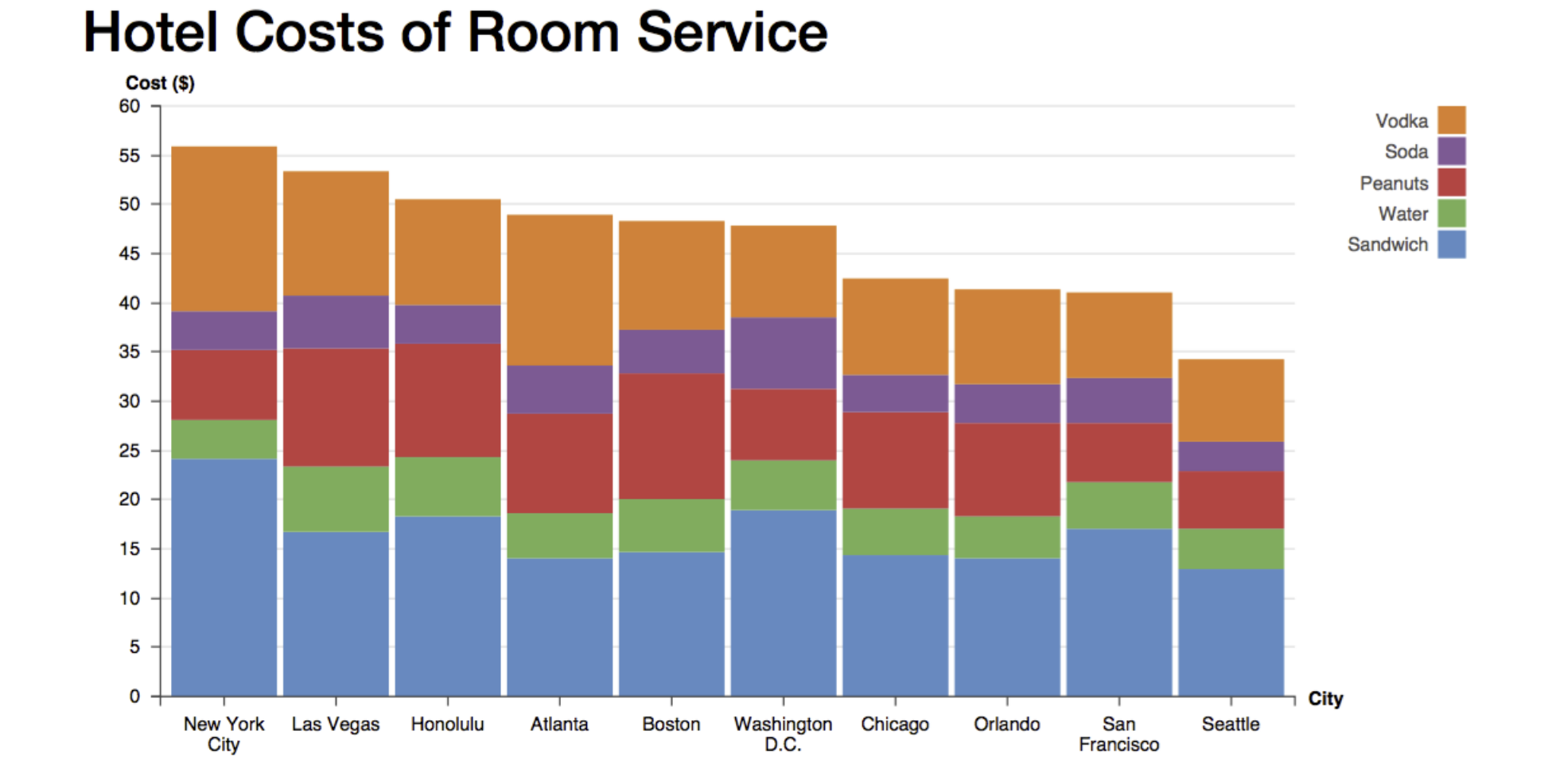}
\QArow{VLAT}{What is the weight for the person who is 165.1 cm tall?}{70.5 kg}{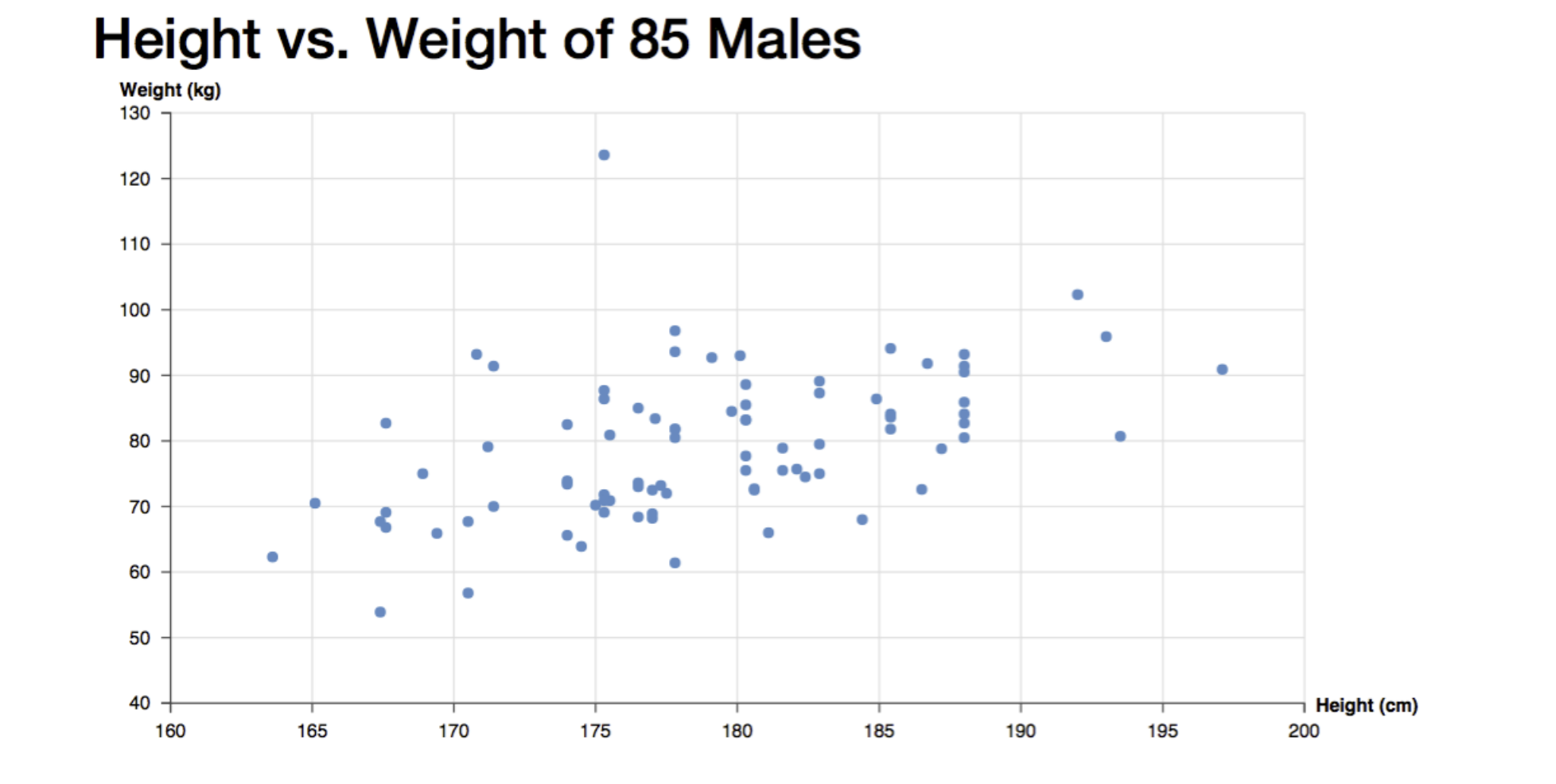}
\QArow{VLAT}{What is the range in weight for the 85 males?}{53.9 - 123.6 kg}{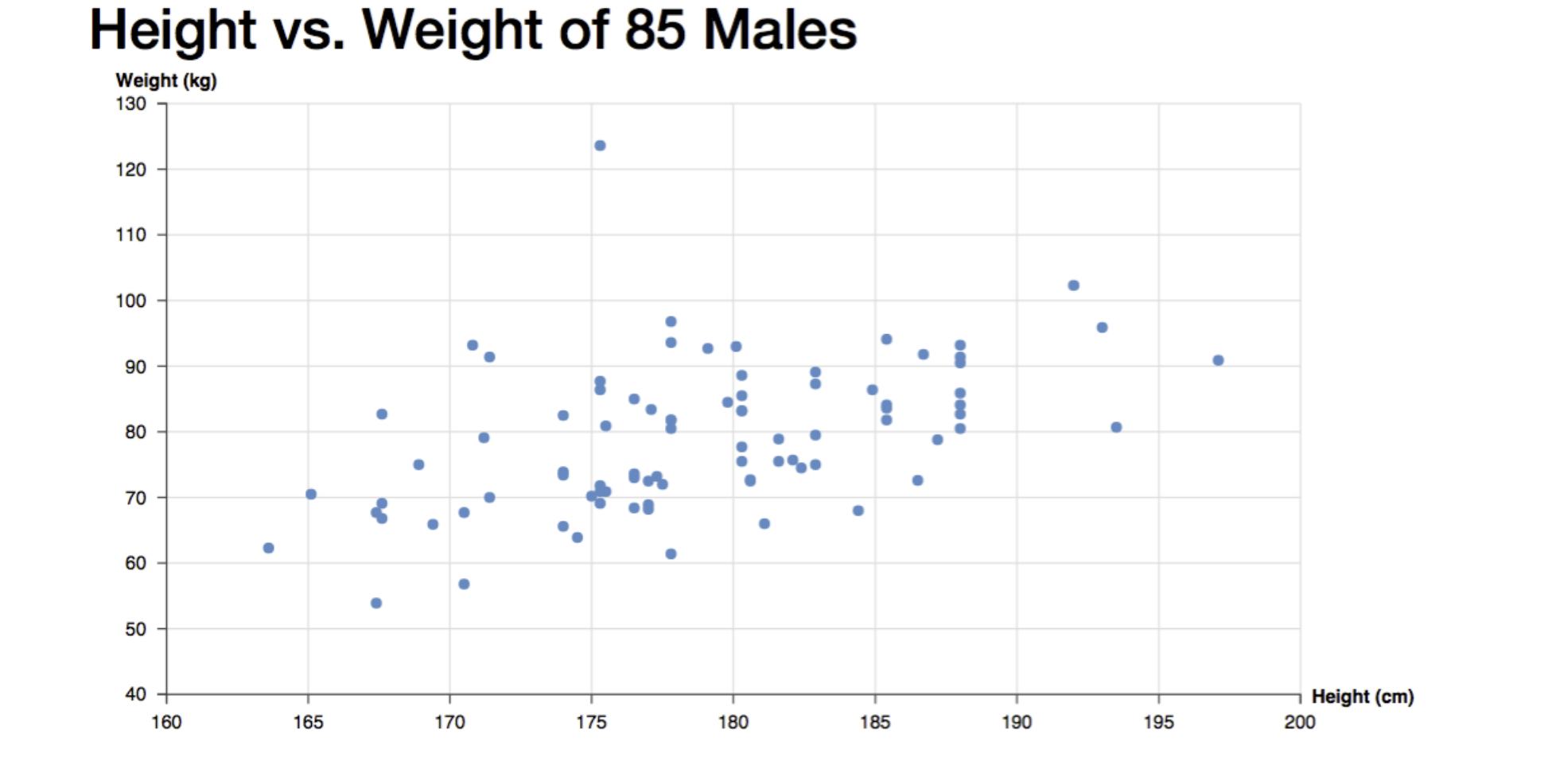}
\QArow{VLAT}{What is the height for a person who lies outside the others the most?}{175.3 cm}{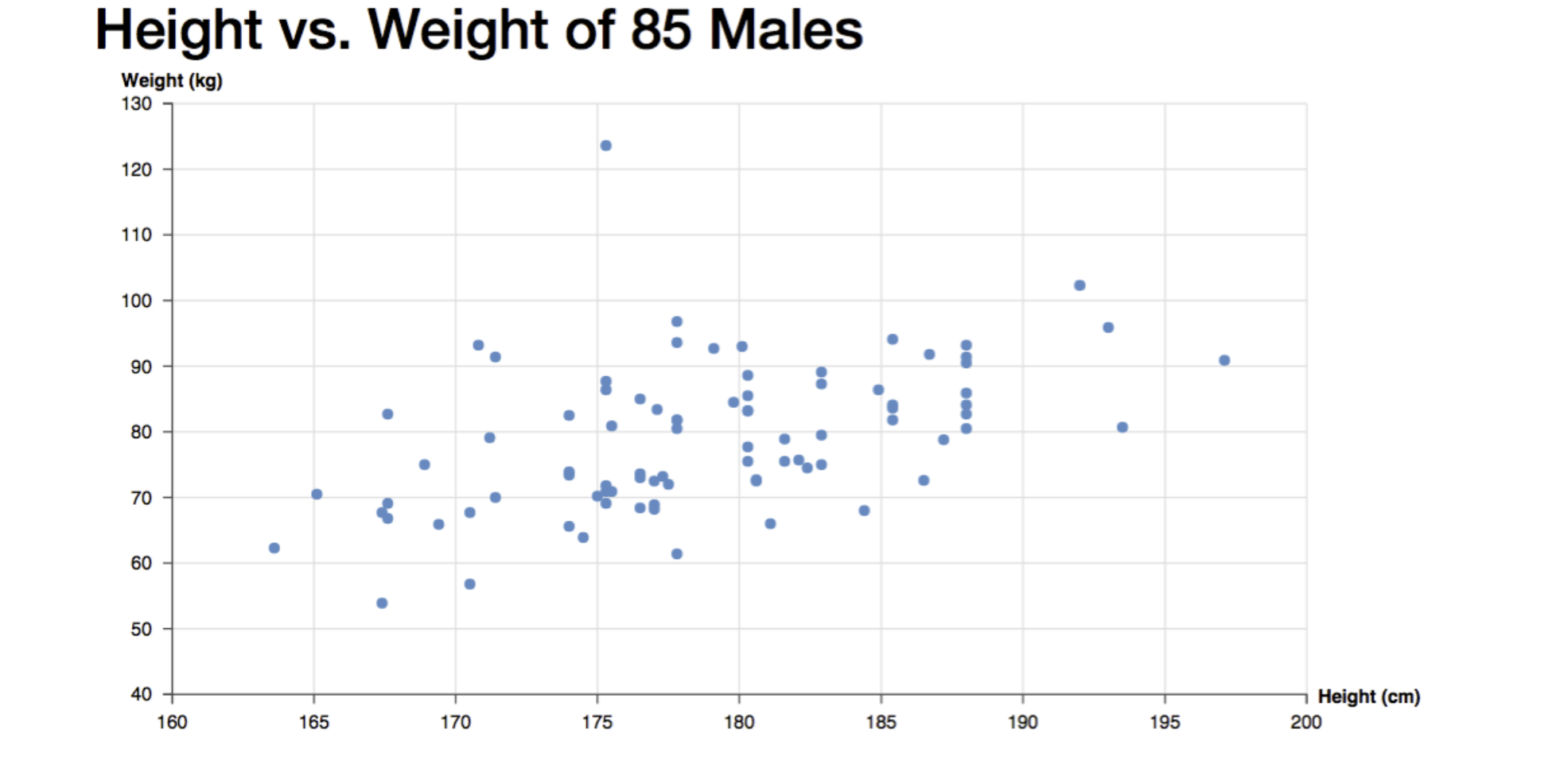}
\QArow{VLAT}{When was the average price of a pound of coffee beans at minimum?}{December 2014}{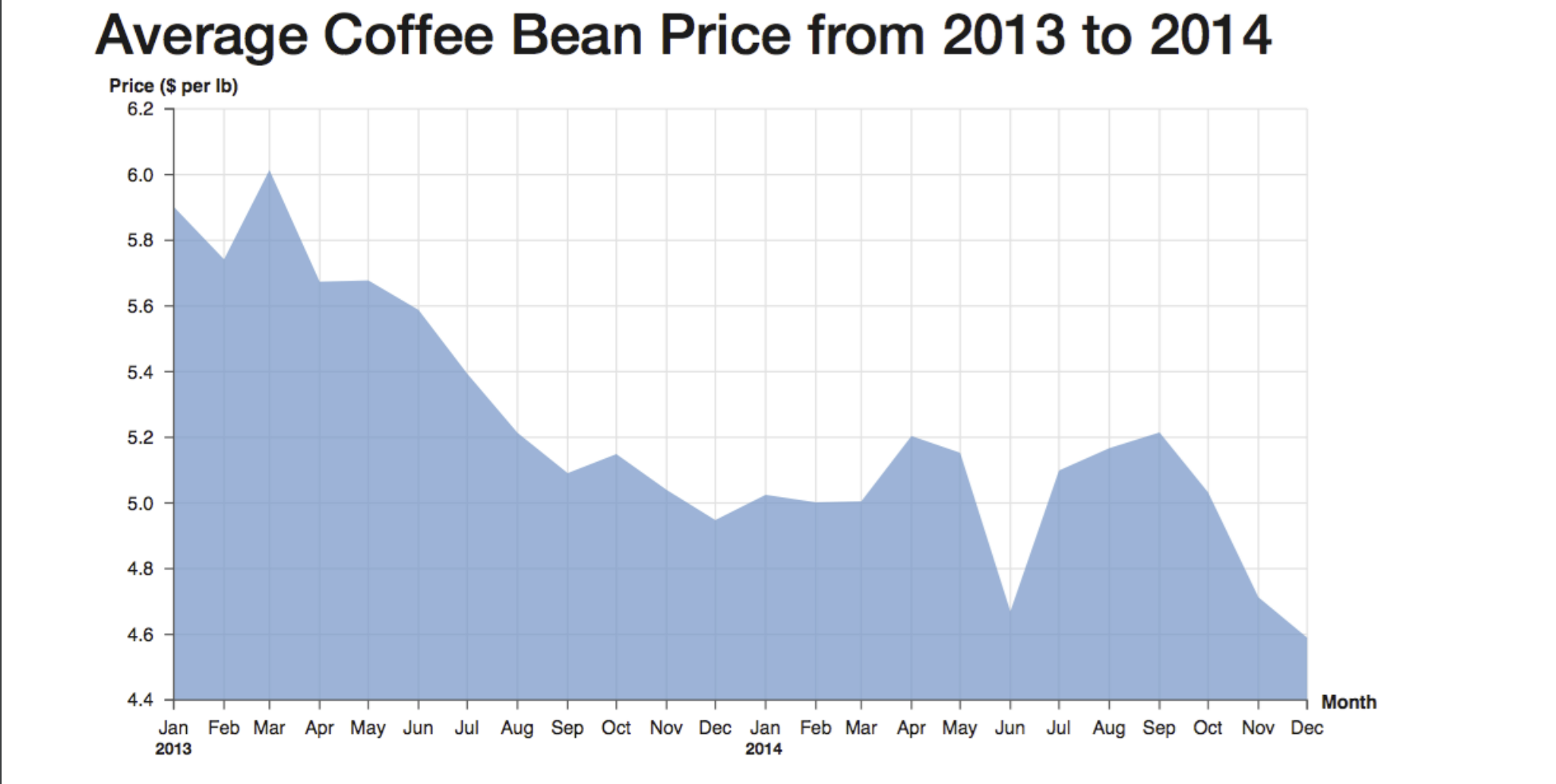}
\QArow{VLAT}{In the UK, the number of girls named 'Amelia' in 2014 was more than it was in 2013.}{False}{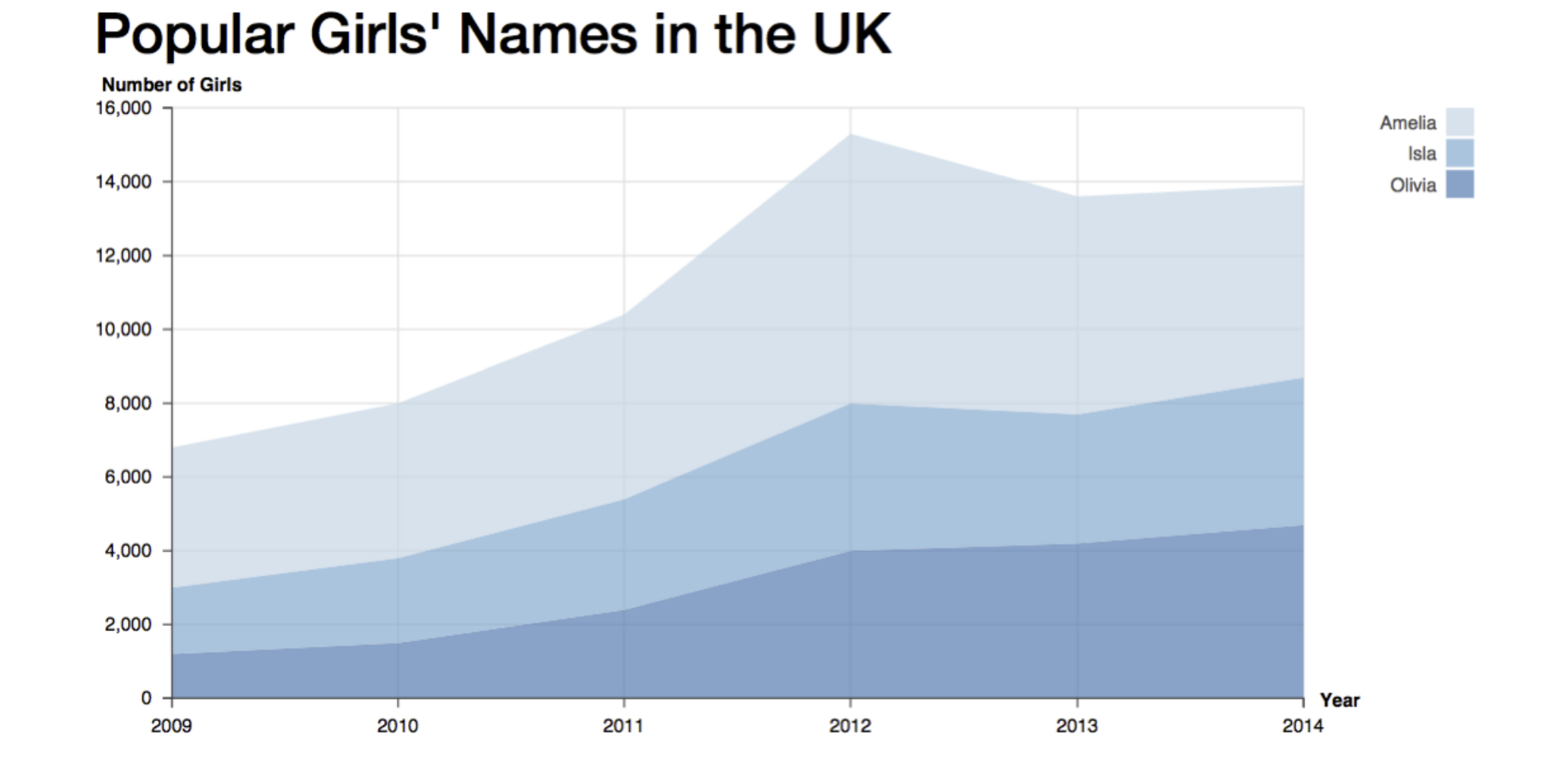}
\QArow{VLAT}{In general, the ridership of the metro system increases as the number of stations increases.}{False}{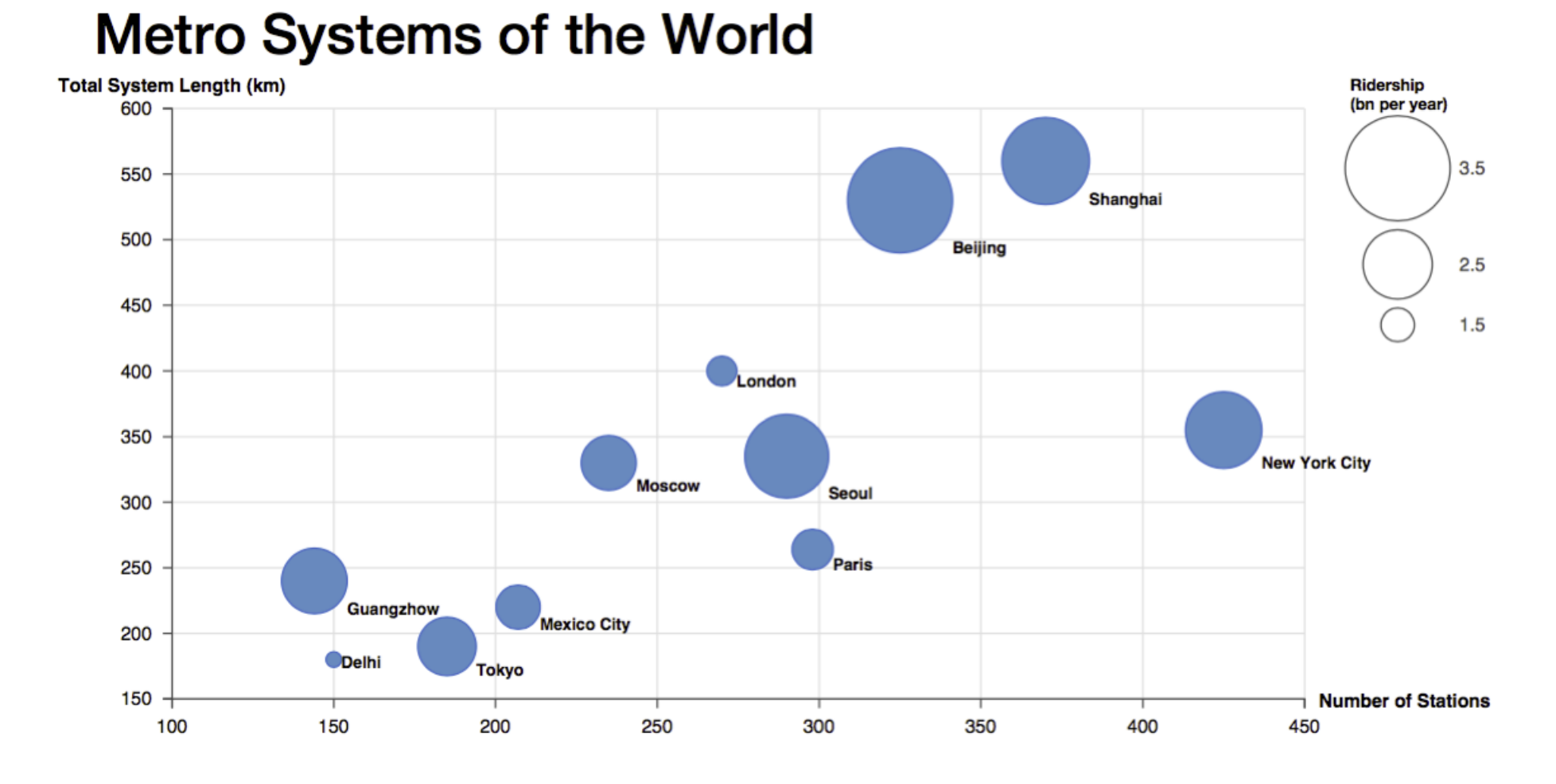}
\QArow{VLAT}{What was the unemployment rate for Indiana (IN) in 2015?}{4.0\% - 5.0\%}{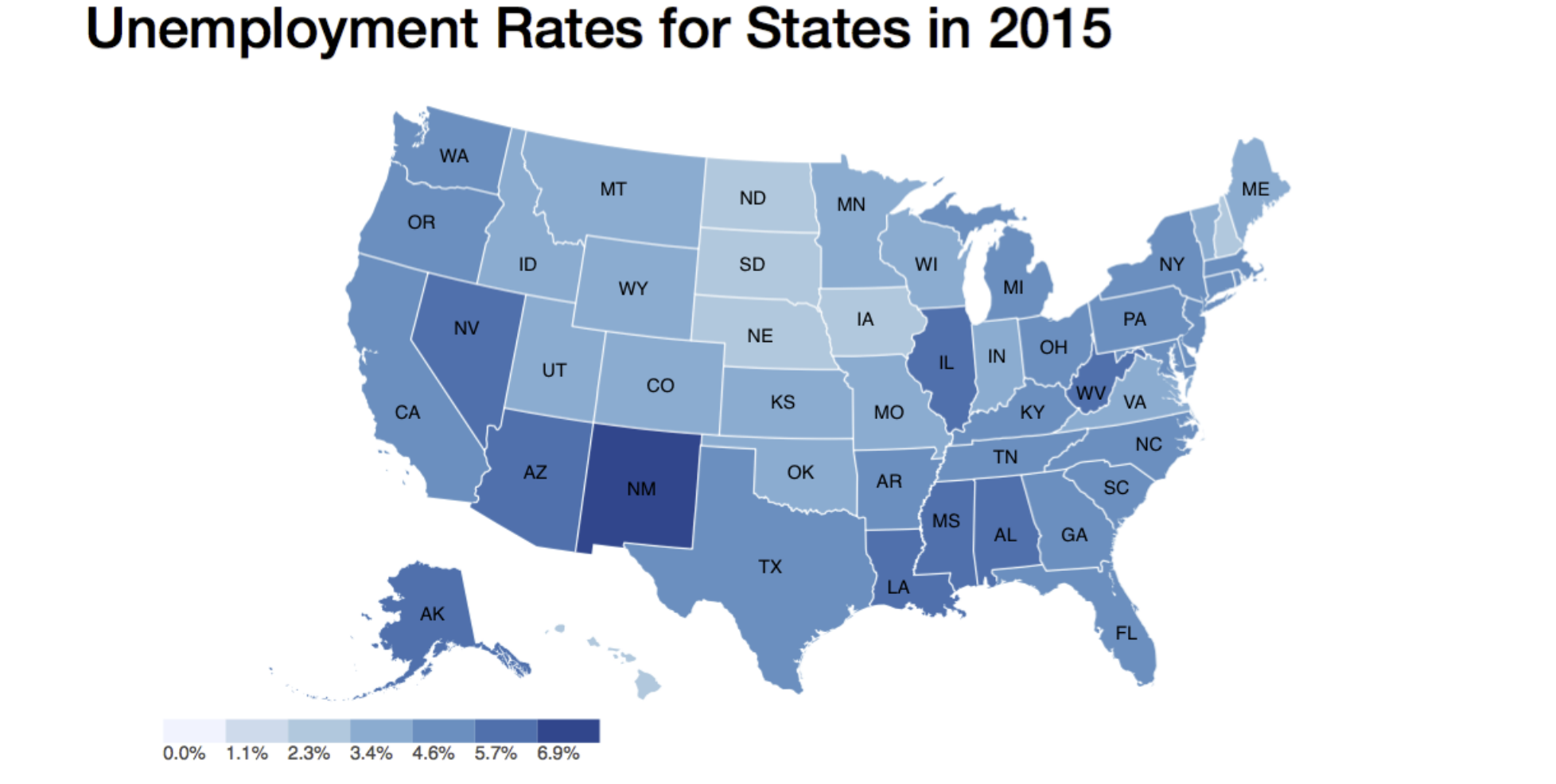}
\QArow{VLAT}{In which state was the unemployment rate the highest in 2015?}{New Mexico (NM)}{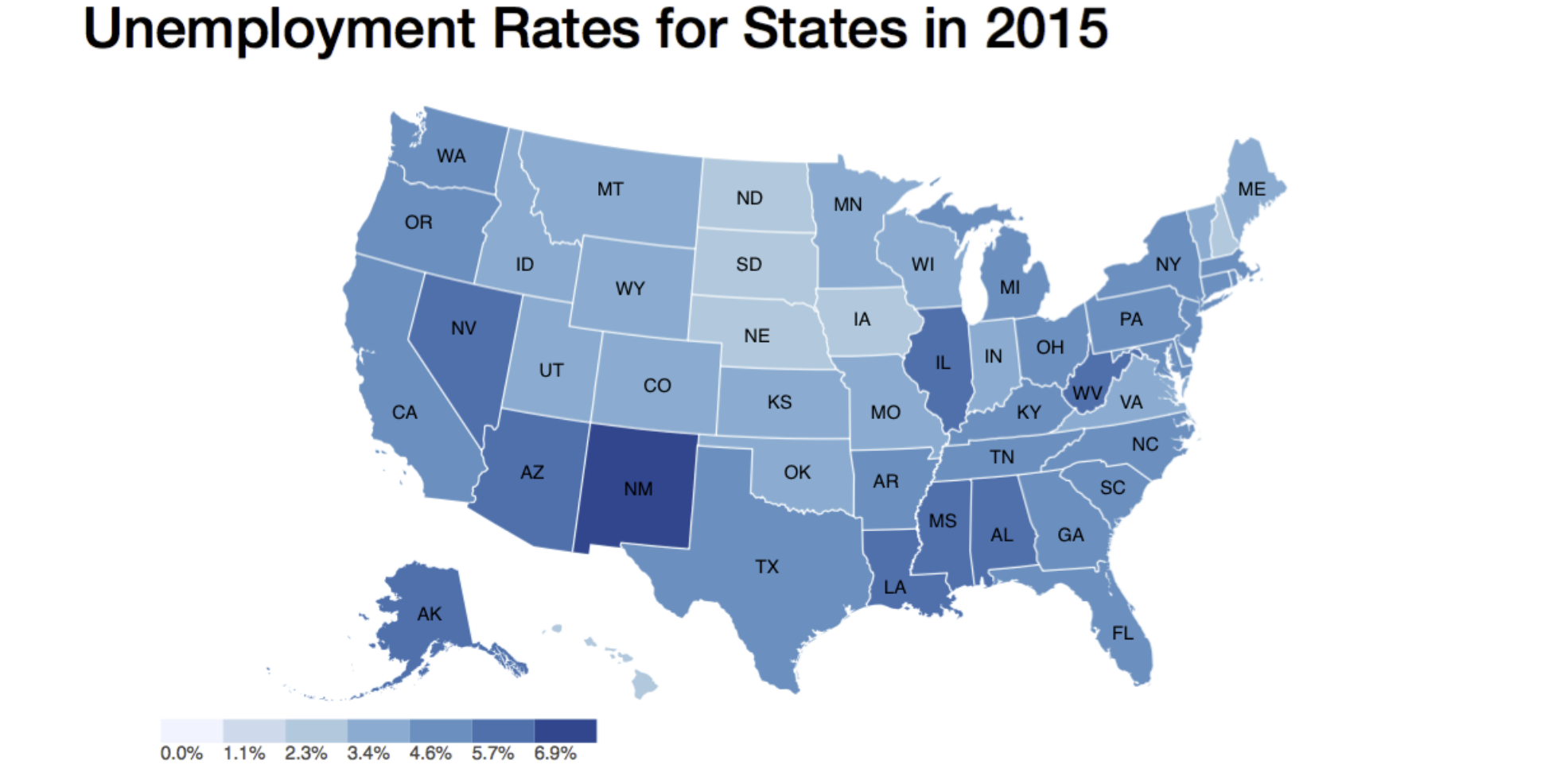}

\QArow{CLAVI-Standard}{Which of the following statements is true about the customers at the three bike shops in city Z?,"On Jul 02, shop B had more customers than A and C combined.",The number of customers at shop C decreased from Jul 01 to Jul 03.,The number of customers at shop A increased from Jul 01 to Jul 03.,None of the above.}{On Jul 02, shop B had more customers than A and C combined.}{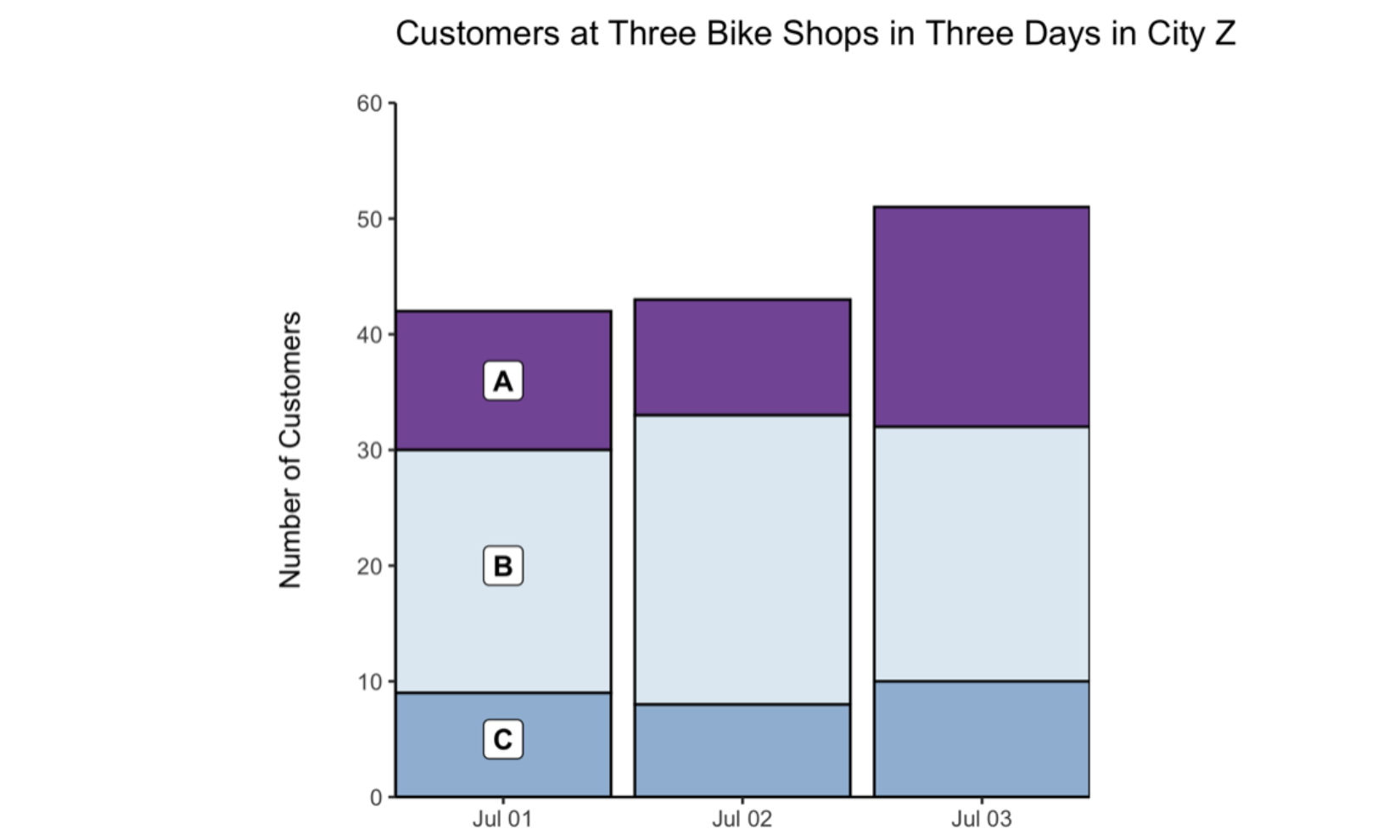}
\QArow{CLAVI-Standard}{Which of the following is true about the four brands of perfume in city Z?,The market share of brand D is on an increasing trend from 2011 to 2015.,The market share of brand A is higher in 2011 than in 2012.,"In 2013, brand B had the largest market share among all four brands.",None of the above.}{In 2013, brand B had the largest market share among all four brands.}{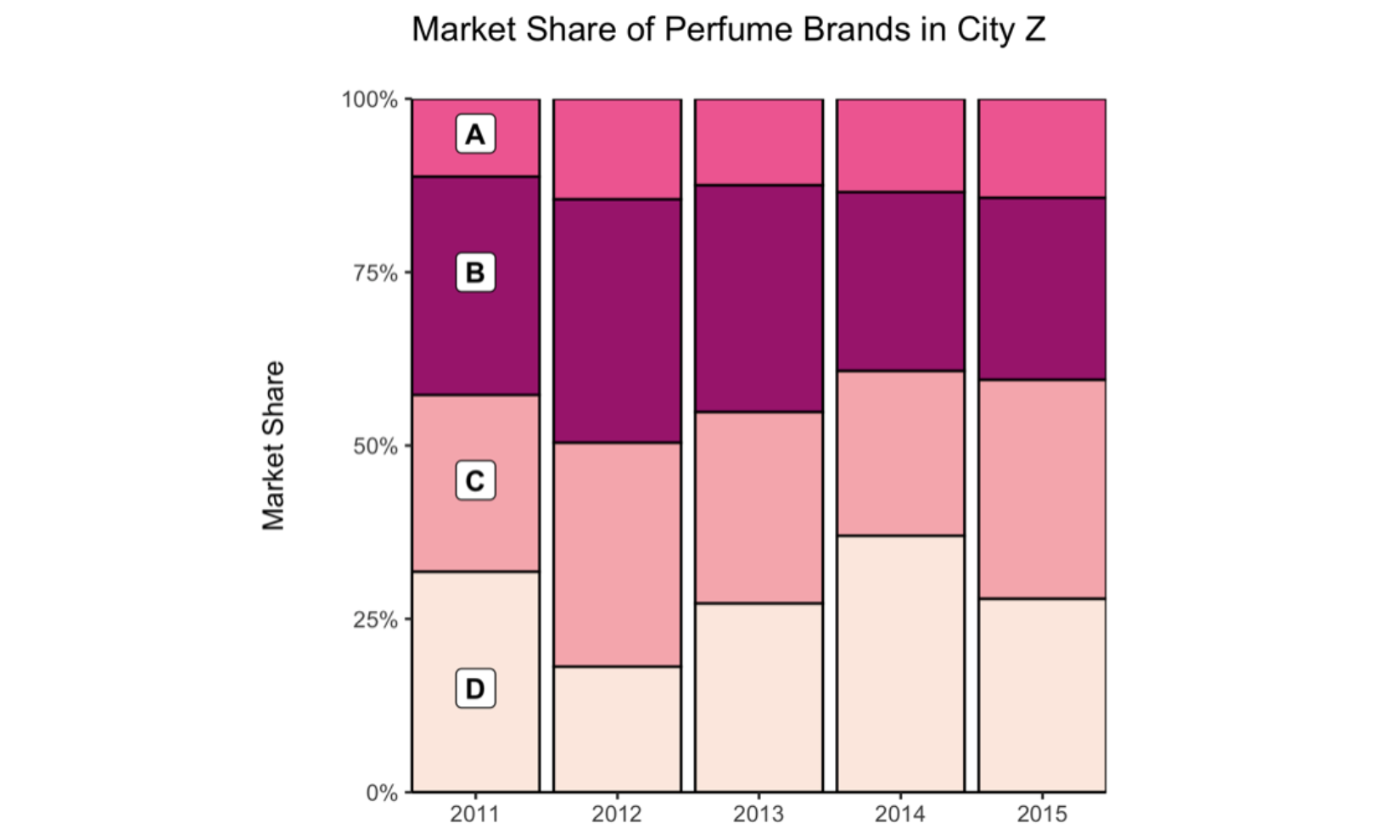}
\QArow{CLAVI-Standard}{Which of the following about the monthly number of new clients is true?,It decreased from Mar to Jun for company A.,It is higher in Feb for company B than company A.,It increased from Sep to Dec for company B.,None of the above.}{It increased from Sep to Dec for company B.}{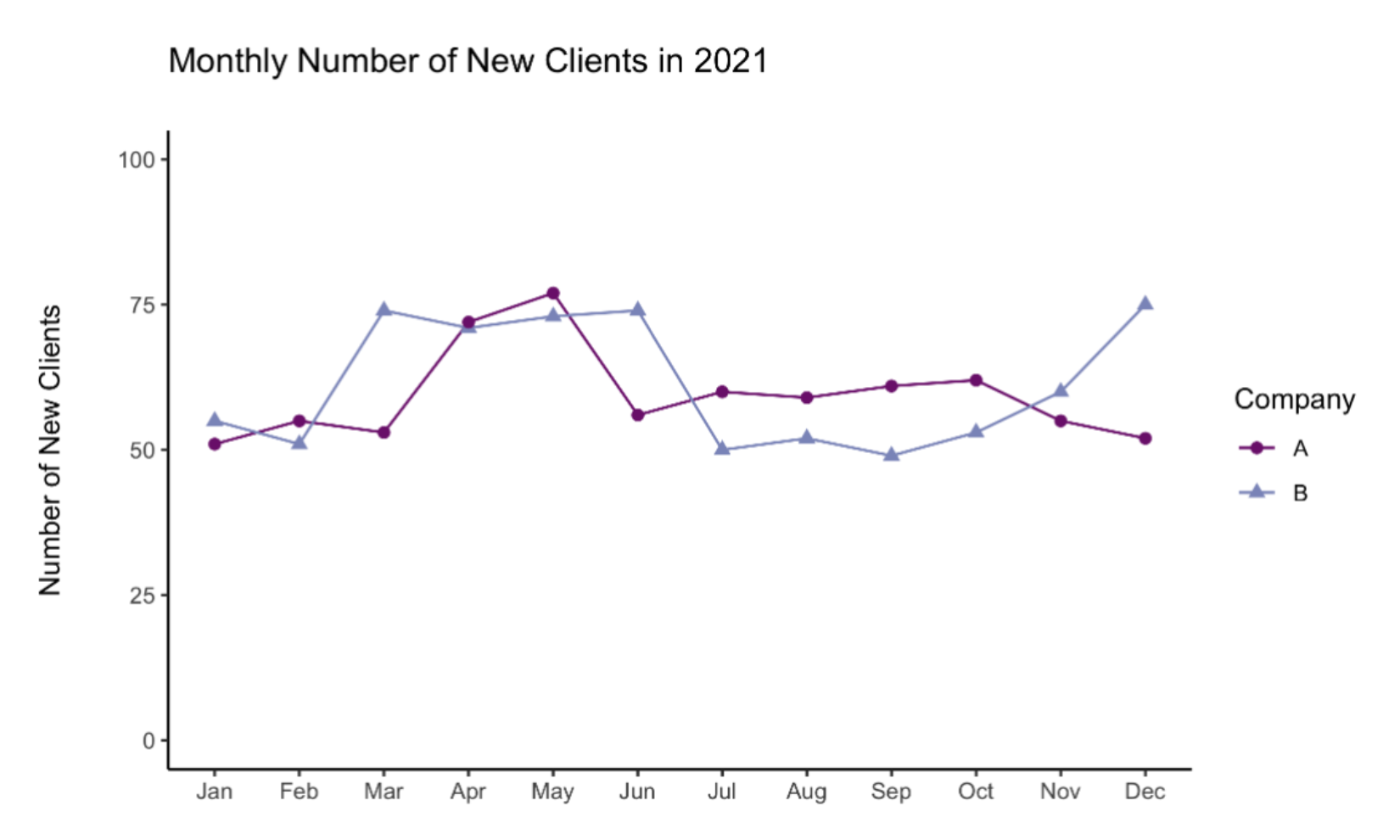}
\QArow{CLAVI-Standard}{Which of the following is true about the education level of residents in city Z?,More than 60\% of residents have a Bachelor's degree or above.,The majority of residents have a graduate degree.,There are fewer residents whose education level is Associate degree than those with a high school degree.,None of the above.}{More than 60\% of residents have a Bachelor's degree or above.}{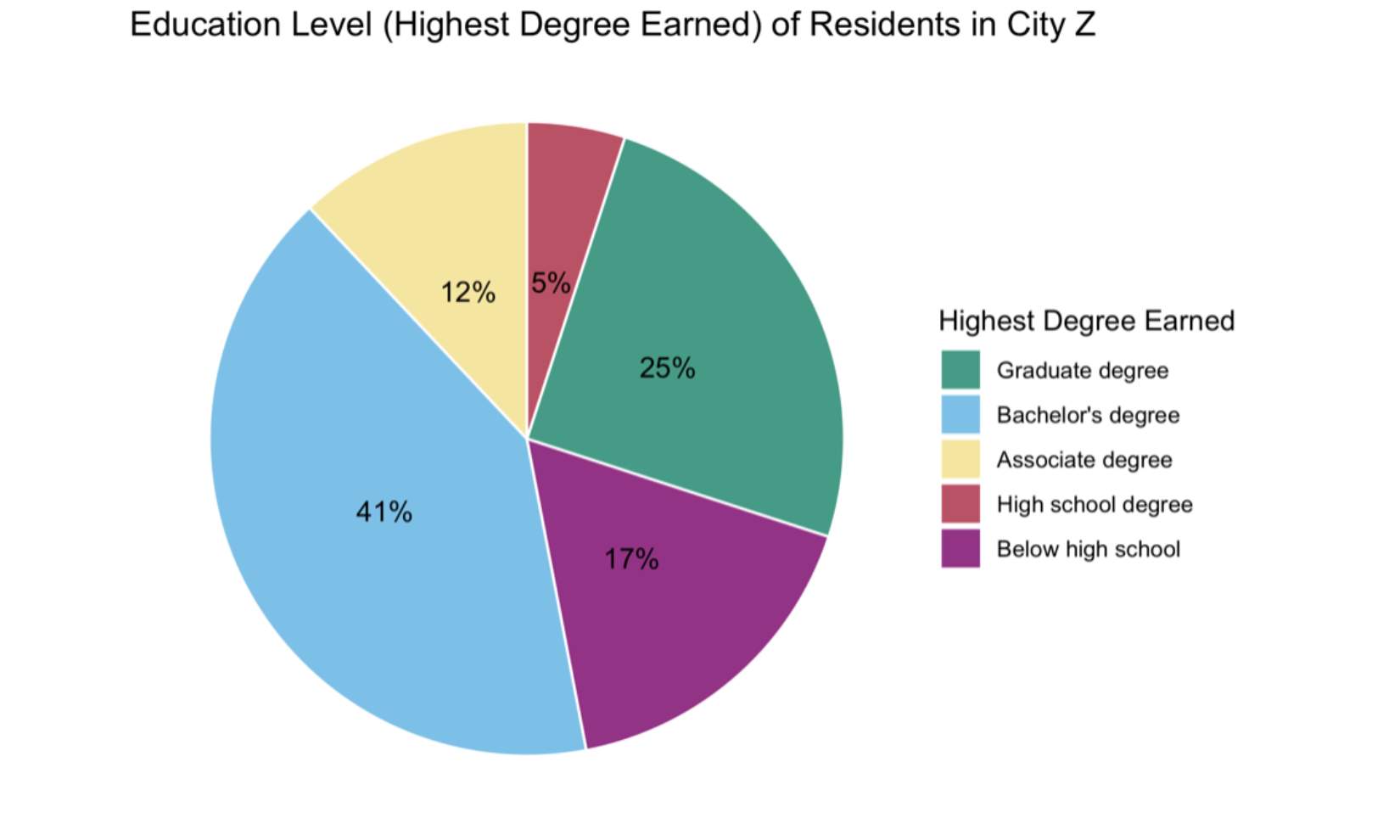}
\QArow{CLAVI-Standard}{Which of the following is true about Restaurant R?,It is more popular in region E than in D.,It is less popular in region A than in C.,It is more popular in region B than in C.,It is less popular in region C than in D.}{It is less popular in region A than in C.}{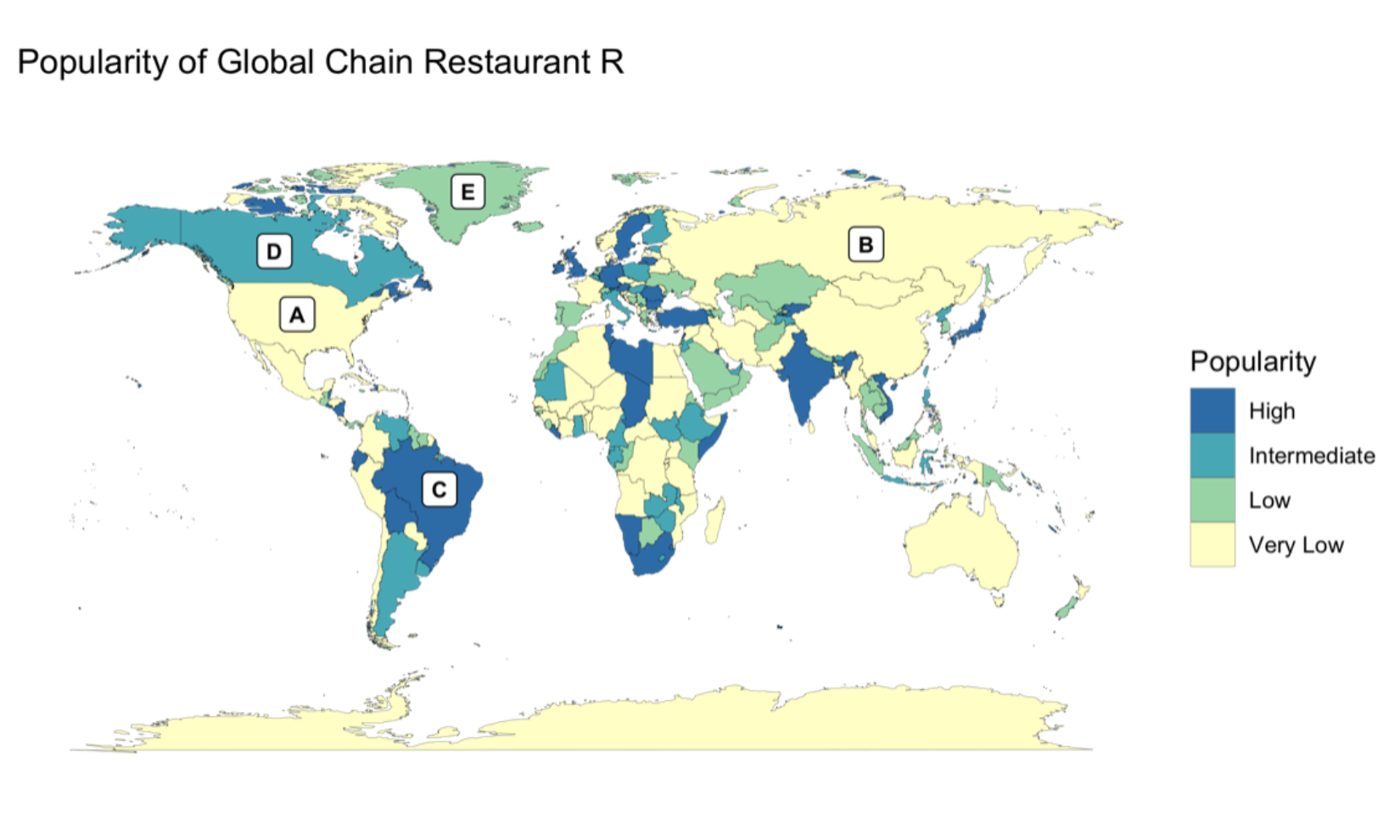}

\QArow{CLAVI-Trick}{Do U.S. states have a more similar number of sunny days in the summer or in the winter?,Summer,Winter,Cannot be inferred / Inadequate Information,}{Cannot be inferred / Inadequate Information}{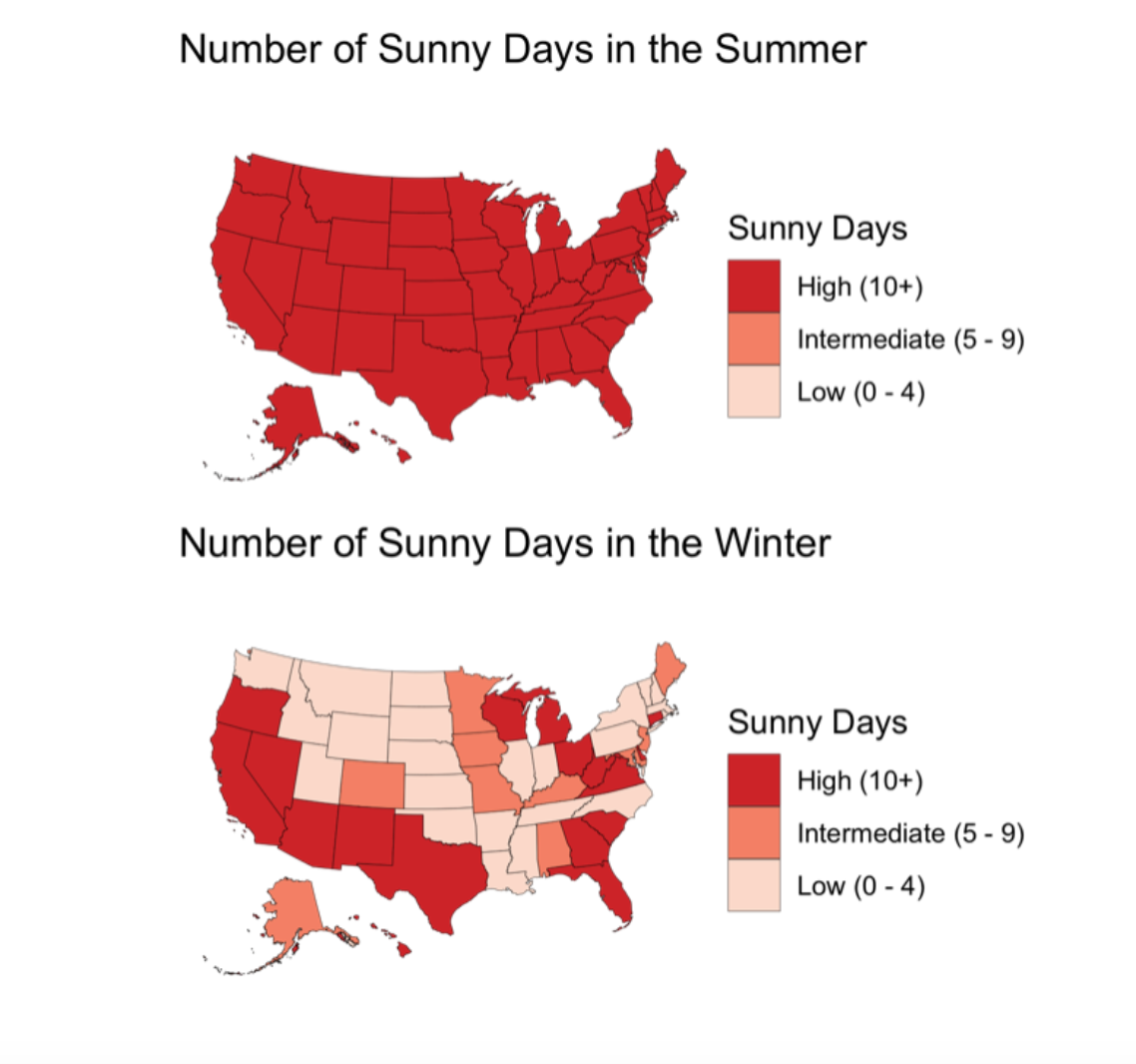}
\QArow{CLAVI-Trick}{Do U.S. states have a more similar number of windy days in the summer or in the winter?,Summer,Winter,Cannot be inferred / Inadequate Information,}{Cannot be inferred / Inadequate Information}{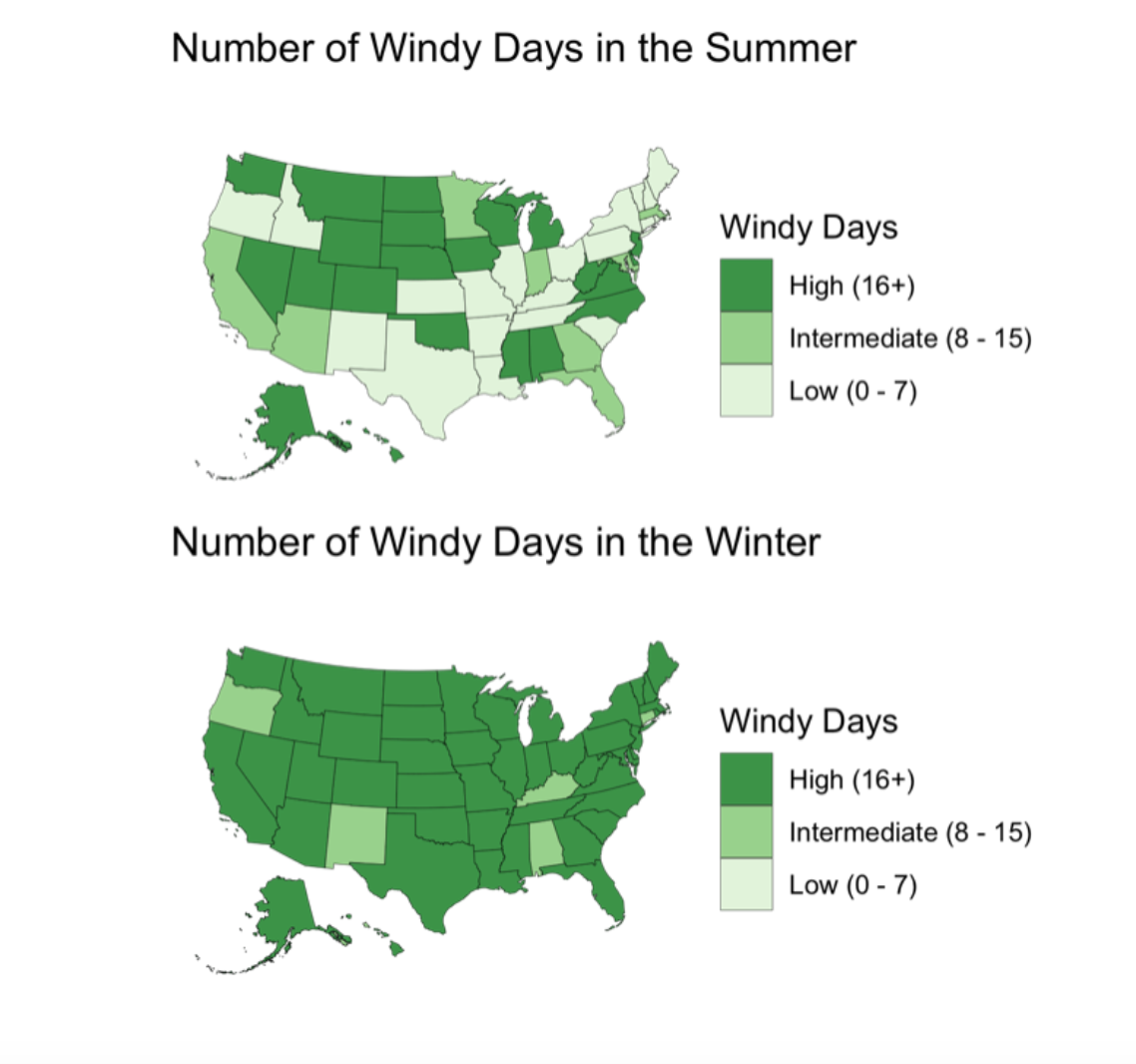}
\QArow{CLAVI-Trick}{What is the number of concert tickets sold on Aug 10 as a proportion of that on Aug 11?,20\%,40\%,60\%,80\%}{80\%}{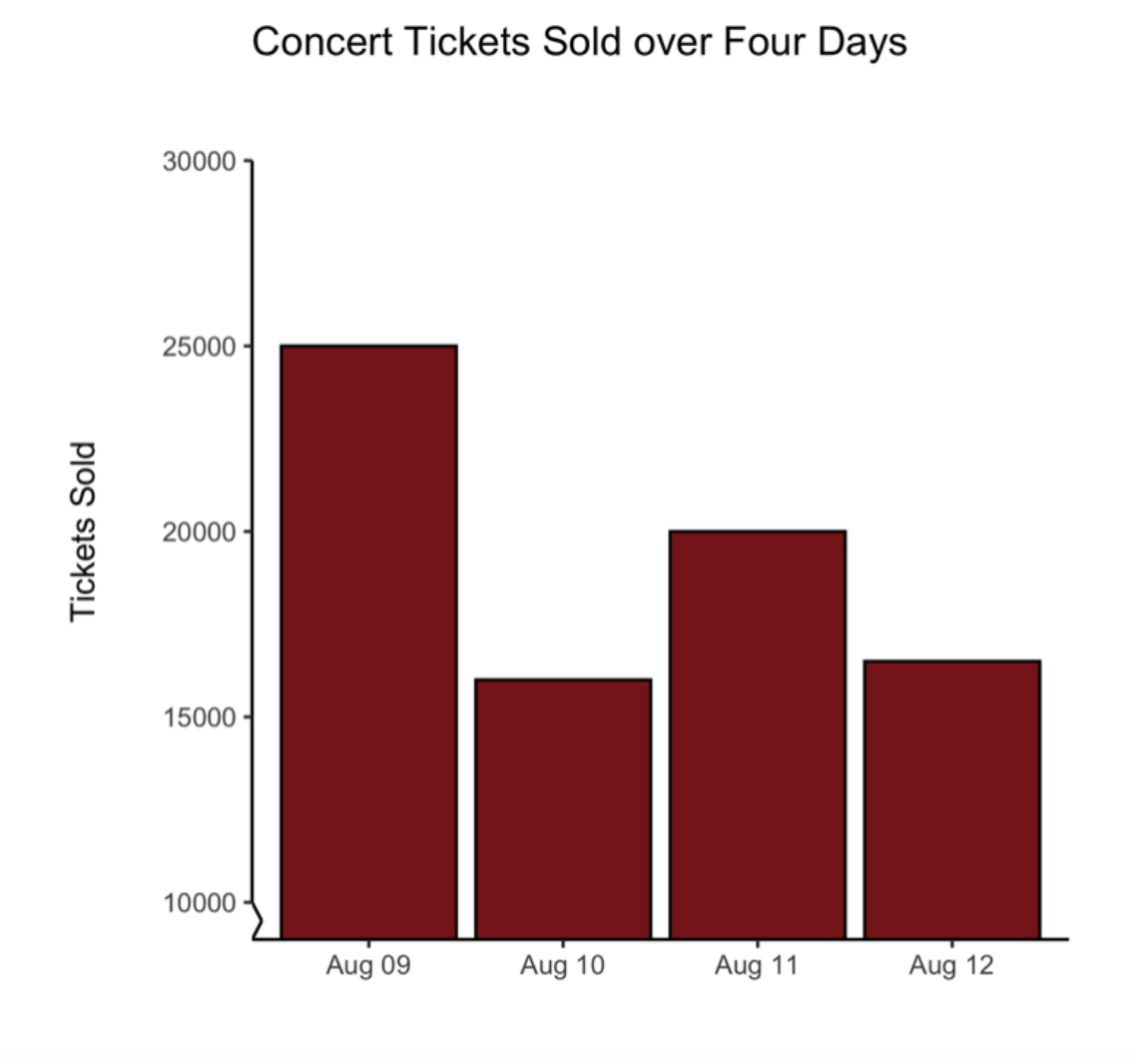}
\QArow{CLAVI-Trick}{What is the trend for the percentage of bacteria Z from 1970 to 2020?,Generally increasing,Generally decreasing,Generally constant,Cannot be inferred / Inadequate Information}{Generally increasing}{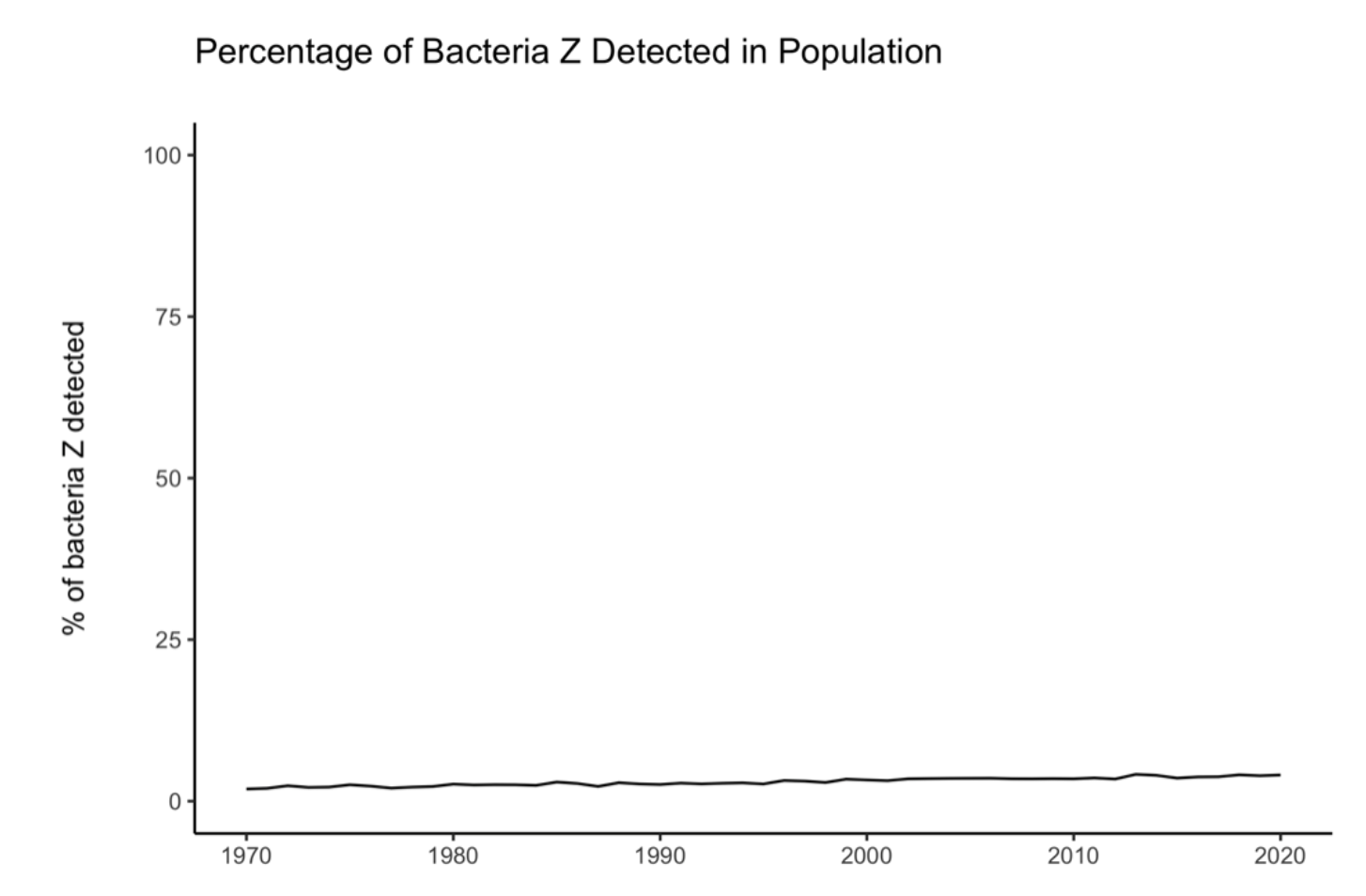}
\QArow{CLAVI-Trick}{"Approximately, what is brand C’s total sales as a proportion of brand B’s total sales from Jan 01 to Mar 15?",20\%,50\%,60\%,90\%}{60\%}{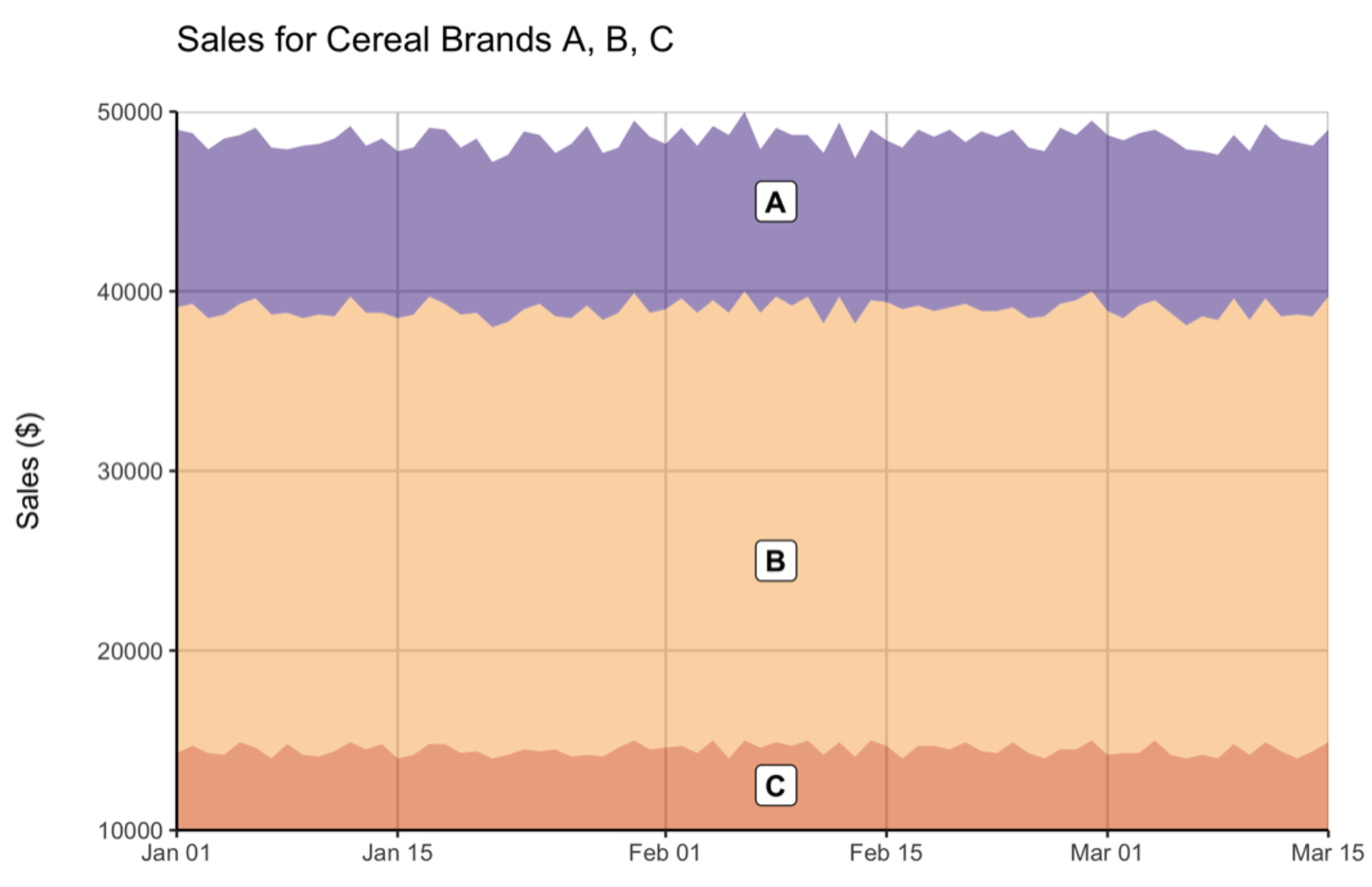}
\QArow{CLAVI-Trick}{What is the trend of the total sales for item X from Apr to Aug?,"Decreasing, then increasing","Increasing, then decreasing",Cannot be inferred / Inadequate Information}{Increasing, then decreasing}{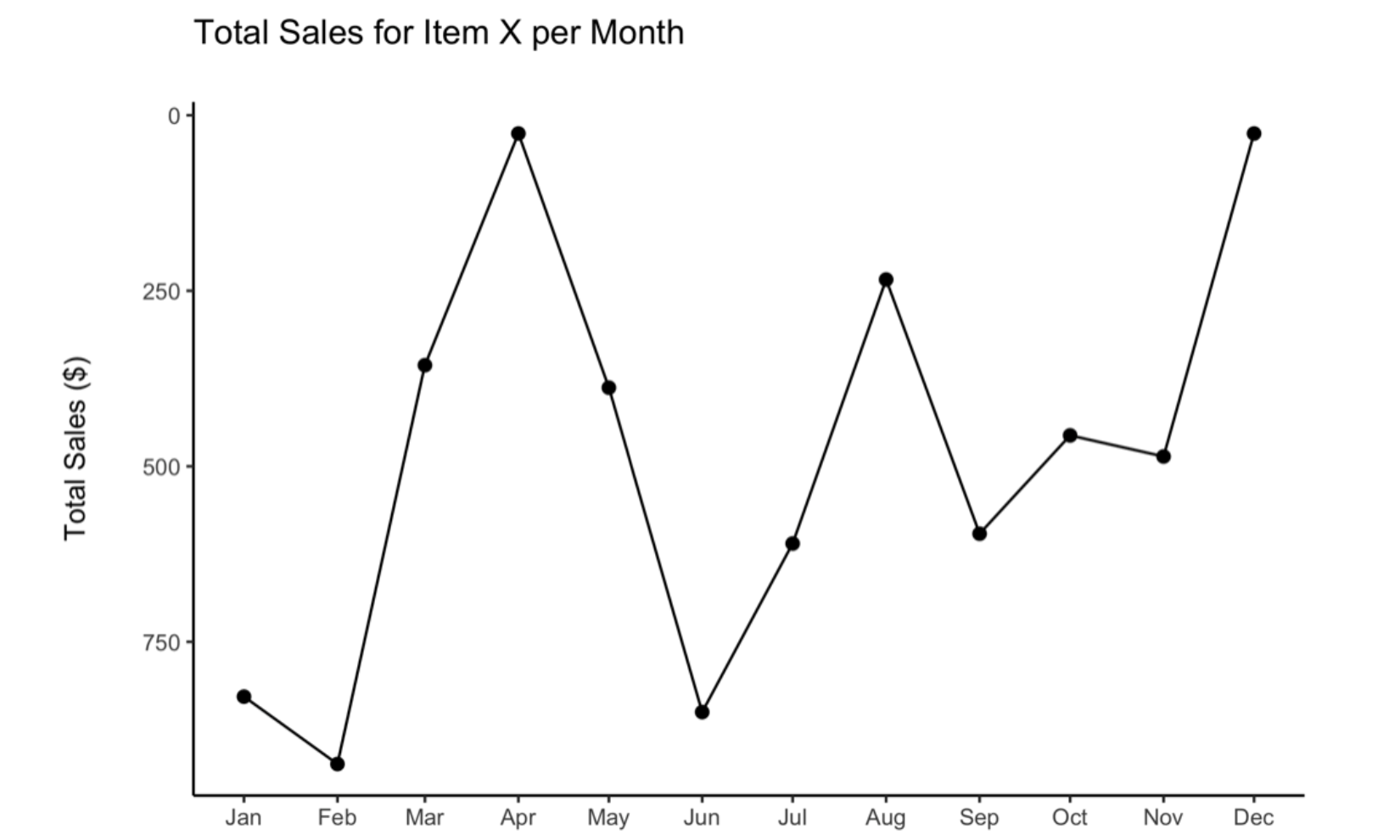}
\QArow{CLAVI-Trick}{What is the trend of the total number of apples harvested in farm X from Jan 11 to Jan 14?,"Increasing, then decreasing","Decreasing, then increasing",Cannot be inferred / Inadequate Information}{Decreasing, then increasing}{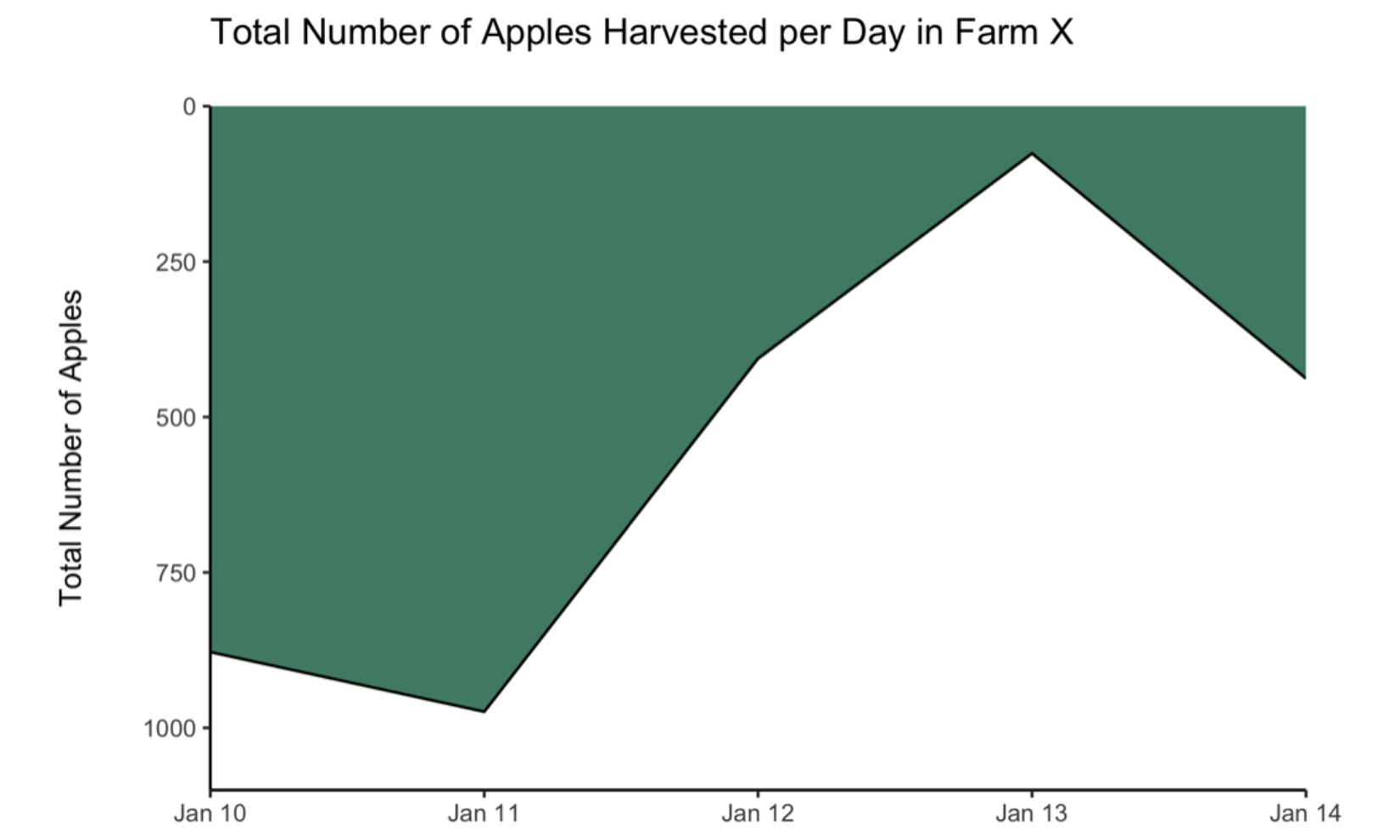}
\QArow{CLAVI-Trick}{What is the trend of the growth for plant species P as the number of hours of sunlight exposure per month increases?,Generally increases,Generally decreases,Generally constant,Cannot be inferred / Inadequate Information}{Generally decreases}{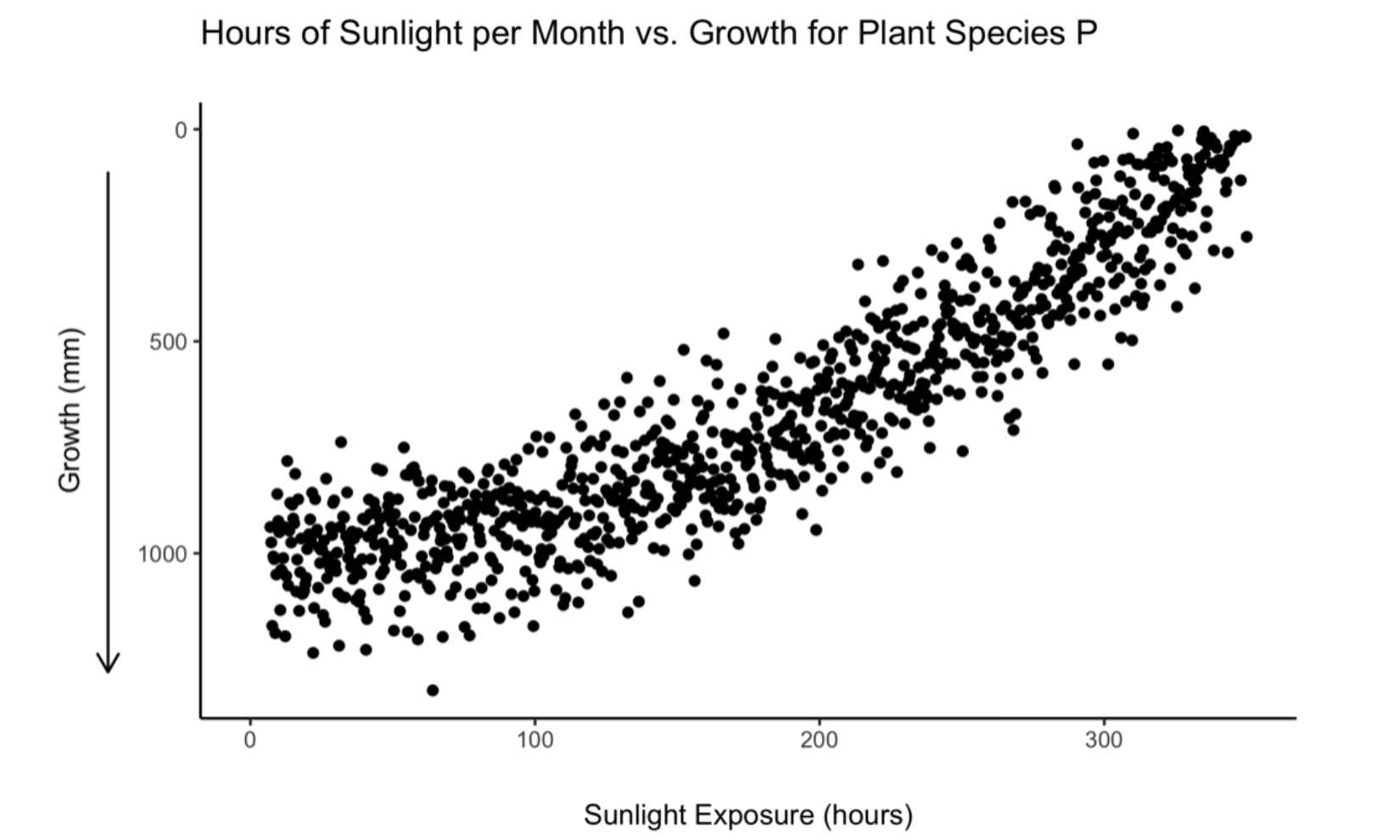}
\QArow{CLAVI-Trick}{What is the trend of the survival rate as weight increases for species Z?,Generally increases,Generally decreases,Generally constant,Cannot be inferred / Inadequate Information}{Generally increases}{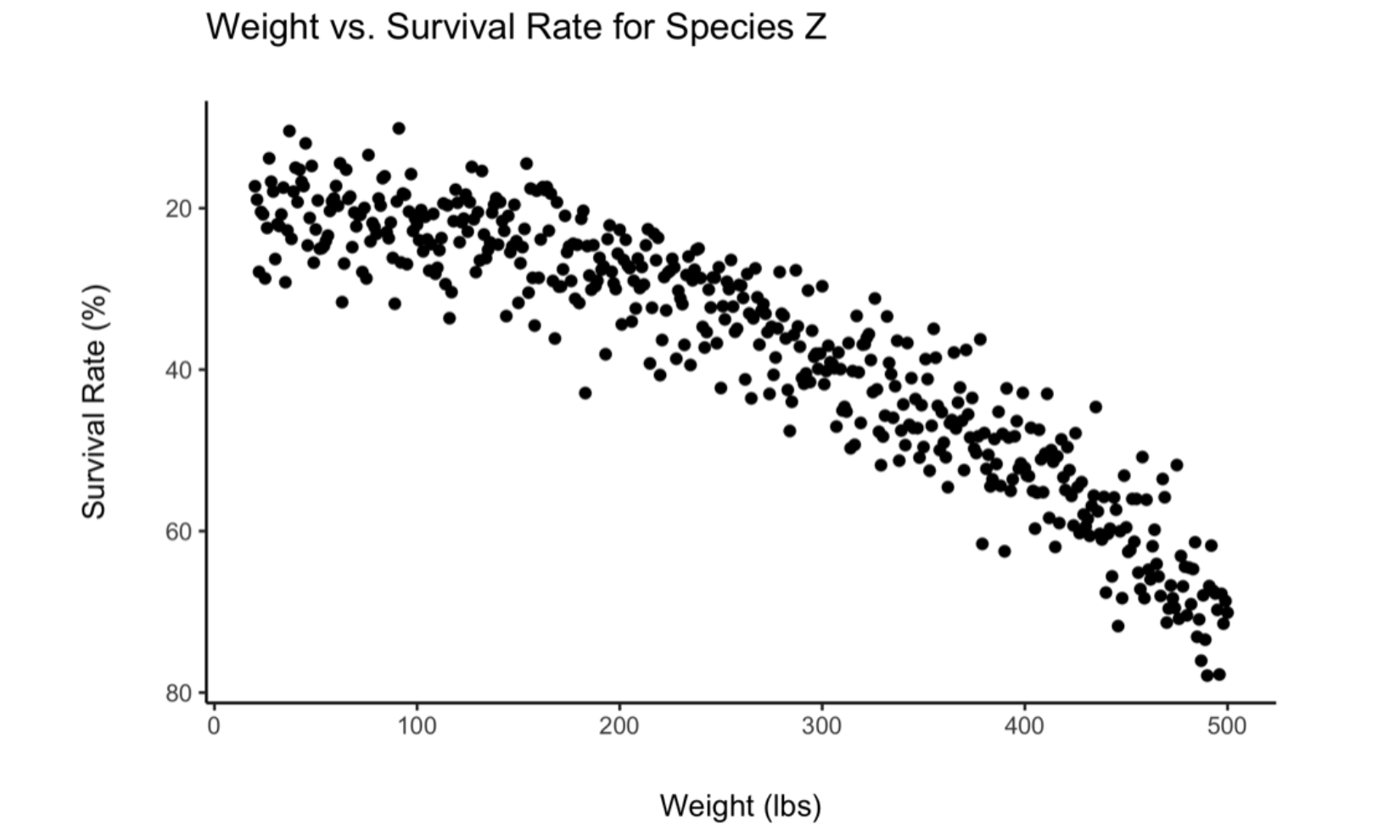}
\QArow{CLAVI-Trick}{"Assuming today is May 10 and given the chart below, which statement about rainfall (mm) is true for the past 14 days?",The amount of rainfall peaked at the end of the past 14 days.,There was more rainfall at the beginning than at the end of the past 14 days.,None of the days had lower than 40mm of rainfall.,The amount of rainfall peaked at the beginning of the past 14 days.}{There was more rainfall at the beginning than at the end of the past 14 days.}{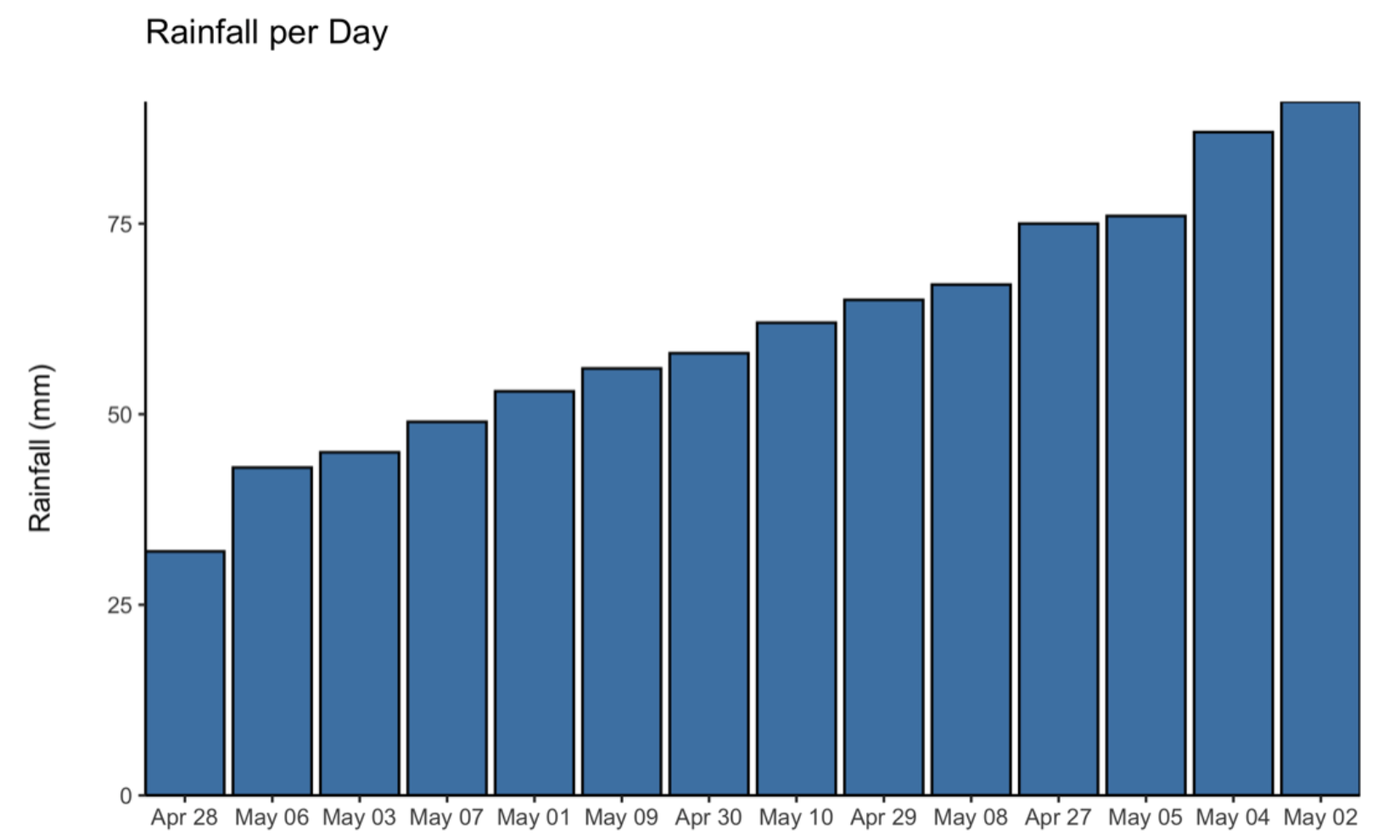}
\QArow{CLAVI-Trick}{Does cell phone brand A have more than half of the total market share?,Yes,No,Cannot be inferred / Inadequate Information,}{Cannot be inferred / Inadequate Information}{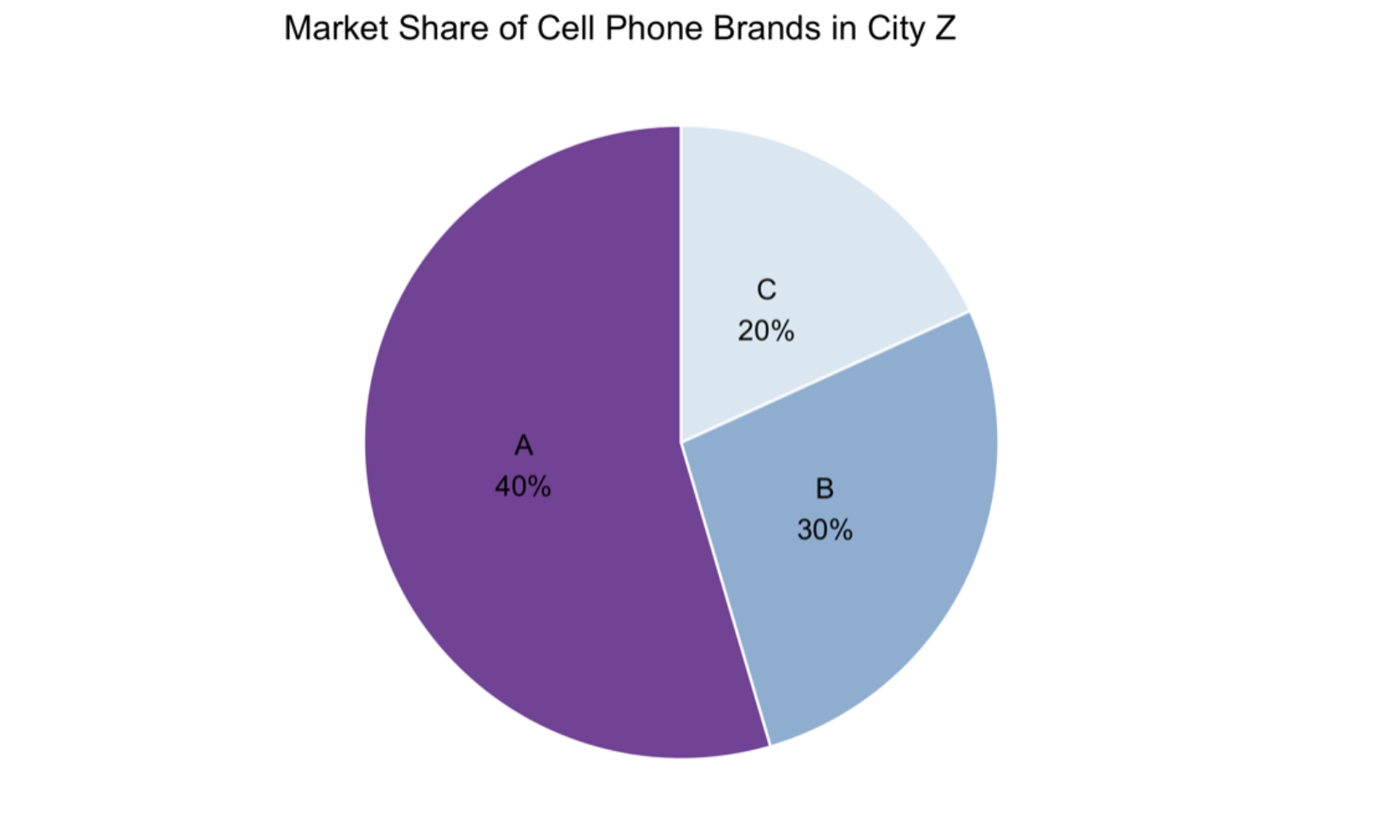}
\QArow{CLAVI-Trick}{Are there more households with exactly 1 car than households with exactly 2 cars?,Yes,No,Cannot be inferred / Inadequate Information,}{Cannot be inferred / Inadequate Information}{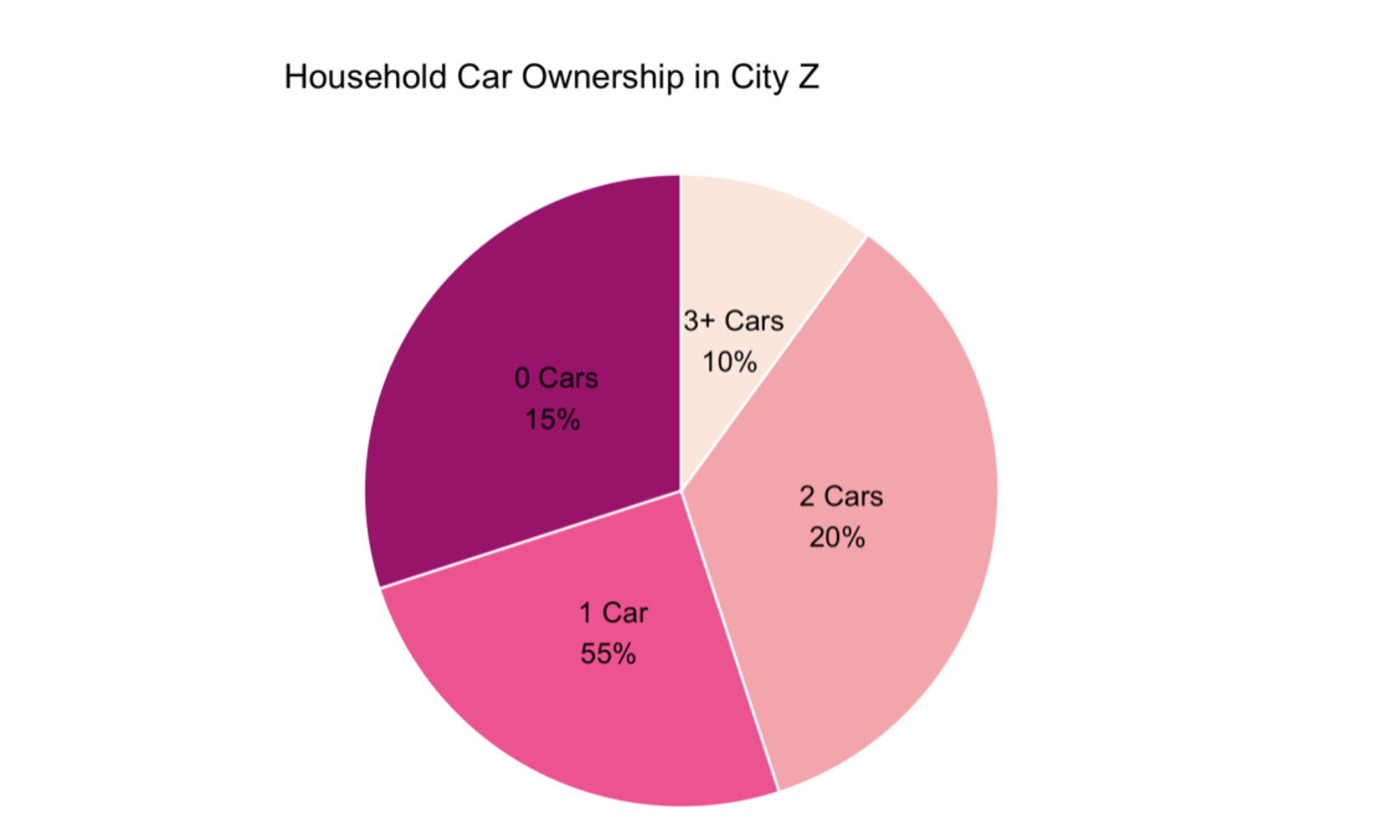}
\QArow{CLAVI-Trick}{Was the average amount of precipitation over 15 years in region D higher than that in region C?,Yes,No,Cannot be inferred / Inadequate Information,}{Cannot be inferred / Inadequate Information}{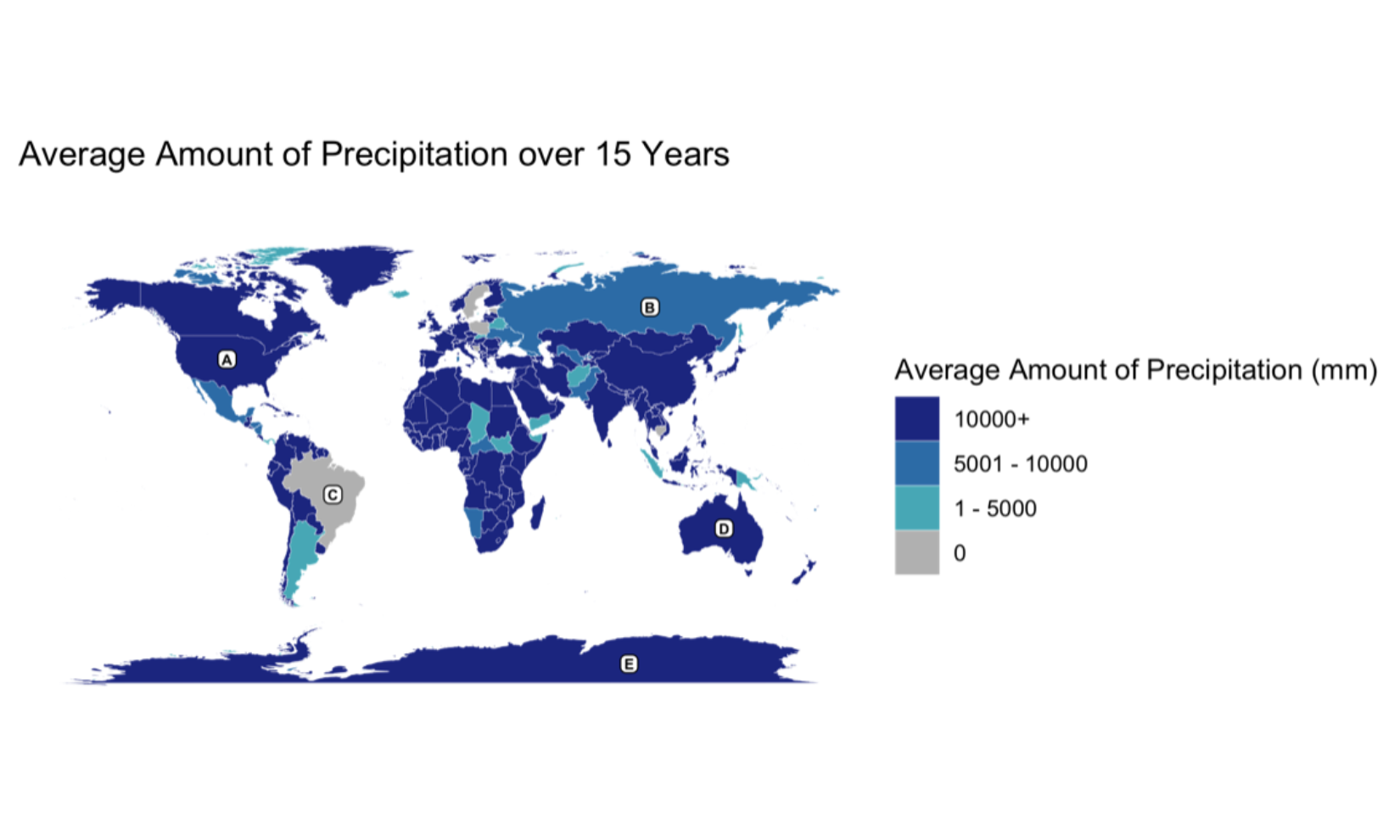}
\QArow{CLAVI-Trick}{"Company A sells a board game. Assuming today is Jun 01, 2022, which of the following statements is true?",2004 has more copies sold than 2003.,Copies sold per year fell sharply in 2022.,Copies sold per year has been strictly increasing since 2003.,None of the above.}{None of the above.}{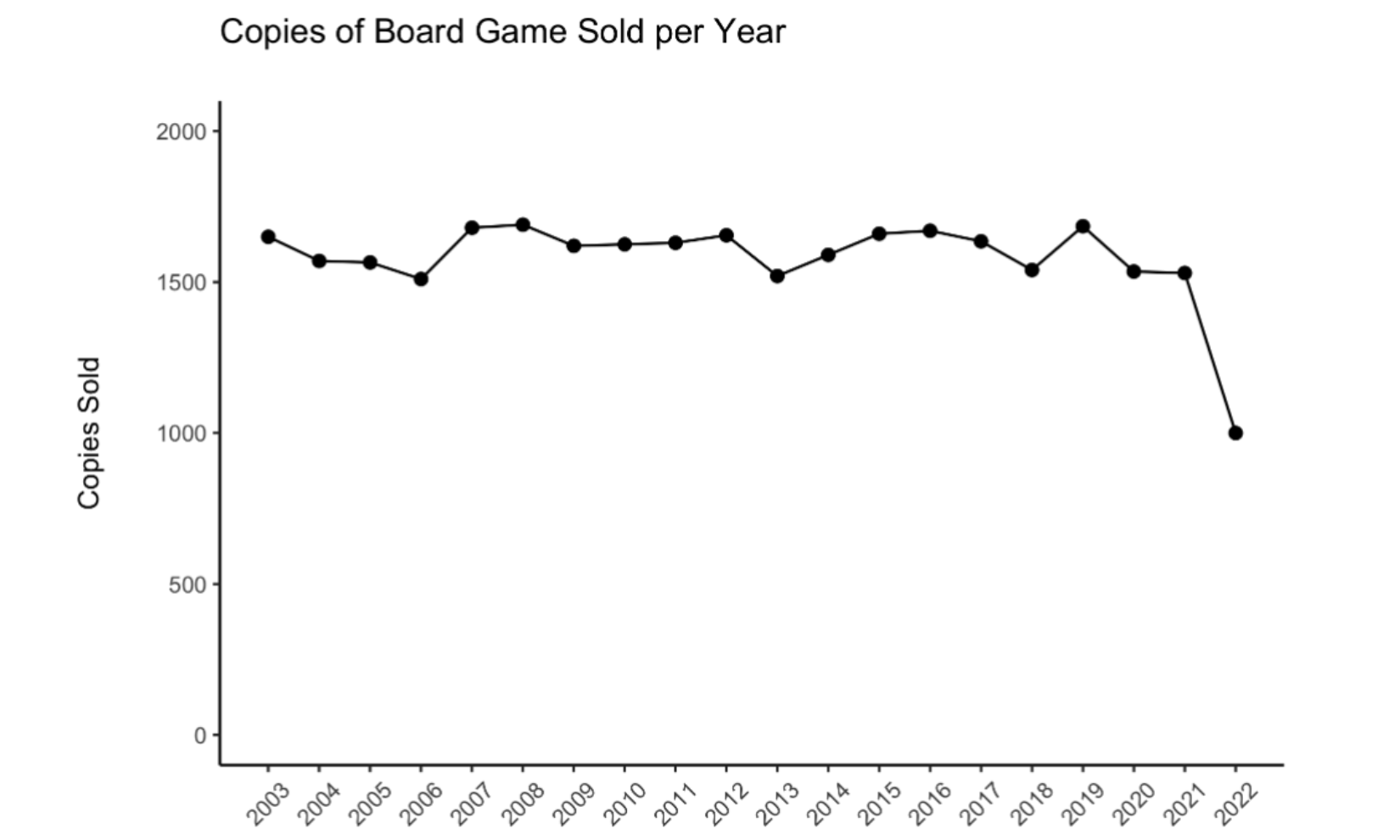}
\QArow{CLAVI-Trick}{"Assuming today is Feb 20, 2022, which of the following is true about the vegetables planted by the three farms?",All three farms planted more vegetables in 2022 than they did in the previous years.,All three farms planted fewer vegetables in 2022 than they did in the previous years.,All three farms planted about the same amount of vegetables in 2022 than they did in the previous years.,None of the above.}{None of the above.}{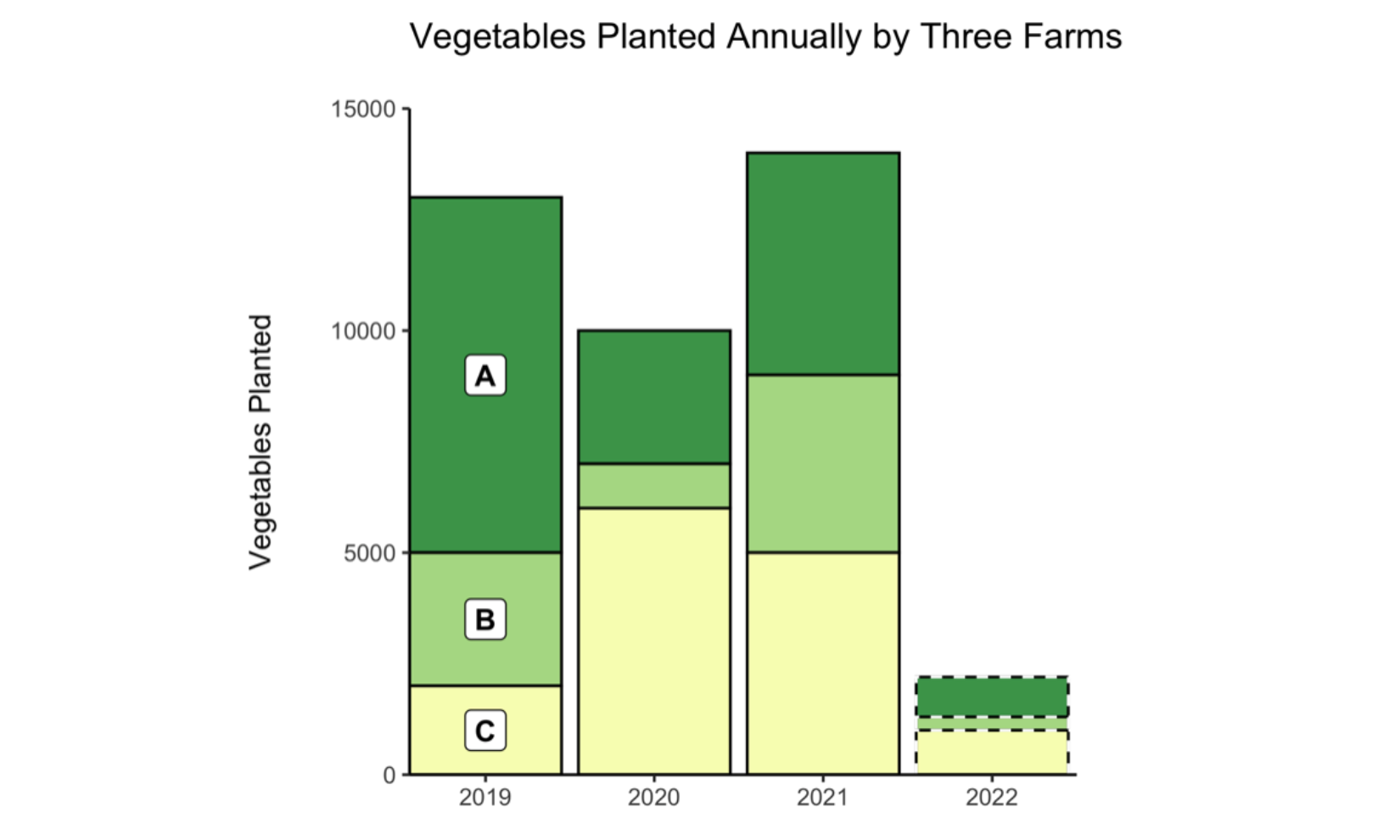}
\QArow{CLAVI-Trick}{"Assuming today is Apr 10, 2022, did the total number of cars parked in the three parking lots fall sharply in 2022?",Yes,No,Cannot be inferred / Inadequate Information,}{Cannot be inferred / Inadequate Information}{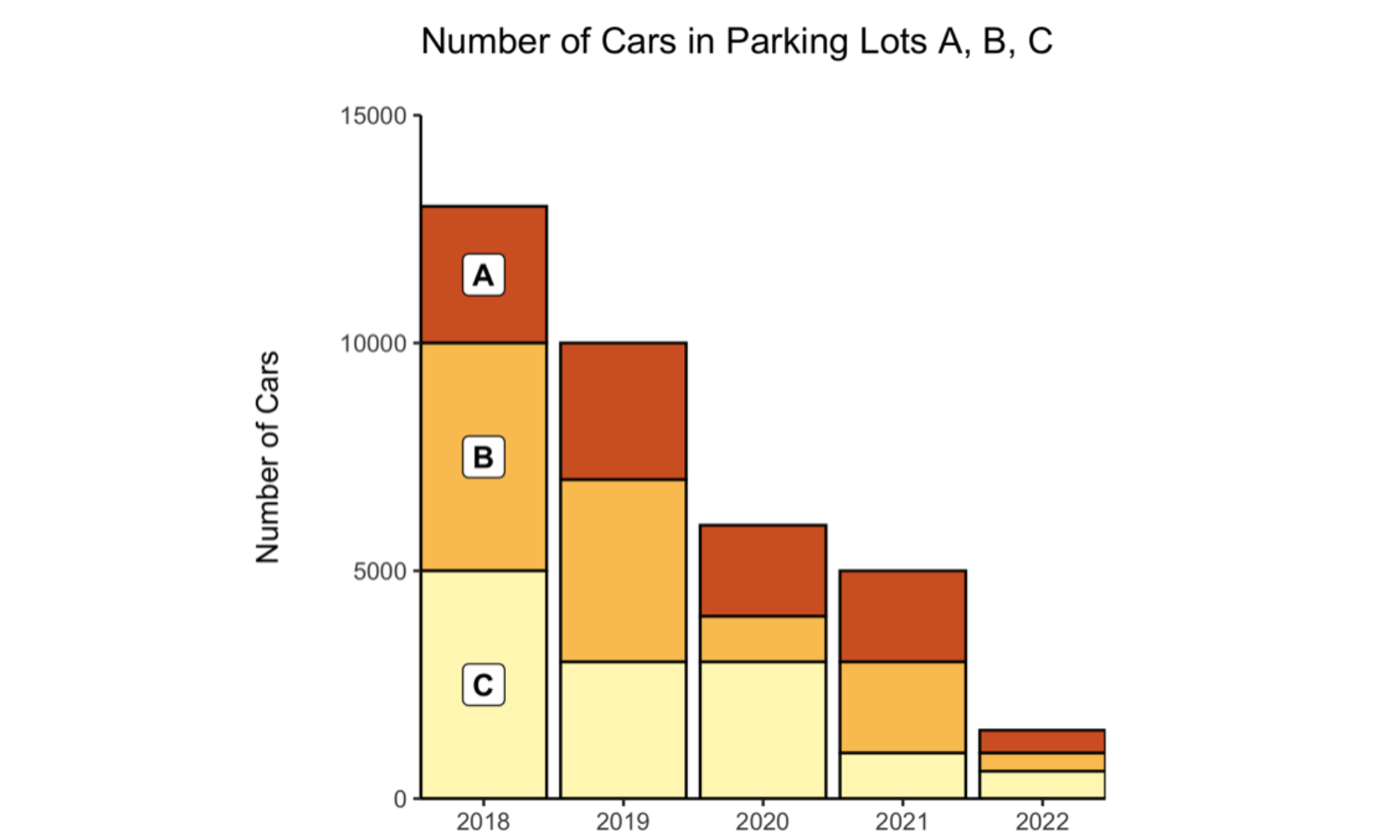}
\QArow{CLAVI-Trick}{"Assuming today is May 01, 2022, which of the following statements is true?",Company Z had more new employees in 2018 than in 2020.,The number of new employees per year fell sharply in 2022.,The total number of employees in company Z decreased in 2022.,None of the above.}{None of the above.}{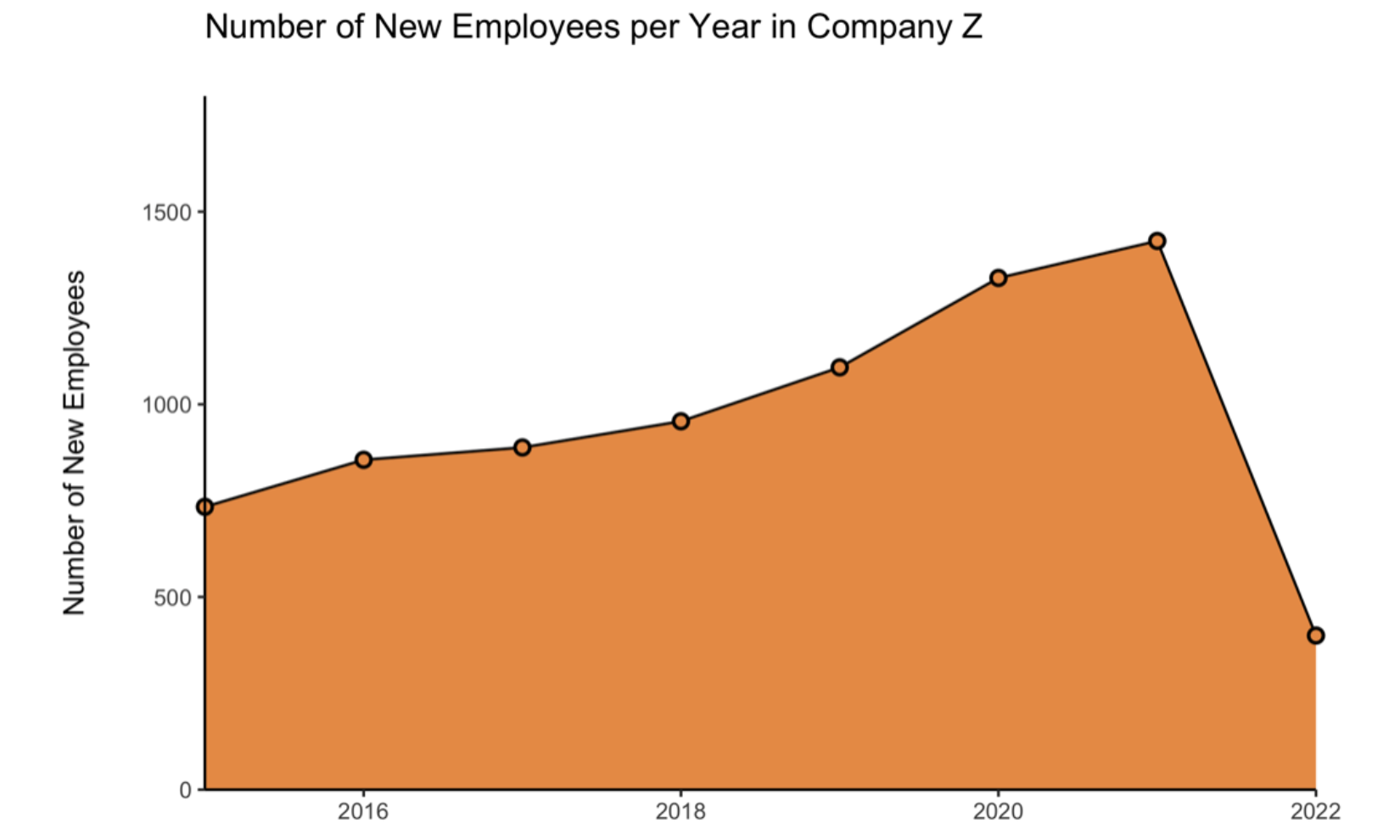}
\QArow{CLAVI-Trick}{"True or False: Within each town, is there a negative relationship between average daily working hours and happiness score?",TRUE,FALSE,Cannot be inferred / Inadequate Information,}{FALSE}{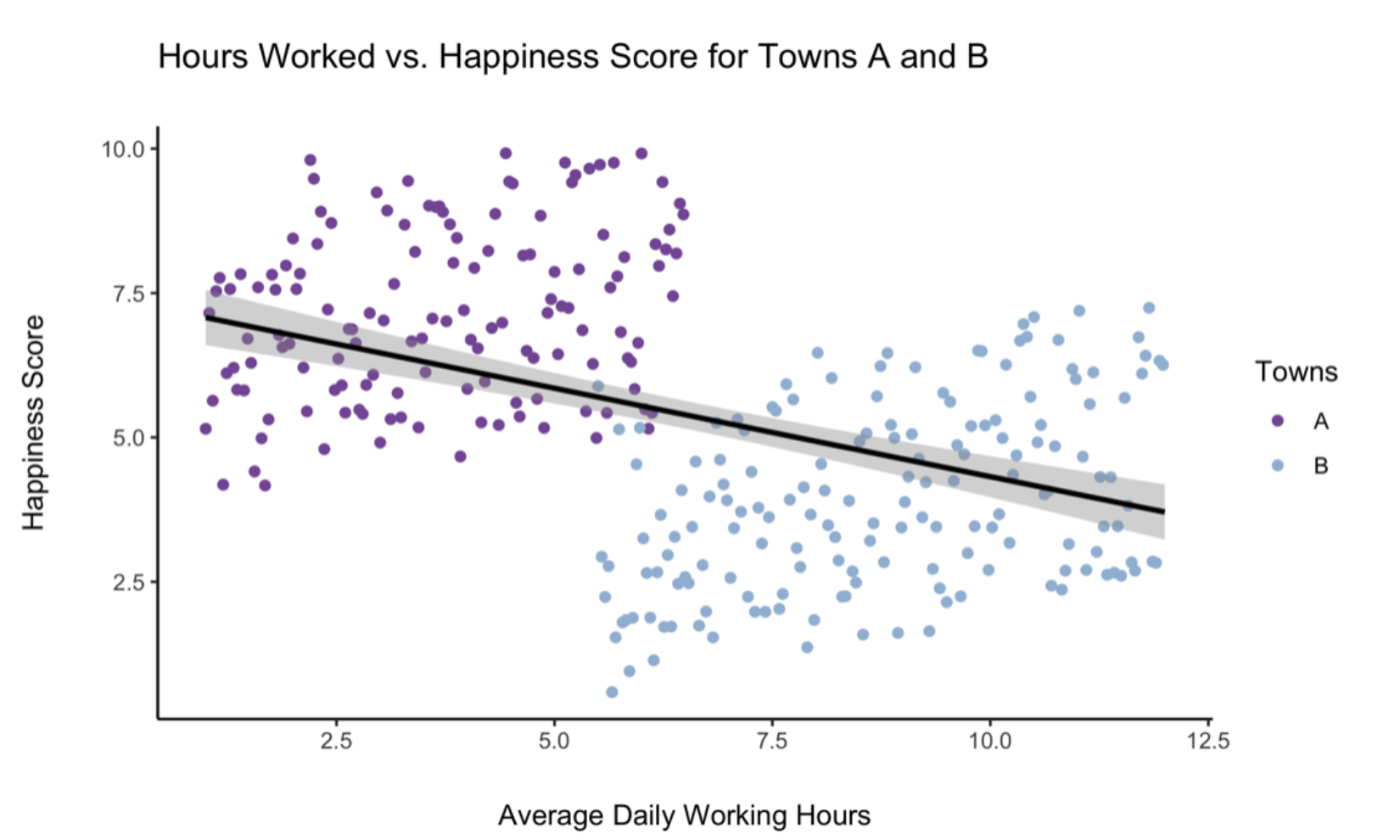}
\QArow{CLAVI-Trick}{"Within each city, is there a positive relationship between salary and average daily working hours?",Yes,No,Cannot be inferred / Inadequate Information,}{Cannot be inferred / Inadequate Information}{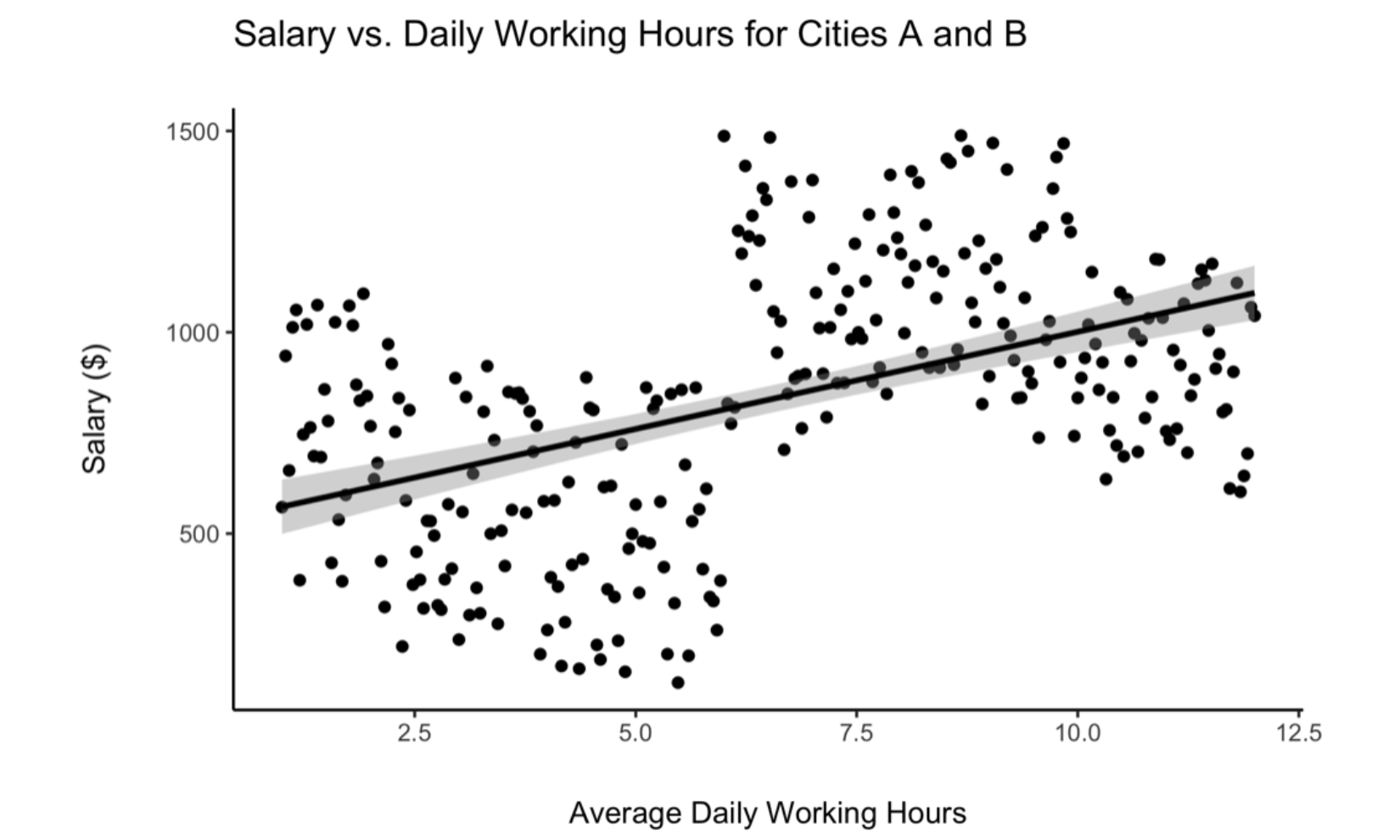}
\QArow{CLAVI-Trick}{Does any members of species C weigh more than 5 lbs?,Yes,No,Cannot be inferred / Inadequate Information,}{Cannot be inferred / Inadequate Information}{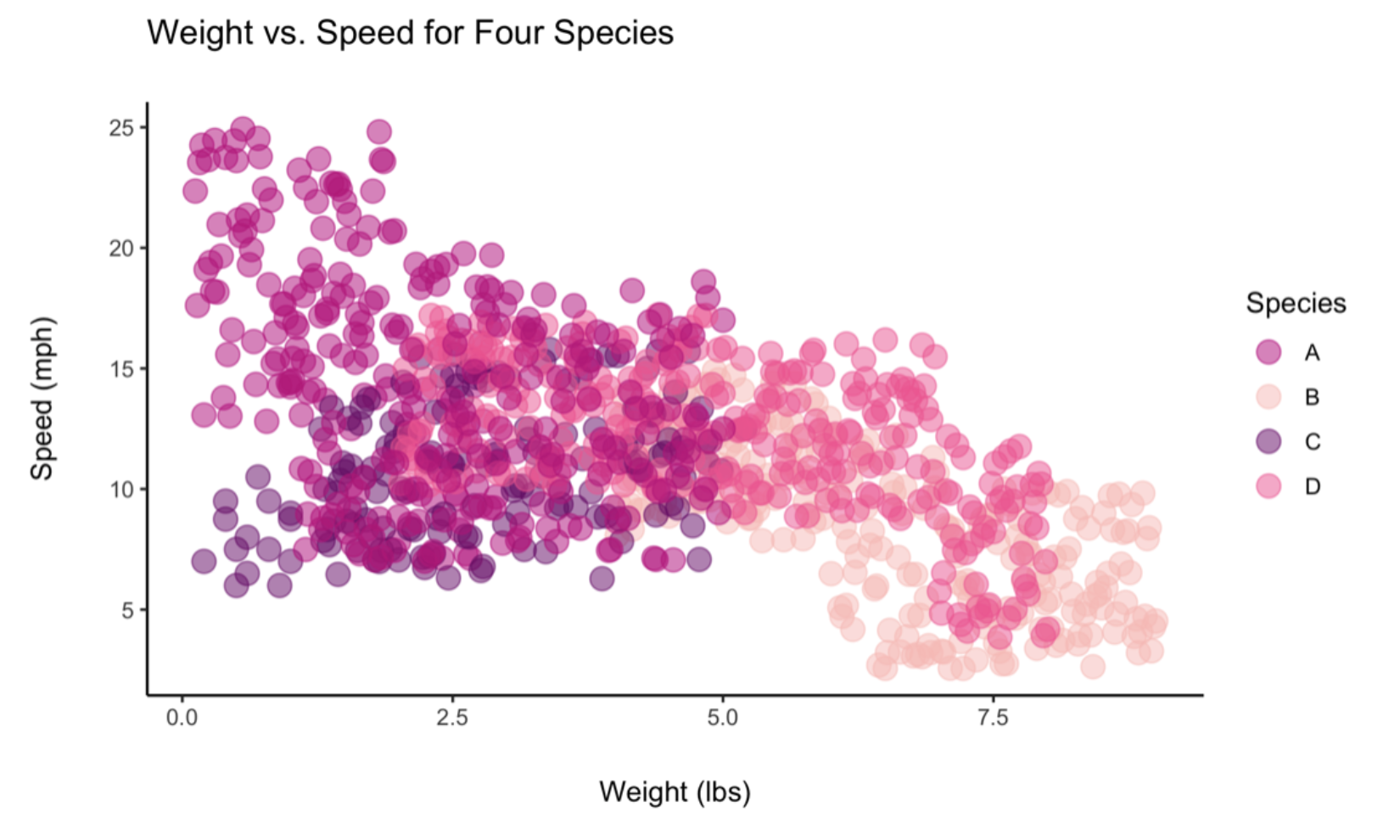}

\QArow{HOLF}{How many miles per hour faster was the fastest storm compared to the slowest storm?}{21.61}{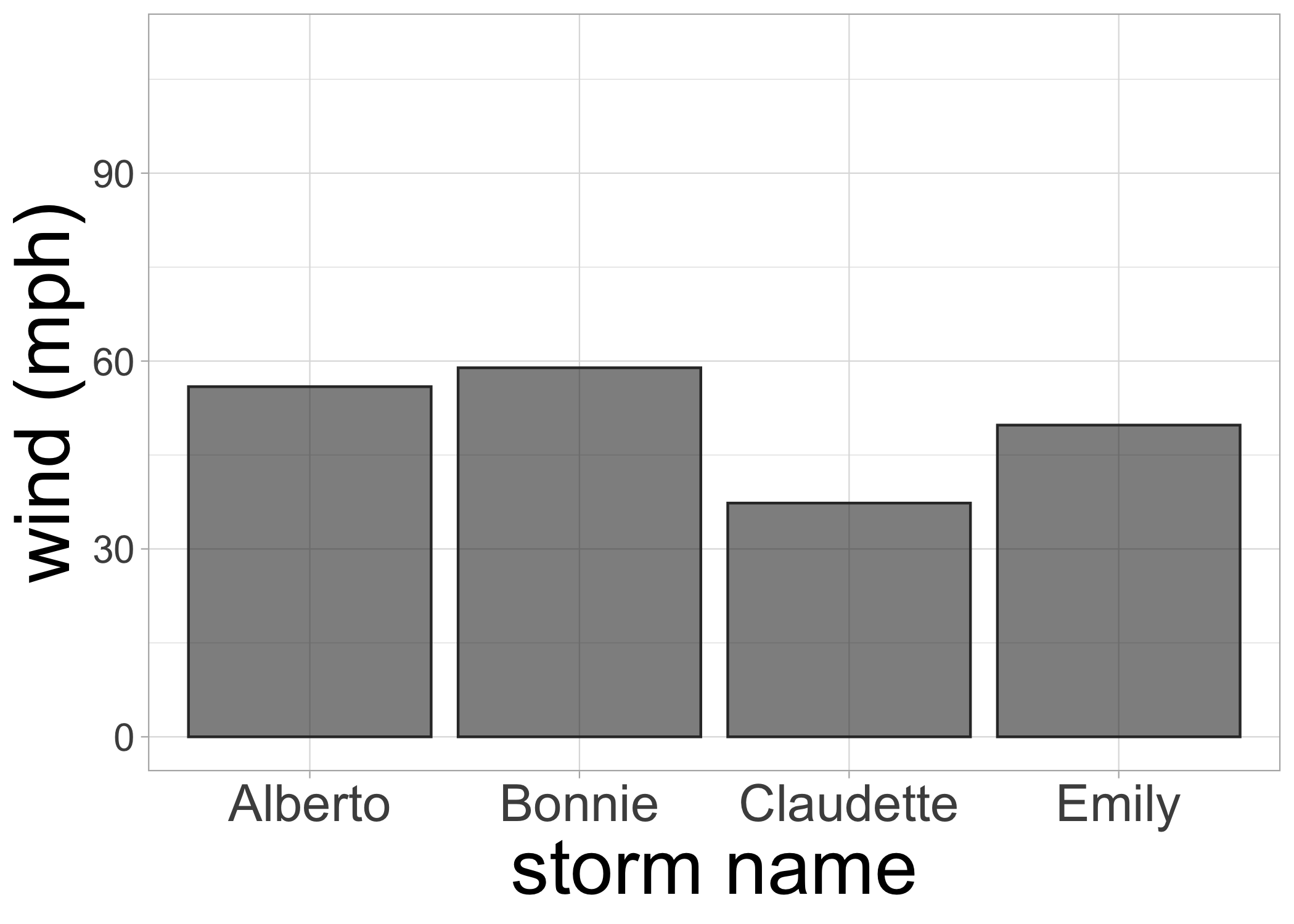}
\QArow{HOLF}{On average, how many miles per hour faster are storms categorized as hurricanes compared to those categorized as tropical depressions?}{56.29}{all_figures/storms_x1name_facetnone_reorderalph_colorgrey20.png}
\QArow{HOLF}{On average, how many miles per hour was Emily when categorized as a hurricane?}{85.0}{all_figures/storms_x1name_facetnone_reorderalph_colorgrey20.png}
\QArow{HOLF}{Within storms categorized as hurricanes, how many more miles per hour was the fastest storm compared to the slowest storm?}{12.08}{all_figures/storms_x1name_facetnone_reorderalph_colorgrey20.png}
\QArow{HOLF}{On average, how many miles per hour faster was Hurricane Alberto compared to Hurricane Claudette?}{2.0}{all_figures/storms_x1name_facetnone_reorderalph_colorgrey20.png}
\QArow{HOLF}{On average, how many miles per hour are storms categorized as tropical storms?}{43.82}{all_figures/storms_x1name_facetnone_reorderalph_colorgrey20.png}

\QArow{HOLF}{On average, what is injury severity level of drivers?}{1.78}{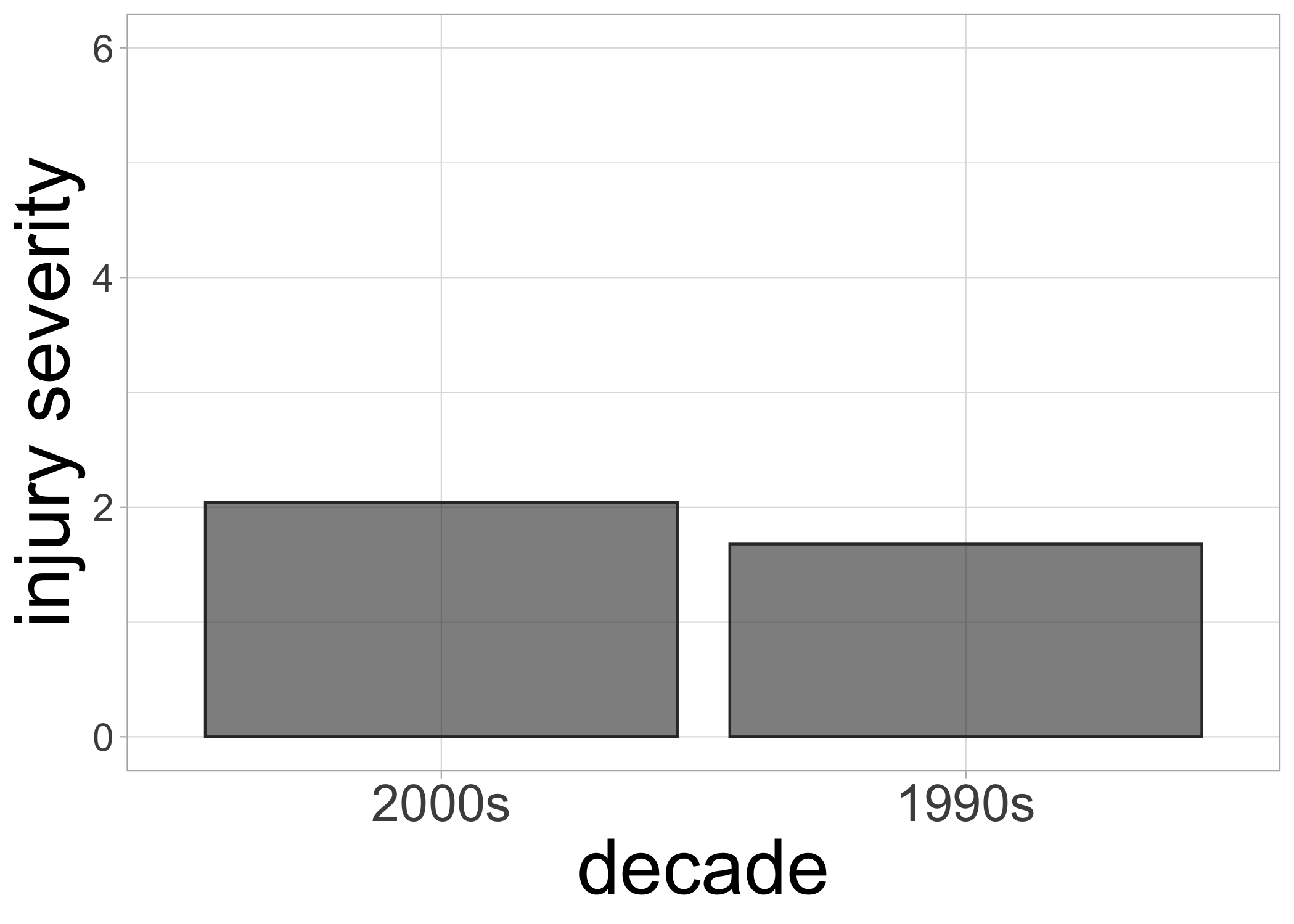}
\QArow{HOLF}{Within of injuries involving backseat passengers, how much higher is the injury severity level in the decade with the higher severity compared to the decade with the lower severity?}{0.96}{all_figures/accidents_x1decade_facetnone_reorderlarge_to_small_colorgrey20.png}
\QArow{HOLF}{On average, what was the injury severity level of drivers in the 1990s?}{1.71}{all_figures/accidents_x1decade_facetnone_reorderlarge_to_small_colorgrey20.png}
\QArow{HOLF}{How much higher is the injury severity level of accidents in the decade with the higher severity compared to the lower severity?}{0.36}{all_figures/accidents_x1decade_facetnone_reorderlarge_to_small_colorgrey20.png}
\QArow{HOLF}{On average, how much higher are the injury severity levels of drivers compared to backseat passengers in the 1990s?}{0.21}{all_figures/accidents_x1decade_facetnone_reorderlarge_to_small_colorgrey20.png}
\QArow{HOLF}{On average, how much higher is the injury severity level of backseat passengers compared to drivers?}{0.31}{all_figures/accidents_x1decade_facetnone_reorderlarge_to_small_colorgrey20.png}

\QArow{HOLF}{How many miles per hour faster was the fastest storm compared to the slowest storm?}{21.61}{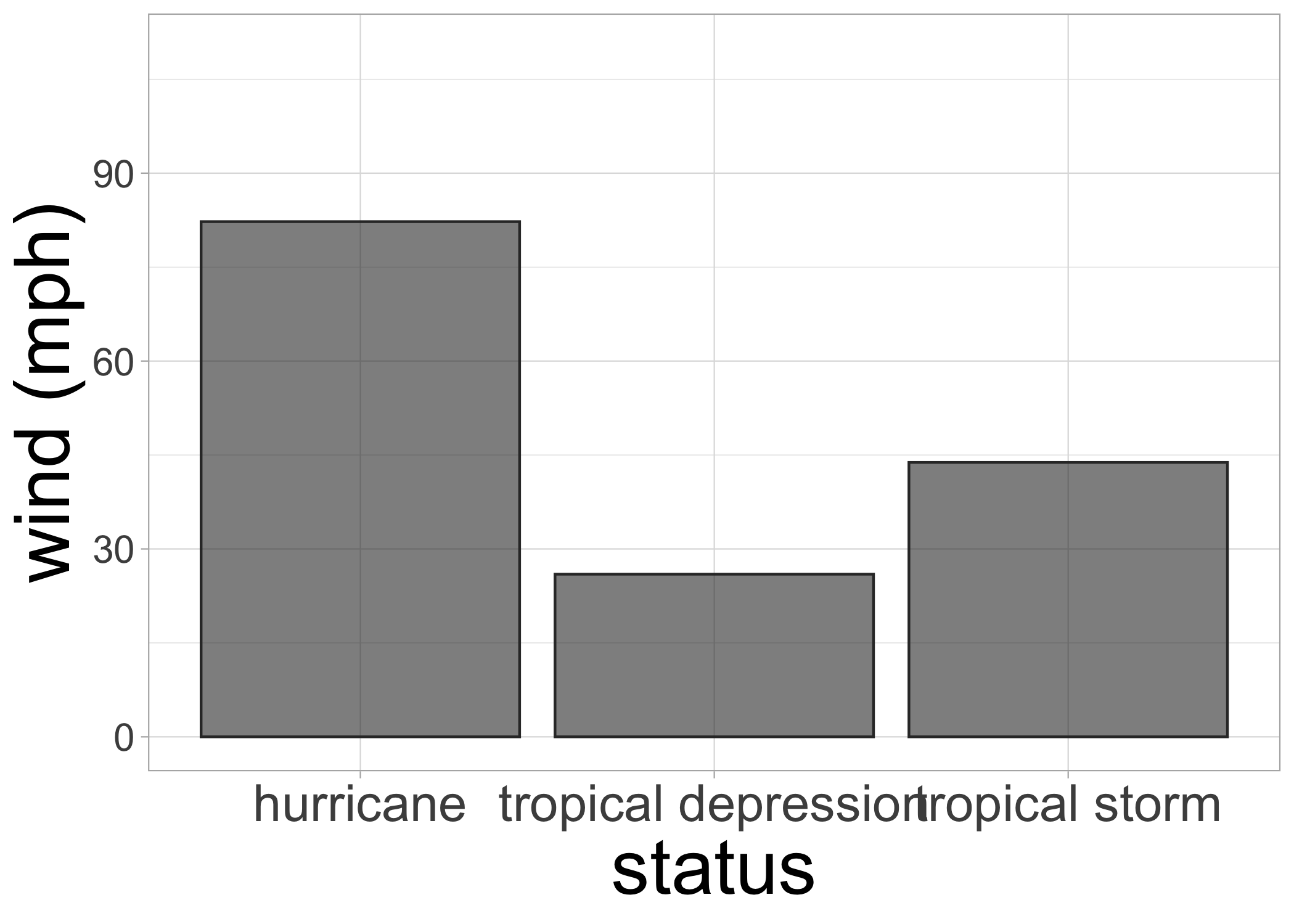}
\QArow{HOLF}{On average, how many miles per hour are storms categorized as tropical storms?}{43.82}{all_figures/storms_x1status_facetnone_reorderalph_colorgrey20.png}
\QArow{HOLF}{On average, how many miles per hour faster are storms categorized as hurricanes compared to those categorized as tropical depressions?}{56.29}{all_figures/storms_x1status_facetnone_reorderalph_colorgrey20.png}
\QArow{HOLF}{Within storms categorized as hurricanes, how many more miles per hour was the fastest storm compared to the slowest storm?}{12.08}{all_figures/storms_x1status_facetnone_reorderalph_colorgrey20.png}
\QArow{HOLF}{On average, how many miles per hour was Emily when categorized as a hurricane?}{85.0}{all_figures/storms_x1status_facetnone_reorderalph_colorgrey20.png}
\QArow{HOLF}{On average, how many miles per hour faster was Hurricane Alberto compared to Hurricane Claudette?}{2.0}{all_figures/storms_x1status_facetnone_reorderalph_colorgrey20.png}

\QArow{HOLF}{On average, how much longer does it take women to complete the race relative to men?}{0.25}{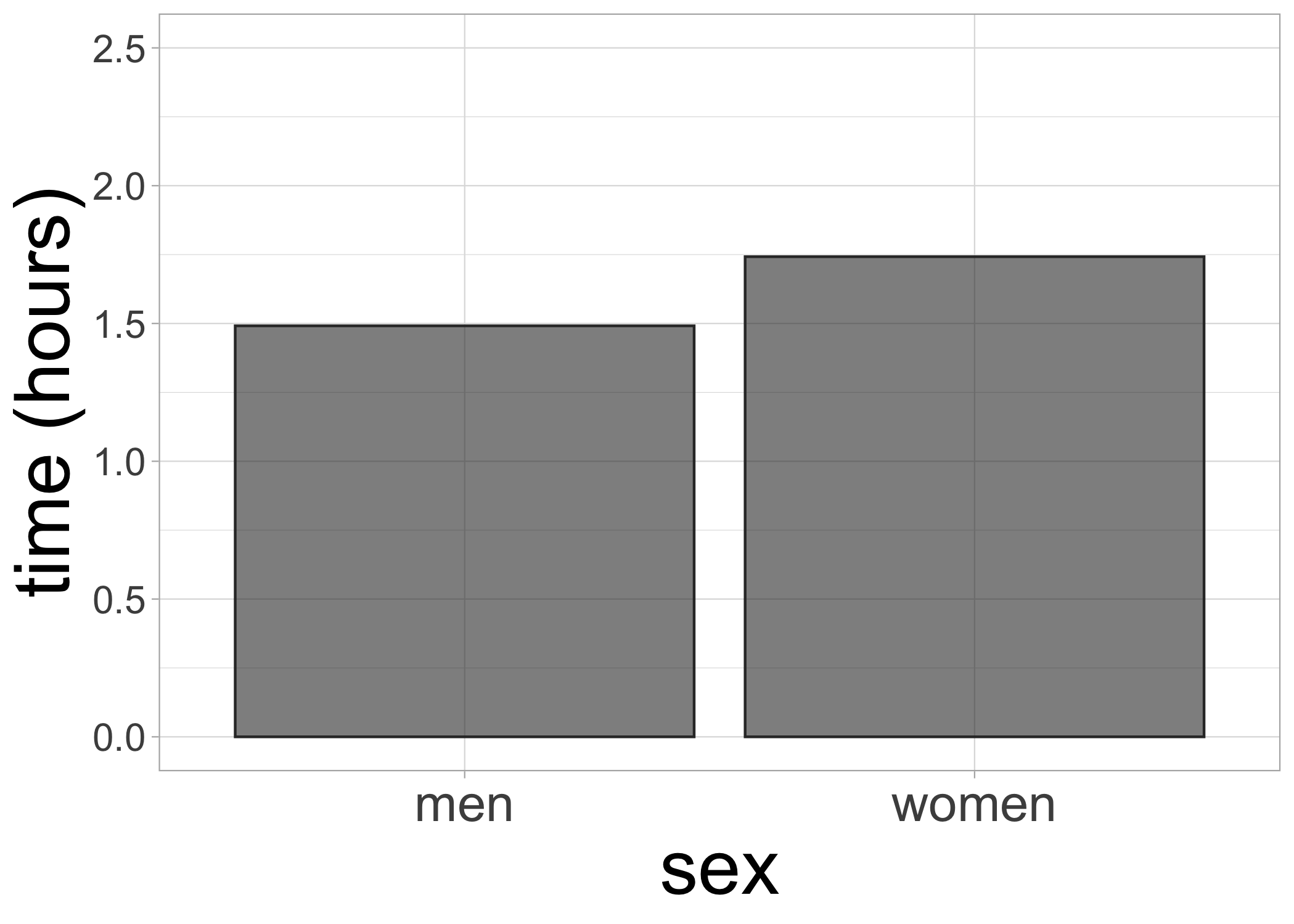}
\QArow{HOLF}{How much longer does it take runners from the state with the longest times to complete the race compared to runners from the state with the fastest times?}{0.05}{all_figures/race_x1sex_facetnone_reorderalph_colorgrey20.png}
\QArow{HOLF}{On average, how long does it take women from DC to complete the race?}{1.75}{all_figures/race_x1sex_facetnone_reorderalph_colorgrey20.png}
\QArow{HOLF}{On average, how long does it take women to complete the race?}{1.74}{all_figures/race_x1sex_facetnone_reorderalph_colorgrey20.png}
\QArow{HOLF}{On average, how much longer does it take women from VA to complete the race compared to women from MD?}{0.06}{all_figures/race_x1sex_facetnone_reorderalph_colorgrey20.png}
\QArow{HOLF}{Within women, how much longer does it take runners from the state with the longest times to complete the race compared to runners from the state with the fastest times?}{0.06}{all_figures/race_x1sex_facetnone_reorderalph_colorgrey20.png}

\QArow{HOLF}{On average, how long does it take women from DC to complete the race?}{1.75}{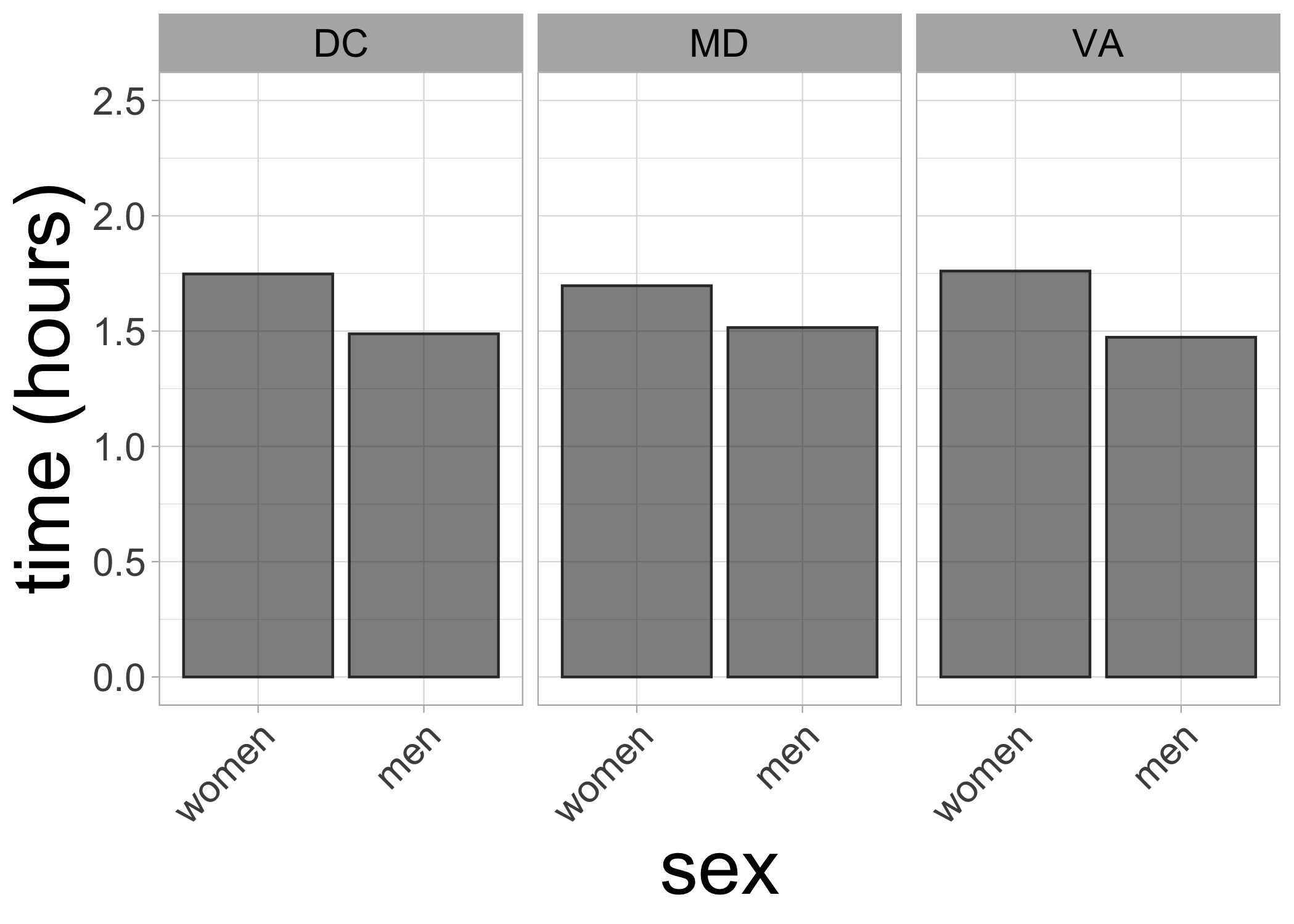}
\QArow{HOLF}{How much longer does it take runners from the state with the longest times to complete the race compared to runners from the state with the fastest times?}{0.05}{all_figures/race_x1sex_facetstate_reorderlarge_to_small_colorgrey20.png}
\QArow{HOLF}{On average, how much longer does it take women from VA to complete the race compared to women from MD?}{0.06}{all_figures/race_x1sex_facetstate_reorderlarge_to_small_colorgrey20.png}
\QArow{HOLF}{Within women, how much longer does it take runners from the state with the longest times to complete the race compared to runners from the state with the fastest times?}{0.06}{all_figures/race_x1sex_facetstate_reorderlarge_to_small_colorgrey20.png}
\QArow{HOLF}{On average, how long does it take women to complete the race?}{1.74}{all_figures/race_x1sex_facetstate_reorderlarge_to_small_colorgrey20.png}
\QArow{HOLF}{On average, how much longer does it take women to complete the race relative to men?}{0.25}{all_figures/race_x1sex_facetstate_reorderlarge_to_small_colorgrey20.png}

\QArow{HOLF-Multi}{Imagine that you came across a random bird strike event involving skies with no clouds in the history books, which involved an aircraft flying at extremely high altitude (17.5K miles). What would you predict its speed to have been at the time of the bird strike?}{273.333}{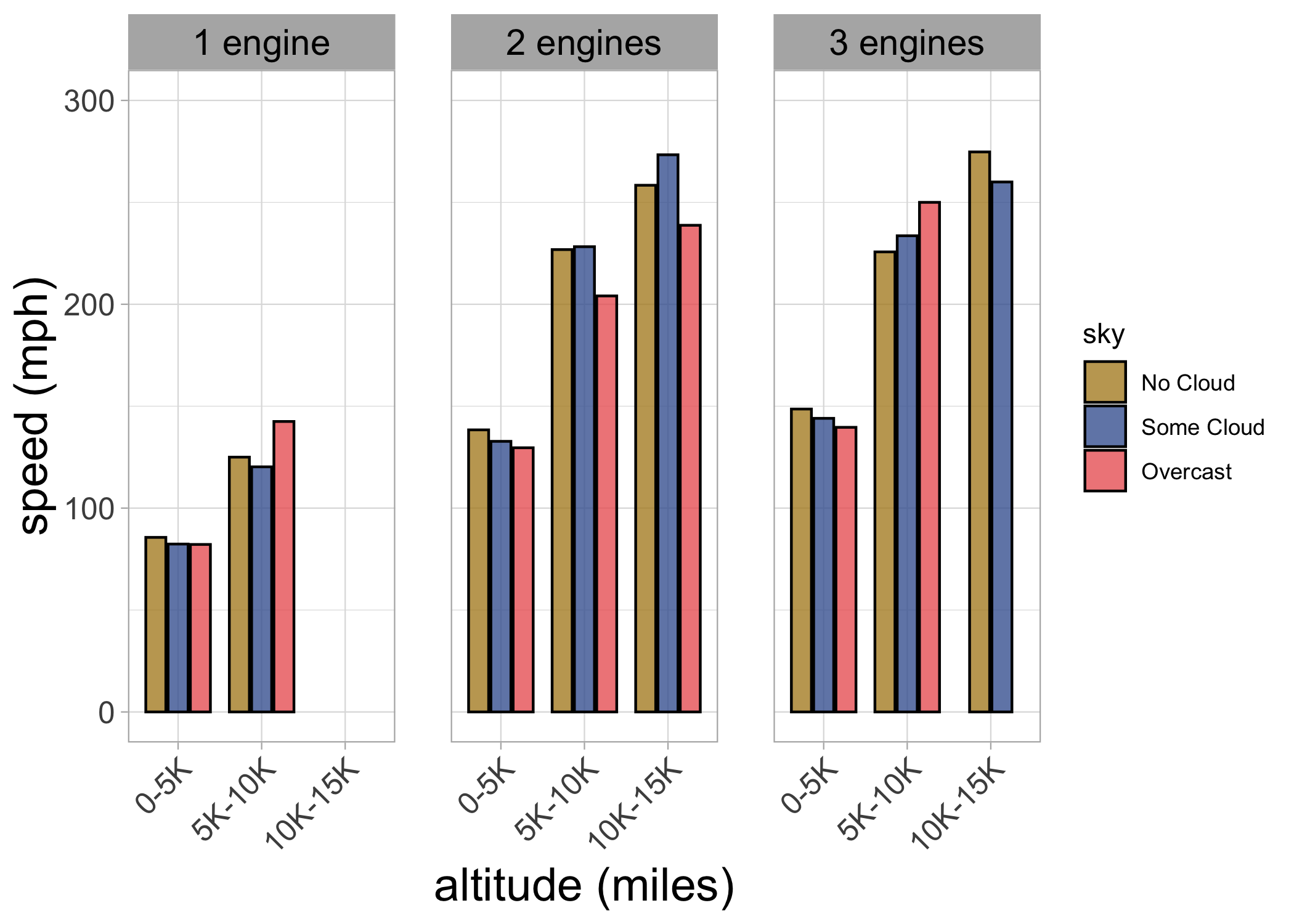}
\QArow{HOLF-Multi}{How much higher is the average speed of aircrafts flying with some cloud skies that encountered bird strikes at 10K-15K miles compared to 5K-10K miles?}{44.476}{all_figures/bar_bird_4_v2heightc_v3sky_v4engine_v5none.png}
\QArow{HOLF-Multi}{What is the average speed of aircrafts flying in overcast skies that encountered bird strikes at 0-5K miles?}{127.338}{all_figures/bar_bird_4_v2heightc_v3sky_v4engine_v5none.png}

\QArow{HOLF-Multi}{What is the average number of fatalities in northeast + midwest states with populations of 1M-1.5M people?}{271.0}{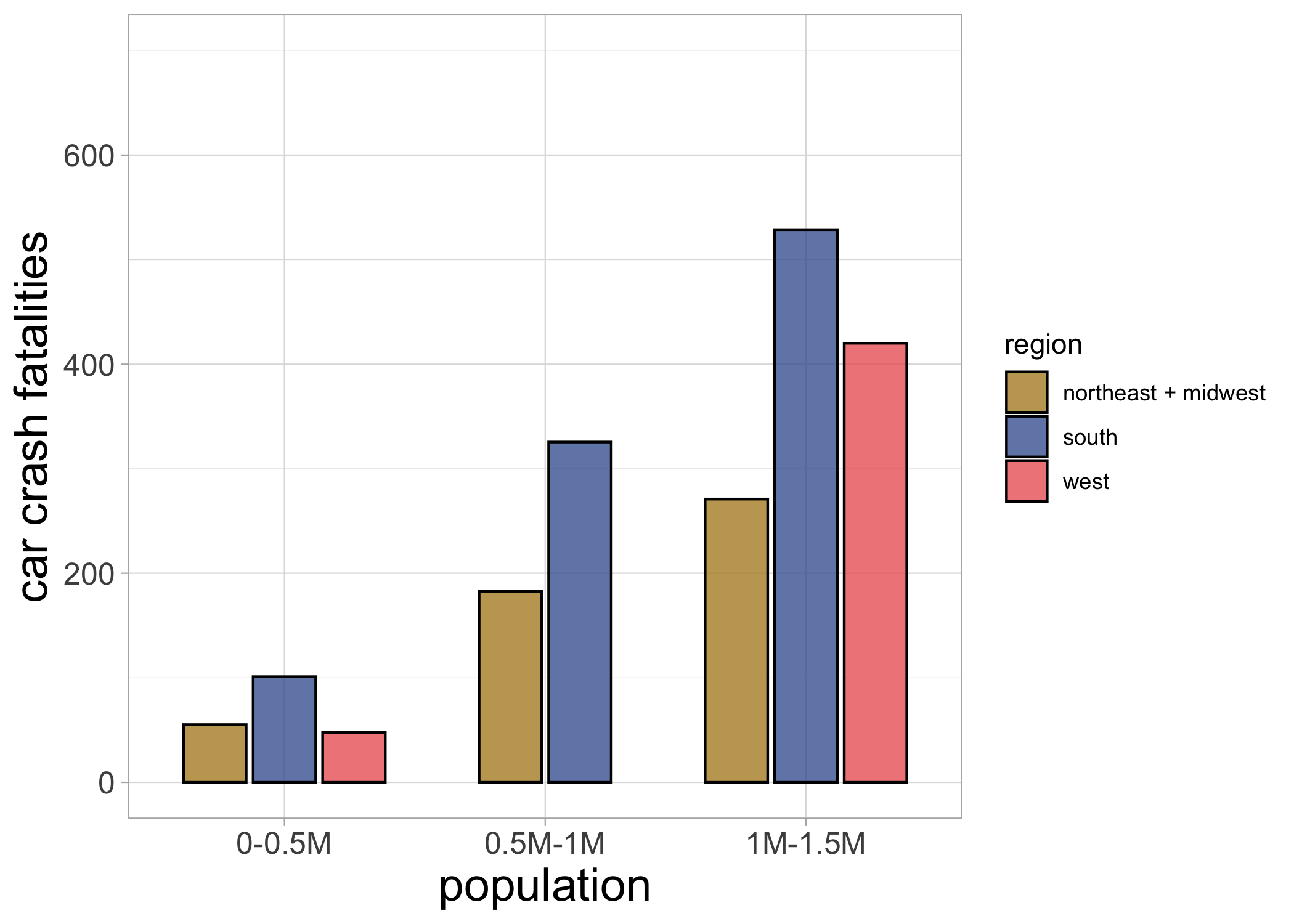}
\QArow{HOLF-Multi}{How much higher is the average number of fatalities in southern states with populations of 0.5M-1M people compared to states with populations of 0-0.5M people?}{224.5}{all_figures/bar_fatal_3_v2popc_v3region_v4none_v5none.png}
\QArow{HOLF-Multi}{Imagine that you came across a random west coast state, which had a population of 1.75M people. What would you predict its number of crash fatalities to be?}{738.286}{all_figures/bar_fatal_3_v2popc_v3region_v4none_v5none.png}

\QArow{HOLF-Multi}{What is the average number of fatalities in northeast + midwest states with populations of 1M-1.5M people?}{271.0}{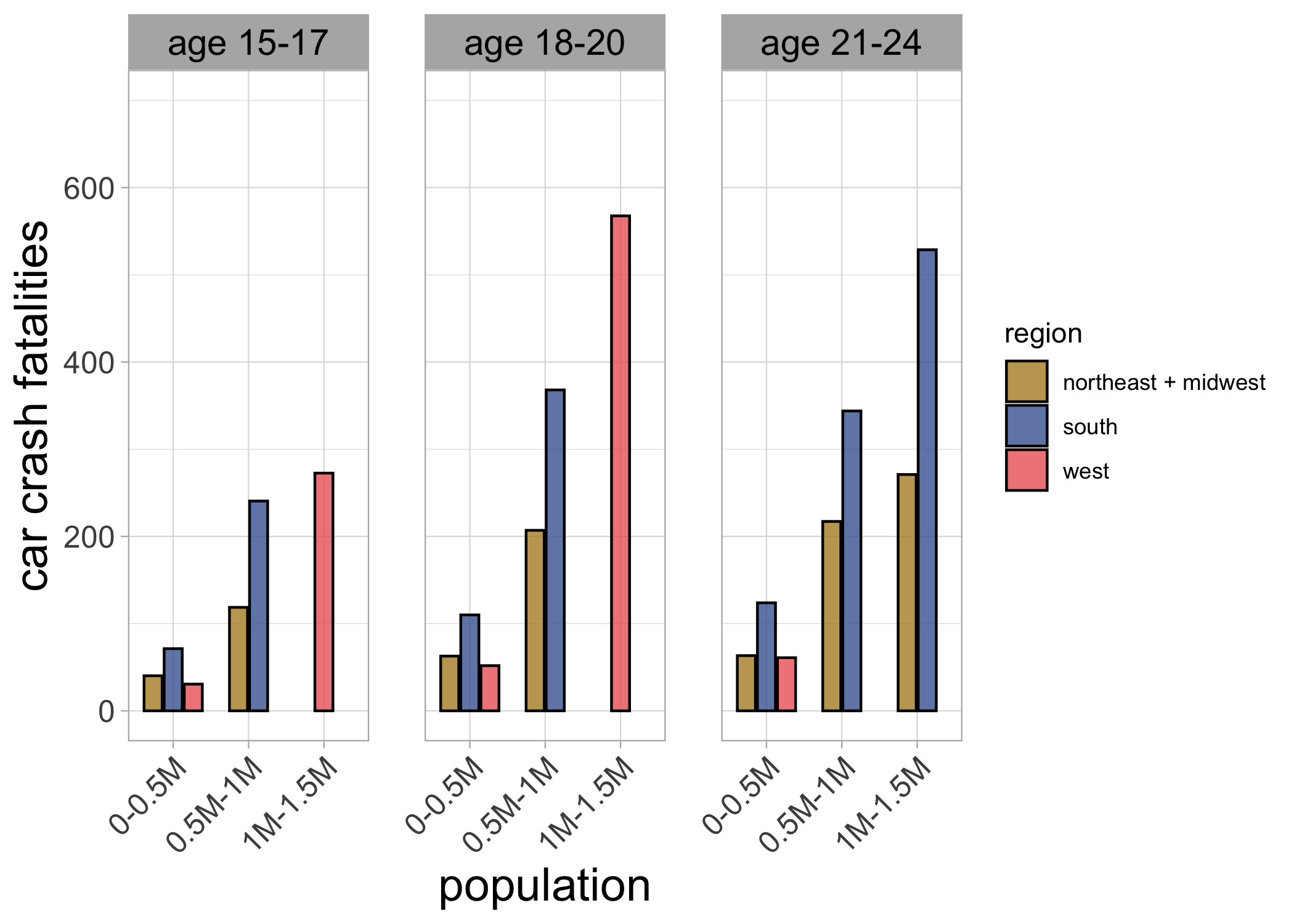}
\QArow{HOLF-Multi}{How much higher is the average number of fatalities in southern states with populations of 0.5M-1M people compared to states with populations of 0-0.5M people?}{224.5}{all_figures/bar_fatal_4_v2popc_v3region_v4age_v5none.png}
\QArow{HOLF-Multi}{Imagine that you came across a random west coast state, which had a population of 1.75M people. What would you predict its number of crash fatalities to be?}{738.286}{all_figures/bar_fatal_4_v2popc_v3region_v4age_v5none.png}

\QArow{HOLF-Multi}{What is the average number of fatalities in northeast + midwest states with populations of 1M-1.5M people?}{271.0}{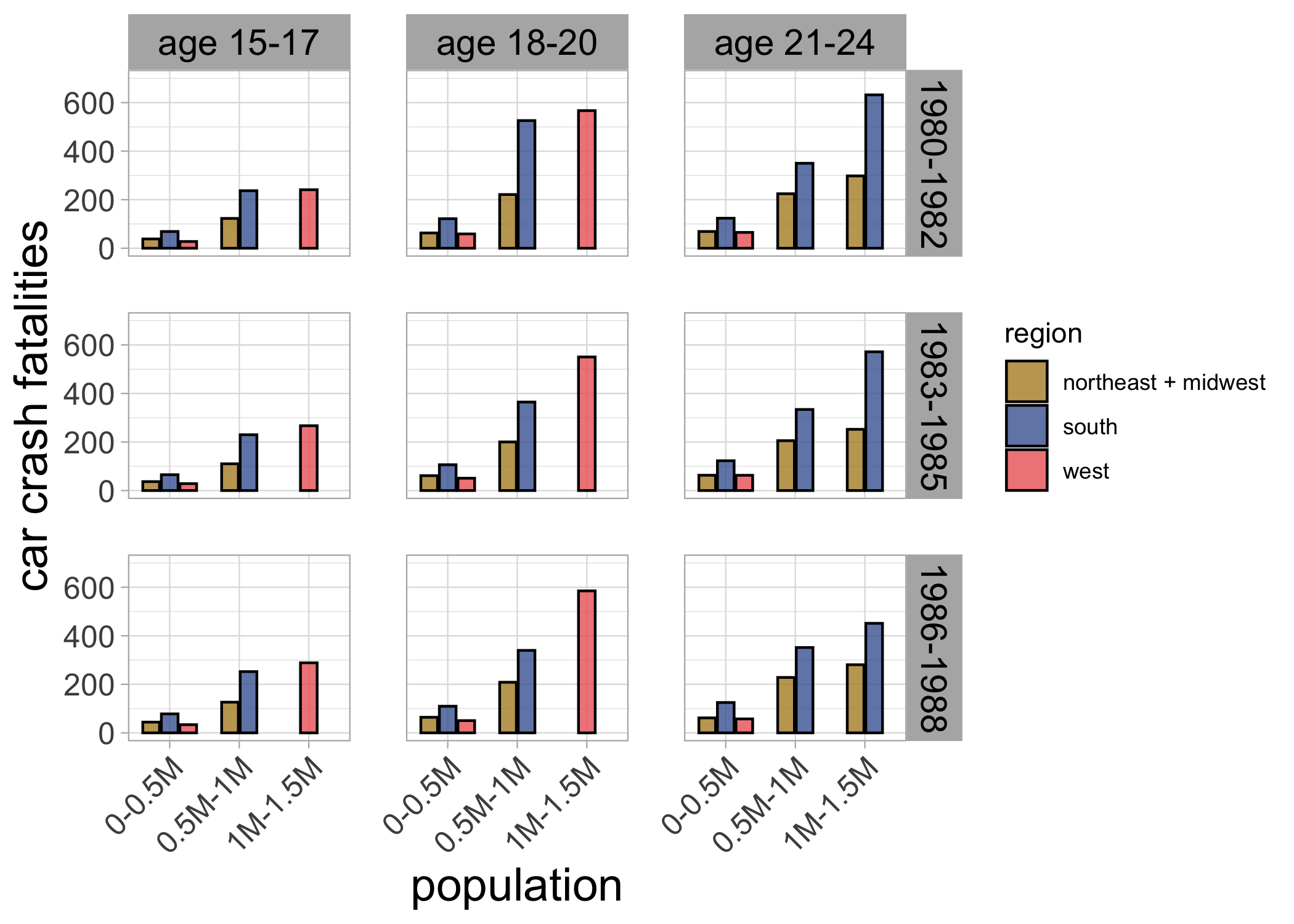}
\QArow{HOLF-Multi}{Imagine that you came across a random west coast state, which had a population of 1.75M people. What would you predict its number of crash fatalities to be?}{738.286}{all_figures/bar_fatal_5_v2popc_v3region_v4age_v5years.png}
\QArow{HOLF-Multi}{How much higher is the average number of fatalities in southern states with populations of 0.5M-1M people compared to states with populations of 0-0.5M people?}{224.5}{all_figures/bar_fatal_5_v2popc_v3region_v4age_v5years.png}

\QArow{HOLF-Multi}{Imagine that you came across a random bird strike event involving skies with no clouds in the history books, which involved an aircraft flying at extremely high altitude (17.5K miles). What would you predict its speed to have been at the time of the bird strike?}{273.333}{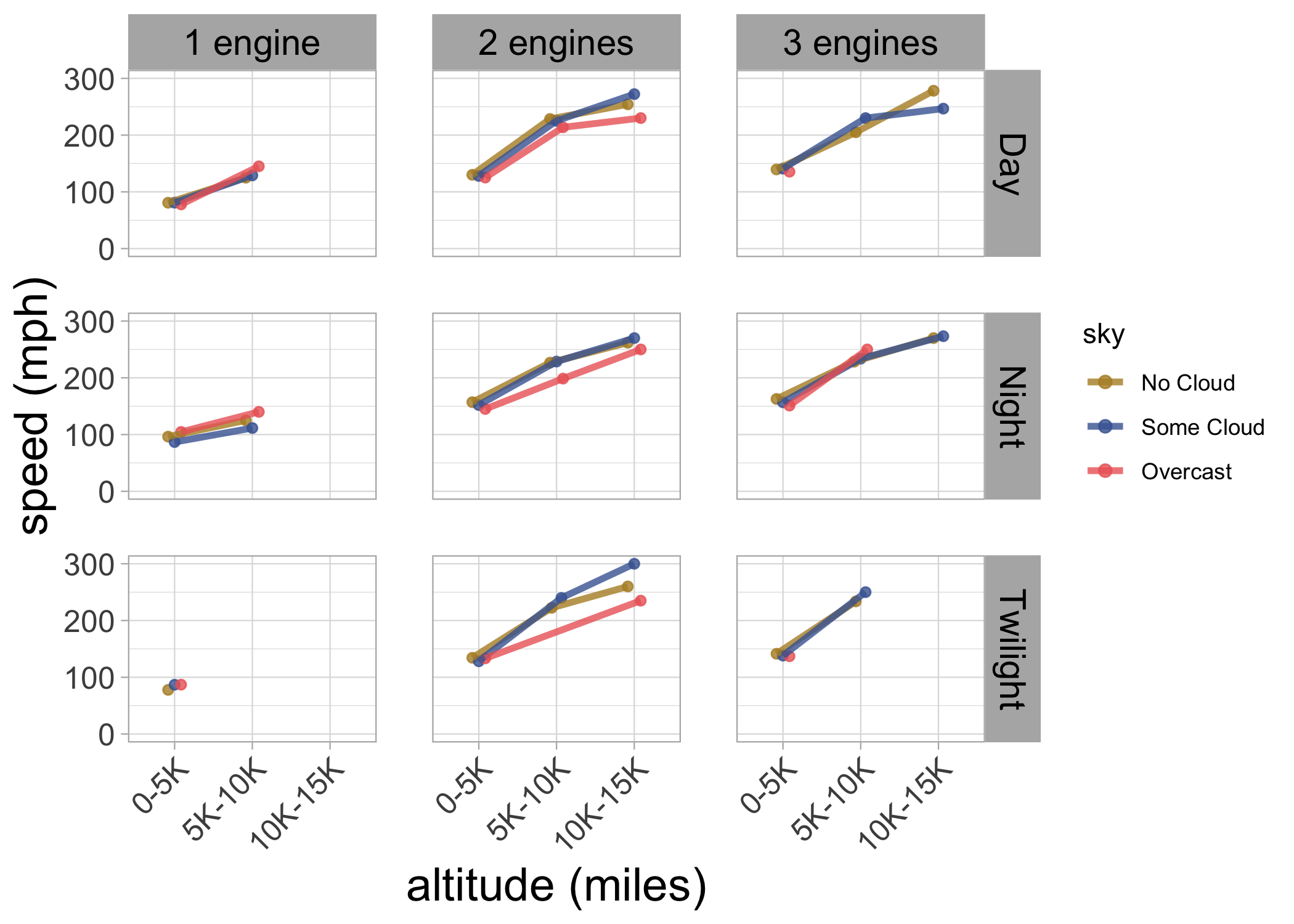}
\QArow{HOLF-Multi}{What is the average speed of aircrafts flying in overcast skies that encountered bird strikes at 0-5K miles?}{127.338}{all_figures/line_bird_5_v2heightc_v3sky_v4engine_v5time.png}
\QArow{HOLF-Multi}{How much higher is the average speed of aircrafts flying with some cloud skies that encountered bird strikes at 10K-15K miles compared to 5K-10K miles?}{44.476}{all_figures/line_bird_5_v2heightc_v3sky_v4engine_v5time.png}

\QArow{HOLF-Multi}{How much higher is the average number of fatalities in southern states with populations of 0.5M-1M people compared to states with populations of 0-0.5M people?}{224.5}{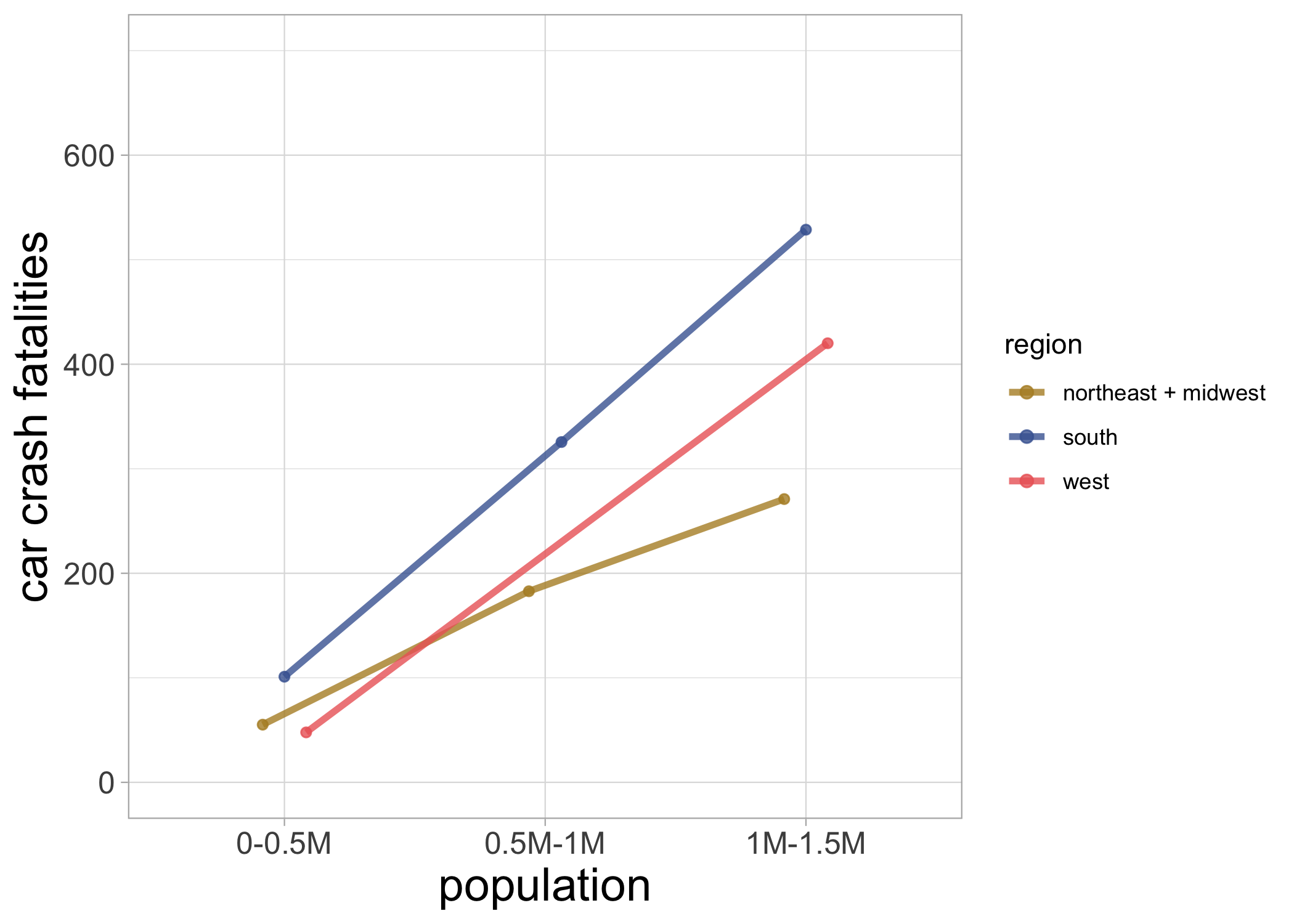}
\QArow{HOLF-Multi}{What is the average number of fatalities in northeast + midwest states with populations of 1M-1.5M people?}{271.0}{all_figures/line_fatal_3_v2popc_v3region_v4none_v5none.png}
\QArow{HOLF-Multi}{Imagine that you came across a random west coast state, which had a population of 1.75M people. What would you predict its number of crash fatalities to be?}{738.286}{all_figures/line_fatal_3_v2popc_v3region_v4none_v5none.png}

\QArow{HOLF-Multi}{How much higher is the average number of fatalities in southern states with populations of 0.5M-1M people compared to states with populations of 0-0.5M people?}{224.5}{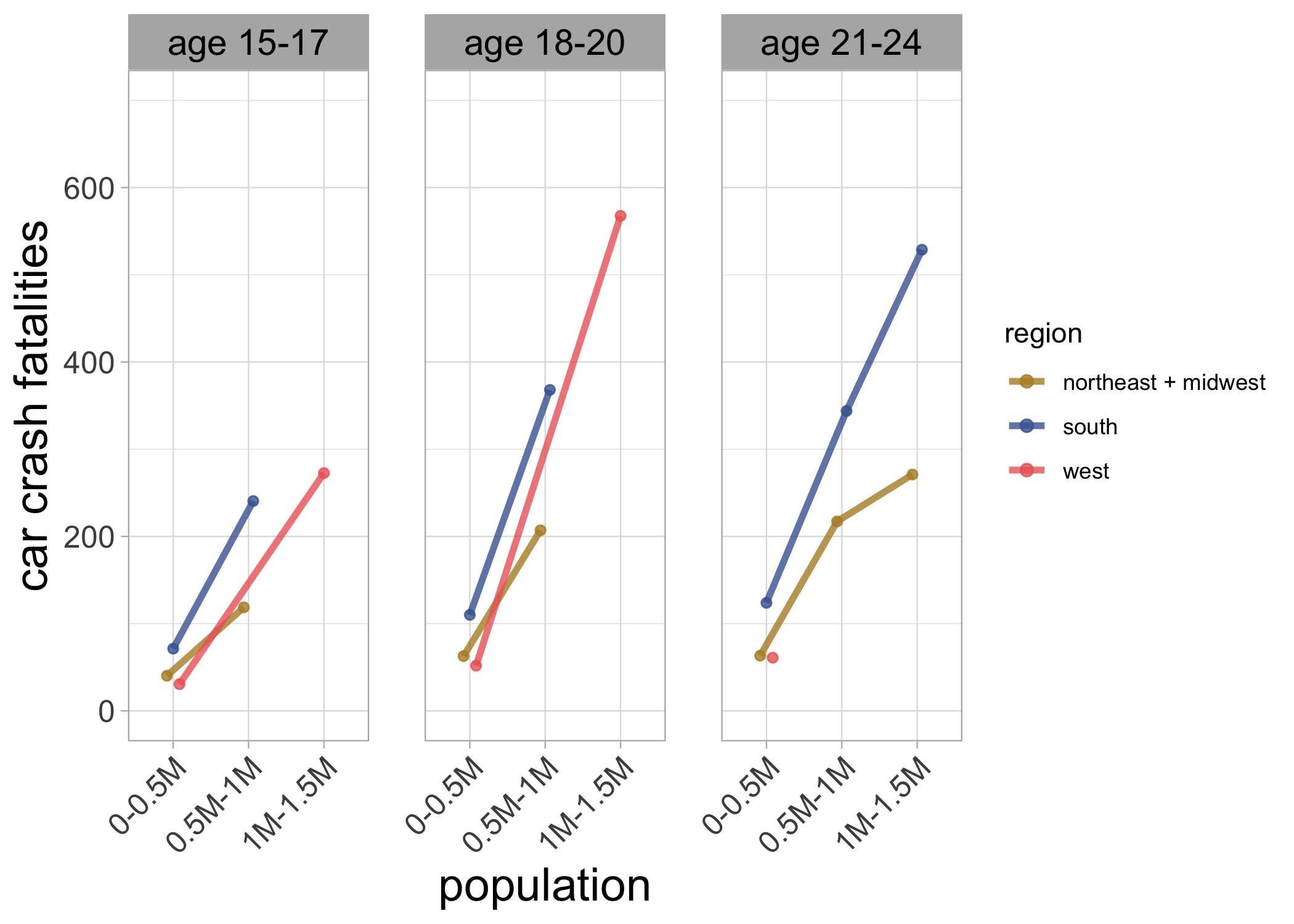}
\QArow{HOLF-Multi}{What is the average number of fatalities in northeast + midwest states with populations of 1M-1.5M people?}{271.0}{all_figures/line_fatal_4_v2popc_v3region_v4age_v5none.png}
\QArow{HOLF-Multi}{Imagine that you came across a random west coast state, which had a population of 1.75M people. What would you predict its number of crash fatalities to be?}{738.286}{all_figures/line_fatal_4_v2popc_v3region_v4age_v5none.png}

\QArow{HOLF-Multi}{How much higher is the average speed of aircrafts flying with some cloud skies that encountered bird strikes at 10K-15K miles compared to 5K-10K miles?}{44.476}{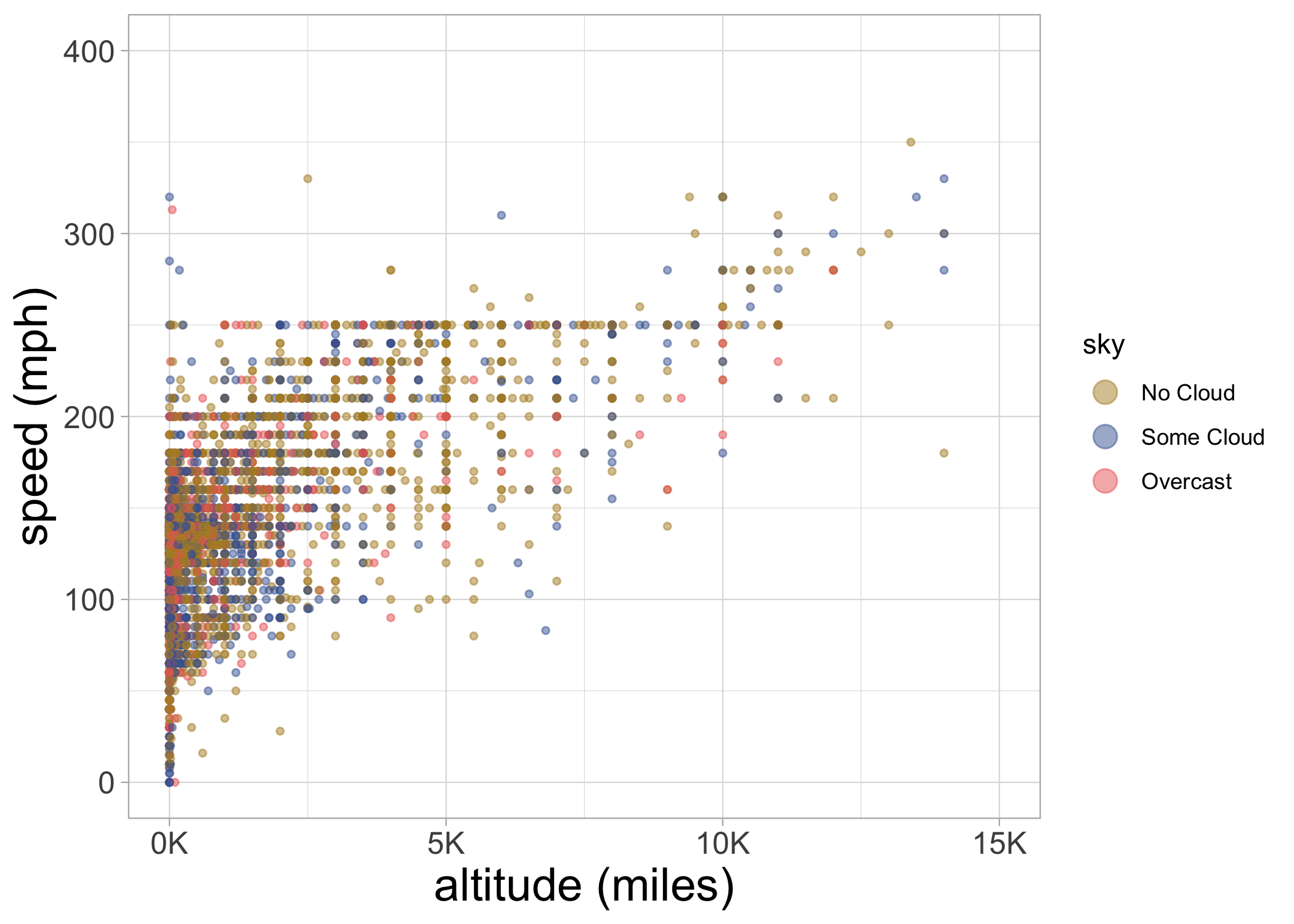}
\QArow{HOLF-Multi}{Imagine that you came across a random bird strike event involving skies with no clouds in the history books, which involved an aircraft flying at extremely high altitude (17.5K miles). What would you predict its speed to have been at the time of the bird strike?}{273.333}{all_figures/scatter_bird_3_v2heightc_v3sky_v4none_v5none.png}
\QArow{HOLF-Multi}{What is the average speed of aircrafts flying in overcast skies that encountered bird strikes at 0-5K miles?}{127.338}{all_figures/scatter_bird_3_v2heightc_v3sky_v4none_v5none.png}

\QArow{HOLF-Multi}{Imagine that you came across a random bird strike event involving skies with no clouds in the history books, which involved an aircraft flying at extremely high altitude (17.5K miles). What would you predict its speed to have been at the time of the bird strike?}{273.333}{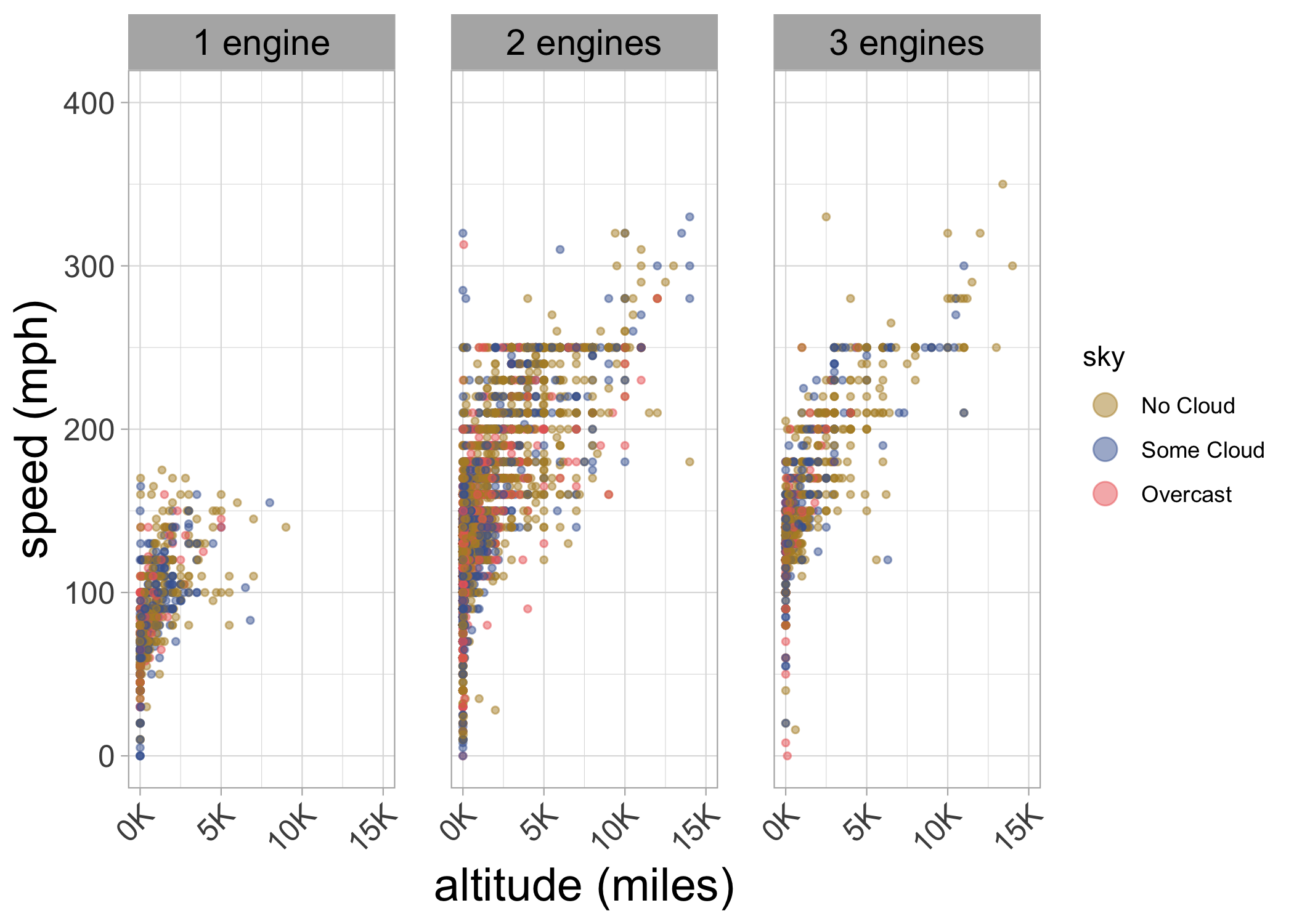}
\QArow{HOLF-Multi}{What is the average speed of aircrafts flying in overcast skies that encountered bird strikes at 0-5K miles?}{127.338}{all_figures/scatter_bird_4_v2heightc_v3sky_v4engine_v5none.png}
\QArow{HOLF-Multi}{How much higher is the average speed of aircrafts flying with some cloud skies that encountered bird strikes at 10K-15K miles compared to 5K-10K miles?}{44.476}{all_figures/scatter_bird_4_v2heightc_v3sky_v4engine_v5none.png}

\QArow{HOLF-Multi}{What is the average speed of aircrafts flying in overcast skies that encountered bird strikes at 0-5K miles?}{127.338}{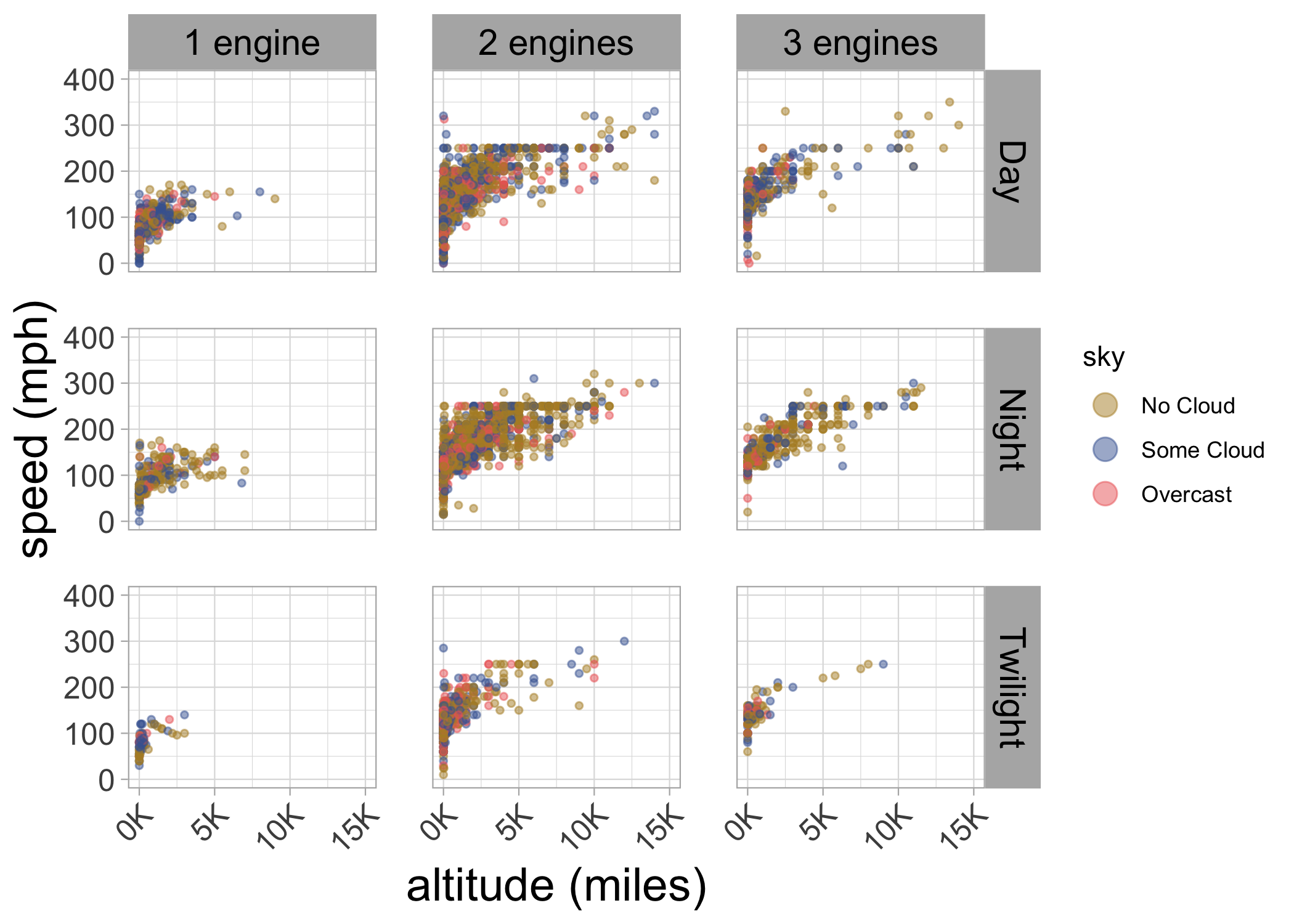}
\QArow{HOLF-Multi}{How much higher is the average speed of aircrafts flying with some cloud skies that encountered bird strikes at 10K-15K miles compared to 5K-10K miles?}{44.476}{all_figures/scatter_bird_5_v2heightc_v3sky_v4engine_v5time.png}
\QArow{HOLF-Multi}{Imagine that you came across a random bird strike event involving skies with no clouds in the history books, which involved an aircraft flying at extremely high altitude (17.5K miles). What would you predict its speed to have been at the time of the bird strike?}{273.333}{all_figures/scatter_bird_5_v2heightc_v3sky_v4engine_v5time.png}

\end{longtable}

\twocolumn

\end{appendices}

\end{document}